\documentclass[final,10pt]{elsarticle}
\usepackage{hyperref}
\usepackage[mathlines,displaymath]{lineno}
\usepackage{graphicx}
\usepackage{subfig}
\usepackage{amsmath,bm}
\usepackage{amsfonts}
\usepackage{amssymb}
\usepackage{amsbsy}
\usepackage{color}
\usepackage{multirow}
\usepackage{framed}
\usepackage{textcomp}
\usepackage{eurosym}
\usepackage{leftidx}
\usepackage{mathtools}
\usepackage{afterpage}
\usepackage{empheq}
\usepackage{booktabs}
\usepackage{enumitem}
\usepackage[top=30mm,left=30mm,body={155mm,210mm}]{geometry}
\usepackage{algorithm,algorithmic}
\usepackage{tikz,pgfplots, pgfplotstable}
\usepackage{mdframed}
\usepackage{newfloat}

\pgfplotsset{compat=1.3}
\usetikzlibrary{external}
\makeatletter
\renewcommand{\ALG@name}{Box}
\makeatother
\DeclareFloatingEnvironment[
  fileext=lob,
  listname={List of Boxes},
  name=Box,
  placement=htp,
]{BOX}
\journal{}
\begin{document}

\bigskip

\begin{frontmatter}

\title{A multiplicative finite strain crystal plasticity formulation based on additive elastic corrector rates: Theory and numerical implementation}

%% or include affiliations in footnotes:.
\author[IMDEAM]{Meijuan Zhang}
\ead{meijuan.zhang@imdea.org}

\author[UPM]{K. Nguyen}
\ead{khanhnguyen.gia@upm.es}

\author[IMDEAM,CAMINOS]{Javier Segurado}
\ead{javier.segurado@upm.es}

\author[UPM,UF]{Francisco J. Mont\'ans \corref{mycorrespondingauthor}}
\ead{fco.montans@upm.es}
\cortext[mycorrespondingauthor]{Corresponding author}

\address[IMDEAM]{IMDEA Materials Institute, Tecnogetafe, Eric Kandel Street, 2, 28906 Getafe, Spain}

\address[UPM]{Escuela T\'ecnica Superior de Ingenier\'ia Aeron\'autica y del Espacio, Universidad Polit\'ecnica de Madrid, Pza. Cardenal Cisneros, 28040 Madrid, Spain}

\address[CAMINOS]{Escuela T\'ecnica Superior de Ingenieros de Caminos, Canales y Puertos, Universidad Polit\'ecnica de Madrid, C/ Prof. Aranguren, 3, 28040 Madrid, Spain}

\address[UF]{Department of Mechanical and Aerospace Engineering, Herbert Wertheim College of Engineering, University of Florida, Gainesville FL 32611, USA}

\begin{abstract}
%\linenumbers

The purpose of continuum plasticity models is to efficiently predict the behavior of structures beyond their elastic limits. The purpose of multiscale materials science models, among them crystal plasticity models,  is to understand the material behavior and design the material for a given target. The current successful continuum hyperelastoplastic models are based in the multiplicative decomposition from crystal plasticity, but significant differences in the computational frameworks of both approaches remain, making comparisons not straightforward.

In previous works we have presented a  theory for multiplicative continuum elastoplasticity which solved many long-standing issues, preserving the appealing structure of additive infinitesimal Wilkins algorithms. In this work we extend the theory to crystal plasticity. We show that the new formulation for crystal plasticity is parallel and comparable in structure to continuum plasticity, preserving the attractive aspects of the  framework: (1) simplicity of the kinematics resulting in  additive strain updates as in the infinitesimal framework; (2) possibility of very large elastic strains and unrestricted type of  hyperelastic behavior; (3) immediate plain backward-Euler algorithmic implementation of the continuum theory avoiding algorithmically motivated  exponential mappings, yet preserving isochoric flow; (4) absence of Mandel-type stresses in the formulation; (5) objectiveness and weak-invariance  by construction due to the use of flow rules in terms of elastic corrector rates. We compare the results of our crystal plasticity formulation with the classical formulation from Kalidindi, Bronkhorst and Anand based on quadratic strains and an exponential mapping update of the plastic deformation gradient.

\end{abstract}

\begin{keyword}
Crystal plasticity, hyperelasticity, elastic rate correctors
\end{keyword}

\end{frontmatter}
%----------------------------------------------------------
%\linenumbers

\section{Introduction}
%\begin{linenumbers}

\label{intro} Plastic deformation is an intrinsic part of the processing and
behavior of metals and alloys, which involve permanent macroscopic changes
to the geometrical shape of components and structures under different
loading types \cite{KhanHuangbook,Kangbook}. The analysis of this
deformation process is performed today mostly through finite elements \cite%
{Bathebook}. In general, the elastoplastic deformation processes and their
numerical simulation are substantially more complicated than the deformation
processes in which only elastic deformations (even when large) are present.
Attending to the purpose of the analysis, the level of detail pursued in
describing the physical process is a balance between accuracy in describing
that physical process and the needed resources. There are two main
approaches to model this behavior: one based on the continuum theory of
plasticity \cite{KhanHuangbook,Kojic2005} (common for engineering structures design)
and the other based on multiscale strategies, among them crystal plasticity
\cite{Rotersbook} (common in materials science and materials design).

Frequently, advances in the continuum theory of plasticity came from
numerical difficulties and inconsistencies found in the numerical
implementation of the theory. For example, hyperelasticity is now
established as the standard approach in continuum elastoplasticity \cite%
{simo2006computational,Kojic2005}.\ It has been popularized in plasticity
due to the complexity of algorithms that incremental objectivity posed in
the formulation; e.g. the Hughes-Winget \cite{hughes1980finite} and
Rolph-Bathe \cite{rolph1984large} algorithms, still available today in many
codes to deal with some anisotropic models. The bypass to these difficulties
came from the proper consideration of the state variables and the fulfillment of the physical
requirement of path independency of the elastic contribution \cite%
{simo1985computational,simo1985unified} (i.e. exact integrability by
fulfilling Bernstein's conditions \cite%
{bernstein1960hypo,bernstein1960relations}). Furthermore, the widely
accepted way to obtain the elastic state variables is currently the use of the
Kr\"oner-Lee \cite{Kroner,Lee1967} multiplicative decomposition, motivated
in crystal plasticity \cite%
{simo1988framework1,simo1988framework2,simo2006computational} (the concepts
behind where introduced previously by Bilby et al \cite{Bilby55}). The
conceptual superiority of the multiplicative decomposition promoted
different formulations which preserved that decomposition in the derivation
of the elastic variables, even for anisotropy and cyclic hardening (e.g.
\cite{simo1985computational,Vladimirov2009,Vladimirov2010,Caminero2011},
among others). This superiority is manifest by the preservation of
ellipticity properties \cite{neff2016loss} and of weak-invariance \cite%
{shutov2014analysis,Shutov2012,Shutov15} during plastic flow (so that simulation results do not depend on the choice of the arbitrary reference configuration), apart from the clear
and well-known physical motivation. Indeed, multiplicative decompositions
and hyperelastic schemes are also important in kinematic (energetic)
hardening to avoid spurious dissipation obtaining stable loops, see e.g.
Figs. 4--9 and 19 in \cite{Brepols2014}. However, Green ansatzes and plastic
metrics are frequently used to obtain the elastic strains (state variable
for hyperelasticity) in complex models, when the multiplicative
decomposition poses mathematical/algorithmic difficulties in traditional
formulations based on the classical plastic flow evolution equations (e.g.
\cite%
{loblein2003application,miehe1998formulation,miehe2002anisotropic,papadopoulos1998general,papadopoulos2001formulation,sansour2003viscoplasticity}%
, among many others).

The exchange of concepts between crystal plasticity
and continuum plasticity continued with the pursue of simplicity in the
computational algorithms, for which the additive structure of infinitesimal
plasticity and of the Wilkins radial return algorithm \cite{Wilkins} seem
optimal. For example, the works of Weber and Anand \cite{weber1990finite}
and Eterovi\'c and Bathe \cite{eterovic1990hyperelastic} in continuum plasticity, brought the
possibility of additive algorithms at finite strains in terms of logarithmic strains preserving the
multiplicative decomposition, thereafter followed by many other authors for
the isotropic case \cite%
{simo1992algorithms,cuitino1992material,SouzaNetoPericBook}, and extended to
the anisotropic elastoplastic case \cite{Caminero2011}. That was possible
due to the use of the exponential mapping in the algorithmic implementation,
which furthermore preserved the volume during plastic flow in a natural way.
Indeed, this conservation was not fulfilled by initial formulations which used flow
rules based on the Lie derivative of the elastic Finger tensor---e.g. Eq.
(9.2.16) in \cite{simo1992associative}; see pp. 385-386 in \cite{SimoChapter}%
. The exponential mapping has been thereafter passed to crystal plasticity
formulations \cite{Miehe1996a}, although not the use of logarithmic strains. The preservation of the additive
structure of algorithms for infinitesimal strains required small or moderate
elastic strains when using the multiplicative decomposition \cite%
{Caminero2011}, both in the continuum plasticity and in the crystal
plasticity formulations \cite{Badreddine2010}. Note that this condition
applies also to trial elastic states, so, in addition, small steps must be
used. In crystal plasticity, since quadratic strains are employed, small elastic strains are often assumed to simplify the algorithms.  Remarkably, the exponential mapping has been introduced and used as an
algorithmic artifact (or ad-hoc algorithmic alternative to the standard
backward-Euler algorithm, despite being motivated in the solution of the
differential equation), not as part of the continuum theory; and as such has
been extended to crystal plasticity; see e.g. Sec. 46 in \cite{SimoChapter},
and how the mapping is introduced in references \cite%
{Caminero2011,weber1990finite,eterovic1990hyperelastic,SouzaNetoPericBook},
among many others.

Departing from the initial works of  Ewing and Rosenhain \cite%
{Ewing1899,Ewing1900} and Taylor and Elam \cite{Taylor1923,Taylor1925},  polycrystalline plasticity was developed \cite{Taylor1938, Bishop1951, Bishop1951a} and was adapted into
the framework of large-strain continuum mechanics by Rice and Hill \cite%
{Rice1971,Hill1972}. Further works extended the framework considering both rate-independent and rate-dependent plasticity \cite{Peirce1982,Peirce1983,Asaro1985,Needleman1985,Rashid1992,Rice1971, Hill1972,Peirce1982,Peirce1983}. Elaborate hardening laws were then proposed in other works, e.g. isotropic
in \cite{Tome1984, Bassani1991}, kinematic in \cite{Meric1991,
Cailletaud1992}, for creep in \cite{Hasija2003, Venkatramani2007}, and
cyclic softening in \cite{Cruzado2017}. Physics-based models rely on the
microscopic physical mechanisms of plastic deformation, e.g. dislocation
densities (considered as internal microstructural state variable), grain
size and shape, second fractions, precipitate morphology, etc. The
dislocation density is considered as the most important variable, and
therefore, it is also treated in many works \cite{Arsenlis2004,Cheong2004,Ma2004,Ma2006,Ma2006a,Dunne2007,
Rodriguez-Galan2015,Shahba2016}. Hence, whereas continuum plasticity and crystal plasticity share most ingredients, there are specific constitutive equations for crystal plasticity mainly related to hardening tied to dislocations density. Indeed, crystal plasticity theory may be
considered a homogenization of the dislocation theory \cite{Berdichevsky,Forhmeister}, where the Burgers vector is $\boldsymbol{b}%
=l_{RVE}\gamma\boldsymbol{s}$, for a Representative Volume Element of
dimension $l_{RVE}$ in the plane direction $\boldsymbol{m}$, and $\gamma$ is
the continuum equivalent slip in direction $\boldsymbol{s}$. Kinematics of
crystal plasticity is of upmost importance in the modelling of both
phenomenological and physics-based micromechanical plastic flow.

Noteworthy, there are many constitutive laws for the slip rates and the
evolution of the internal variables, but few works pay attention to the
development of rigorous numerical implicit implementations directly derived
from kinematics of a continuum theory. As happened in continuum plasticity,
they may unveil a better treatment of both aspects (theory and algorithm).
The initial implicit approaches for rate-dependent formulations, including
efficient and well-posed integration, have been reviewed by Cuiti\~no and
Ortiz \cite{Cuitino1993}. The derivation of a general return-mapping scheme
for rate-independent single crystal models has been proposed by Borja and
Wren \cite{Borja1993} for the infinitesimal theory and by Kalidindi, Bronkhorst and
Anand \cite{Kalidindi1992}, and Miehe \cite{Miehe1996,Miehe1996a} for finite
strains, in which the mentioned exponential mapping has been proposed in the
crystal plasticity context. Currently, one of the better established
frameworks in materials science is the one of Kalidindi, Bronkhorst and Anand \cite%
{Kalidindi1992}. However, whereas some ingredients of the crystal plasticity
theory and its numerical implementation are well established, other
ingredients differ in the literature, which bring different approximations
to simplify either the theoretical or the numerical treatments. For example,
Jaumann rate based formulations are still common (e.g. \cite%
{Kuroda,Izadbakhsh,Zhou19,Sakaguchi,Zecevic}), despite the mentioned issues
regarding integrability. Other works use quadratic stored energies along the
second Piola Kirchhoff stress in the intermediate configuration (e.g. \cite%
{KimKim,Li19,Lu20}, among others), the Mandel stress (e.g. \cite%
{Guo20,KaiserMenzel}), or any other stress tensor under the restriction that
\textit{elastic} strains are small. Whereas this usually holds in metals,
the formulations lack generality. Interestingly, the logarithmic strains advocated for continuum plasticity \cite{Anand79,Anand86} and employed with the exponential mapping \cite{weber1990finite,eterovic1990hyperelastic,simo1992algorithms}, are seldom used in crystal plasticity. Of course, if elastic strains are
considered small, a further simplification is to consider a small strains
framework from the outset, as still employed in many recent works (e.g. \cite%
{Baudoin19,Farooq20}). Indeed, as shown below, this option is also close to
a large strains framework if plastic spin is also considered.

In summary, crystal plasticity and continuum plasticity share most of the
ingredients, and in the pursue of sound, simple and computationally
efficient formulations, developments in one framework have been passed to
the other framework. However, the acceptance of a general fully satisfactory
formulation applicable to both frameworks has still not been achieved. In
view of the above comments, the main aspects that such formulation must take
into account are: (1) both elastic and kinematic hardening behavior must be
exactly integrable (hyperelastic) and general (unrestricted in form); (2)
the continuum theory and the integration algorithm must be objective,
preserve ellipticity properties of elastic energies and be weak-invariant,
which is facilitated if (3) the elastic state variables come from the
multiplicative decomposition and the plastic flow equation is insensitive to
the reference configuration; (4) an implicit computational algorithm must be
conceptually simple, when possible (5) mimicking the additive structure of
the infinitesimal framework; (6) the formulation should not be restricted to
small elastic strains, elastic isotropy, or have any similar limitation; and
(7) both continuum plasticity and crystal plasticity formulations should be parallel, except for specific particularities given by the physics
considered in each approach, like specific flow mechanisms or specific crystal
elastic anisotropy.

Recently we have presented a novel approach to deal with multiplicative flow
kinematics based on elastic strain corrector rates \cite{latorreAMP}.
Motivated in finite strain anisotropic non-equilibrium viscoelasticity \cite%
{latorre2015anisotropic,latorre2016fully}, the approach is based on \textit{%
conventional} flow rules directly written in terms of these rates,
consistently derived from thermodynamics, the dissipation equation, and the
chain rule. Whereas a parallelism is found between these \textit{continuum}
corrector rates and the \textit{algorithmic} strain corrector, the former is
defined as a continuum rate immediately derived from the chain rule. The
plastic flow evolution equation is written directly in terms of this rate
instead of the rate of a plastic measure. The plastic gradient only plays a mapping role when applicable. We have shown that the continuum
theory may be written, in a completely equivalent manner, in terms of any
stress-strain conjugate pair \cite{latorreAMP}; and have shown that for the fully
isotropic case it particularizes to the framework based on the Lie
derivative of the elastic Finger tensor \cite%
{simo1988framework1,simo1988framework2}, but written in a conventional way.
However, when using logarithmic strains, the \textit{plain} backward-Euler
integration algorithm reduces to an additive structure identical to that of
infinitesimal plasticity \cite{Sanz2017}. It does so keeping the multiplicative
decomposition of the deformation gradient, allowing for arbitrarily large
elastic strains, allowing for any isotropic or anisotropic stored energy, and
not needing any approximation for an exponential mapping (which is not
explicitly present since we employ a plain backward-Euler approximation).
Moreover, the plastic spin is completely uncoupled, so no assumption on it
is needed for computing the symmetric part in the continuum plasticity
model. We have extended this approach to model cyclic plasticity at finite
strains without explicitly employing the backstress concept \cite%
{zhang2019simple}, and thereafter developed the (to the authors' knowledge)
first plane stress \textit{projected} algorithm employing the multiplicative
decomposition \cite{nguyen2019}.

The purpose of this work is to extend this novel approach also to crystal
plasticity, showing that a parallel structure to continuum plasticity is
attained, preserving the above-mentioned properties. In particular, the
formulation is not restricted to infinitesimal elastic strains or elastic
isotropy, the exponential mapping is not explicitly employed, and the
integration algorithm has an additive structure similar to the infinitesimal
counterpart. Moreover, the symmetric flow is uncoupled from the skew-symmetric one, which is computed thereafter as an explicit update using the Rodrigues formula. To facilitate comparisons, we summarize in the next section a
typical framework employed in crystal plasticity, namely, the
Kalidindi-Bronkhorst-Anand formulation, in which we also introduce our notation.
Thereafter, in the following sections we introduce the continuum formulation
of our proposal and the implicit integration algorithm, including the
algorithmic tangent. We finish with some examples comparing numerical
results from both approaches and conclusions.
\newpage

\section{Notation}
Most quantities are defined in the text when first used. For fast reference, we summarize here the main notation:
\medskip

\begin{tabular}{cp{1.\textwidth}}
  $t, \Delta t$ & Time and time increment \\
  $^t(\bullet)$ & An object at time $t$ (e.g. a stress or a velocity gradient) \\
$_0^t(\bullet)$ & An object at time $t$ with reference or from time $0$, e.g. a strain from $0$ to $t$ \\
$^{tr}(\bullet)$ & An object referred to a plastic dissipation frozen \\
$^{ct}(\bullet)$ & An object referred to purely internal evolution (external power frozen)
\\
$\otimes$ & Dyadic product, e.g. $[\boldsymbol{a}\otimes\boldsymbol{b}]_{ij}=a_i b_j$ and $[d^2\boldsymbol{T}/(d\boldsymbol{a}\otimes d\boldsymbol{b})]_{ijkl}=d^2T_{ij}/(da_kdb_l) $\\
$\odot$ & Cross-dyadic product, e.g. $[\boldsymbol{A}\odot\boldsymbol{B}]_{ijkl}=A_{ik} B_{jl}$ \\
  $\boldsymbol{I}$ & Second order identity tensor \\
  $_0^t\boldsymbol{X}$ & Deformation gradient from time $0$ to time $t$ \\
  $_0^t\boldsymbol{X}_p$ & Plastic deformation gradient from the Kr\"oner-Lee decomposition of $_0^t\boldsymbol{X} $\\
 $_0^t\boldsymbol{X}_e$ & Elastic deformation gradient from the Kr\"oner-Lee decomposition of $_0^t\boldsymbol{X} $\\
 $_0^t\boldsymbol{U}_e$ & Right elastic stretch tensor from time $0$ to time $t$\\
 $_0^t\boldsymbol{R}_e$ & Elastic rotation tensor (including rigid body motions) \\
  $^t\boldsymbol{l}_p$ & ($=\,_0^t\boldsymbol{\dot X}_p\,_0^t\boldsymbol{X}_p^{-1}$)Plastic velocity gradient in the intermediate configuration\\
  $^t\boldsymbol{l}_e$ & ($=\,_0^t\boldsymbol{\dot X}_e\,_0^t\boldsymbol{X}_e^{-1}$) Elastic velocity gradient in the spatial configuration\\
 $_0^t\boldsymbol{C}_e$ & ($=\,_0^t\boldsymbol{X}_e^T \,_0^t\boldsymbol{X}_e$) Right elastic Cauchy-Green deformation tensor from time $0$ to time $t$
\\
  $_0^t\boldsymbol{A}$ & ($=\tfrac{1}{2}(_0^t\boldsymbol{C}_e-\boldsymbol{I})$) Green-Lagrange deformation tensor from time $0$ to time $t$
\\
  $_0^t\boldsymbol{E}$ & Logarithmic strain tensor in the reference configuration \\
  $_0^t\boldsymbol{E_e}$ & ($=\tfrac{1}{2}\log{_0^t\boldsymbol{C}_e}$) Elastic logarithmic strain tensor in the intermediate configuration
\\
  $\mathcal{P}$ & Stress power
\\
  $\mathcal{D}$ & Plastic disipation
\\
  $\Psi$ & Strain energy
\\
  $^t\boldsymbol{S}$ & Second Piola-Kirchhoff stress tensor at time $t$
\\
  $^t\boldsymbol{S^{|e}}$ & ($=d\Psi/d_0^t\boldsymbol{{A}}_e$) Second Piola-Kirchhoff stress in the intermediate configuration
\\
  $^t\boldsymbol{\Xi}$ & ($=\boldsymbol{C}_e\boldsymbol{S^{|e}}$) Unsymmetric Mandel stress tensor
\\
  $^t\boldsymbol{T^{|e}}$ & ($=d\Psi/d_0^t\boldsymbol{{E}}_e$) Generalized Kirchhoff stress tensor in the intermediate configuration\\
  $\mathbb{C}^{|e}$ & ($=d^2\Psi/(d_0^t\boldsymbol{{A}}_e\otimes d_0^t\boldsymbol{{A}}_e$)) Tensor of elastic constants (quadratic strains)\\
  $^t\mathbb{A}^{|e}$ & ($=d^2\Psi/(d_0^t\boldsymbol{{E}}_e\otimes d_0^t\boldsymbol{{E}}_e$)) Tensor of elastic constants (Logarithmic strains)\\
  $\mathbb{M}|^{\bullet}_\star$ & A mapping tensor for, e.g. performing push-forward or pull-back operations\\
  $\boldsymbol{s}_g$ & Schmidt slip direction for mechanism $g$\\
  $\boldsymbol{m}_g$ & Schmidt slip plane for mechanism $g$  \\
  $\bar\kappa_g$ & ($=\boldsymbol{s}_g\cdot\boldsymbol{\Xi}\cdot%
\boldsymbol{m}_g$) Resolved shear stress for the classical framework for mechanism $g$ \\
 $\dot\gamma_g^p$ & Plastic slip rate for the classical framework for mechanism $g$
\\
  $\tau_g^{|e}$ & (=$\boldsymbol{s}_g\cdot\boldsymbol{T}^{|e}\cdot%
\boldsymbol{m}_g$) Resolved shear stress for the proposed framework for mechanism $g$ \\
 $\dot\gamma_g$ & Plastic slip rate for the proposed framework for mechanism $g$
\\

\end{tabular}\\
\section{Summary of the conventional crystal plasticity formulation}

A large amount of crystal plasticity models used in materials science and
their numerical implementations are based on the Kalidindi-Bornkhorst-Anand framework
\cite{Kalidindi1992} (see also \cite{Kalidindi1993,KalidindiKotari}), or
variations of this approach. The Kalidindi-Bornkhorst-Anand framework is developed
using Green-Lagrange strains in the intermediate configuration and their
associated second Piola-Kirchhoff stresses.

Using the notation of our previous works and e.g. \cite{Bathebook,Kojic2005,DvorkinBook}, we denote the deformation gradient $%
\boldsymbol{X}$ from time $t=0$ to time $t$ as $\,_0^t\boldsymbol{X}$. The
Kr\"oner-Lee multiplicative decomposition of the deformation gradient in
elastic (plus rigid body rotations) and plastic part is
\begin{equation}
\,_0^t\boldsymbol{X}=\,_0^t\boldsymbol{X}_e\,_0^t\boldsymbol{X}_p\;\;\text{
so }\;\,_0^t\boldsymbol{X}_e=\,_0^t\boldsymbol{X}\,_0^t\boldsymbol{X}_p^{-1}
\label{eqLee}
\end{equation}
and the plastic velocity gradient in the intermediate configuration by $%
\,_0^t\boldsymbol{X}_p$ is denoted by $^t\boldsymbol{l}_p=\,_0^t\boldsymbol{%
\dot X}_p\,_0^t\boldsymbol{X}_p^{-1}$. Assuming $G$ possible glide
mechanisms (e.g. 12 in a FCC crystal), from the work of Rice \cite{Rice1971}%
, $\boldsymbol{l}_p$ is assumed as the sum of the contribution of each
mechanism
\begin{equation}
\boldsymbol{l}_{p}\simeq\sum_{g=1}^{G}\boldsymbol{l}_{pg}=\sum_{g=1}^{G}\dot{%
\gamma}_{g}^p\boldsymbol{s}_{g}\otimes \boldsymbol{m}_{g}  \label{lpadditive}
\end{equation}
where $\gamma_g^p$ is the glide amount of the crystal plane $\boldsymbol{m}_g$
(e.g. 4 planes in a FCC crystal) in direction $\boldsymbol{s}_g$ (one of the
3 directions per plane in the FCC crystal), and $\boldsymbol{s}_g\cdot
\boldsymbol{m}_g = 0$ and $|\boldsymbol{s}_g|=|\boldsymbol{m}_g|=1$ in the
intermediate configuration. Introduced by Weber and Anand in the context of
continuum mechanics, and by Kalidindi, Bronkhorst and Anand in the context of crystal
plasticity, most current models use an algorithmic (exponential map) update
motivated in the solution of the differential equation $\boldsymbol{\dot X}_p%
\boldsymbol{X}_p^{-1}=\boldsymbol{l}_p$ for $\boldsymbol{l}_p$ constant, as

\begin{equation}
_{\hspace{3.2ex}0}^{t+\Delta t}\boldsymbol{X}_p\equiv \,_{\hspace{3.2ex}%
t}^{t+\Delta t}\boldsymbol{X}_p\,_{0}^{t}\boldsymbol{X}_p=\exp (^{t+\Delta t}%
\boldsymbol{l}_{p}\Delta t)\text{ }_{0}^{t}\boldsymbol{X}_{p}\text{ }\simeq
\boldsymbol{(I}+\,^{t+\Delta t}\boldsymbol{l}_{p}\Delta t)\text{ }_{0}^{t}%
\boldsymbol{X}_{p}  \label{eqexpupdt}
\end{equation}
where the last approximation used, e.g. by Eterovi\'c and Bathe \cite%
{eterovic1990hyperelastic} (see also \cite{Kalidindi1992}), holds for {small
steps} ($||\,^{t+\Delta t}\boldsymbol{l}_{p}\Delta t||<<1$); a condition
which in any case should be satisfied for accuracy reasons. For computing the
stresses from a hyperelastic relation we need the elastic state variables,
which in these crystal plasticity  models (as a difference with the continuum isotropic ones which use logarithmic strains), is the right elastic Cauchy-Green deformation tensor
(or alternatively the Finger tensor). This tensor is obtained from Eqs. %
\eqref{eqLee}, \eqref{eqexpupdt} as

\begin{equation}
_{\hspace{3.2ex}0}^{t+\Delta t}\boldsymbol{C}_{e}=\,_{\hspace{3.2ex}%
0}^{t+\Delta t}\boldsymbol{X}_{e}^{T}\text{ }_{\hspace{3.2ex}0}^{t+\Delta t}%
\boldsymbol{X}_{e}\simeq (\boldsymbol{I}-\,^{t+\Delta t}\boldsymbol{l}%
_{p}^{T}\Delta t)\text{ }\,^{tr}\boldsymbol{C}_e\text{ }(\boldsymbol{I}%
-\,^{t+\Delta t}\boldsymbol{l}_{p}\Delta t)
\end{equation}
with $\,^{tr}\boldsymbol{X}_e:=\,_{\hspace{3.2ex}0}^{t+\Delta t}\boldsymbol{X%
}\text{ }_{0}^{t}\boldsymbol{X}_{p}^{-1}$ defined as the trial elastic
deformation gradient and $\,^{tr}\boldsymbol{C}_e:=\,^{tr}\boldsymbol{X}_e^T
\,^{tr}\boldsymbol{X}_e$ defined as the trial elastic right Cauchy-Green deformation
tensor. Note that all these definitions are performed in an algorithmic
setting, just as a specific integration framework, without a direct link to
a continuum theory. A further approximation often used to simplify some
algorithm settings is to neglect once more higher order terms with the
condition that the time step size is small,
so
\begin{eqnarray}
_{\hspace{3.2ex}0}^{t+\Delta t}\boldsymbol{C}_{e} &\simeq&^{tr}\boldsymbol{C}%
_{e}-(^{tr}\boldsymbol{C}_{e}\text{ }^{t+\Delta t}\boldsymbol{l}_{p}+\text{ }%
^{t+\Delta t}\boldsymbol{l}_{p}^{T\text{ }tr}\boldsymbol{C}_{e})\Delta t+%
\text{ }^{t+\Delta t}\boldsymbol{l}_{p}^{T\text{ }tr}\boldsymbol{C}_{e}\text{
}^{t+\Delta t}\boldsymbol{l}_{p}\Delta t^{2}  \notag \\
&\simeq &^{tr}\boldsymbol{C}_{e}-(^{tr}\boldsymbol{C}_{e}\text{ }^{t+\Delta
t}\boldsymbol{l}_{p}+\text{ }^{t+\Delta t}\boldsymbol{l}_{p}^{T\text{ }tr}%
\boldsymbol{C}_{e})\Delta t  \notag
\end{eqnarray}
The elastic Green-Lagrange strains in the intermediate configuration
(pull-back of the Almansi strains by $^{t+\Delta t}\boldsymbol{X}_e$) are
obtained immediately from this tensor as
\begin{equation}
_{\hspace{3.2ex}0}^{t+\Delta t}\boldsymbol{A}_e:=\tfrac{1}{2}\left( _{%
\hspace{3.2ex}0}^{t+\Delta t}\boldsymbol{C}_e-\boldsymbol{I}\right)
\end{equation}
The hyperelastic relation in these models is restricted to quadratic forms
of the type $\Psi(\boldsymbol{A}_e,...)=\tfrac{1}{2}\boldsymbol{A}_e:\mathbb{%
C}^{|e}:\boldsymbol{A}_e$ where by the ellipsis we indicate crystal symmetry
group information contained in the tensor of \textit{constants} $\mathbb{C}%
^{|e}$. Then, the second Piola-Kirchhoff stress tensor in the intermediate
configuration is (evaluated at the trial or at the final state)%
\begin{equation}
_{\hspace{3.2ex}}^{t+\Delta t}\boldsymbol{S}^{|e}=\mathbb{C}^{|e}:\,_{%
\hspace{3.2ex}0}^{t+\Delta t}\boldsymbol{A}_{e} \;\;\text{ and }\;\;^{tr}%
\boldsymbol{S}^{|e}=\mathbb{C}^{|e}:\,^{tr}\boldsymbol{A}_{e}
\end{equation}
The next important issue is the computation of the Schmid resolved stress
for each mechanism. Obviously, it can be defined in any configuration and
using any stress measure, because there is a direct transformation between
them. However, since the plane and direction are orthonormal in the
intermediate configuration, it seems logical to define a Schmidt stress $%
\bar\kappa_g$ as the work conjugate to the slip rate  $\dot\gamma^p_g$, so the dissipated
power, using the classical expression in terms of the \textit{unsymmetric}
Mandel stress tensor $\boldsymbol{\Xi}:=\boldsymbol{C}_e\boldsymbol{S}^{|e}$
and the plastic velocity gradient $\boldsymbol{l}_p$, is ---e.g. cf. Eq.
(42) in \cite{KalidindiKotari}
\begin{equation}
\mathcal{D}_{g}\equiv\boldsymbol{\Xi}:\boldsymbol{l}_{pg}=\underbrace{%
\boldsymbol{s}_g\cdot\boldsymbol{C}_e\boldsymbol{S}^{|e}\cdot\boldsymbol{m}_g%
}_{\bar \kappa_g}\dot\gamma_g^p:=\bar\kappa_g\dot\gamma_g^p
\end{equation}
Note that $\boldsymbol{s}_g\cdot\boldsymbol{C}_e\boldsymbol{S}^{|e}\cdot%
\boldsymbol{m}_g\ne \boldsymbol{m}_g\cdot\boldsymbol{C}_e\boldsymbol{S}%
^{|e}\cdot\boldsymbol{s}_g$ because the Mandel stress tensor is unsymmetric (%
$\boldsymbol{\Xi}\ne\boldsymbol{\Xi}^T$), but it is the work conjugate of $%
\boldsymbol{l}_p$ in the dissipation equation. However, given the elusive
interpretation of $\boldsymbol{\Xi}$, some approximations are often employed
considering that elastic strains are typically infinitesimal, e.g.
\begin{equation}
\boldsymbol{s}_g\cdot\boldsymbol{C}_e\boldsymbol{S}^{|e}\cdot\boldsymbol{m}%
_g\simeq\boldsymbol{s}_g\cdot\boldsymbol{S}^{|e}\cdot\boldsymbol{m}_g\simeq
\boldsymbol{s}_g\cdot\boldsymbol{\tau}_R\cdot\boldsymbol{m}_g\simeq
\boldsymbol{s}_g\cdot\boldsymbol{\sigma}_R\cdot\boldsymbol{m}_g
\end{equation}
where $\boldsymbol{\tau}_R$ are the rotated Kirchhoff stresses and $%
\boldsymbol{\sigma}_R$ are the rotated Cauchy stresses, i.e. $\boldsymbol{%
\sigma}_R=\boldsymbol{R}_e^T\boldsymbol{\sigma}\boldsymbol{R}_e$, where $%
\boldsymbol{\sigma}$ are the Cauchy stresses and $\boldsymbol{R}_e$ are the
rotations from the polar decomposition of the elastic deformation tensor: $%
\boldsymbol{X}_e=\boldsymbol{R}_e\boldsymbol{U}_e$, with $\boldsymbol{U}_e=+%
\sqrt{\boldsymbol{C}_e}$ being the elastic right stretch tensor. The specific
algorithm depends on the authors, and whether it is semi-implicit, fully
implicit, rate-dependent or rate-independent. However, a typical scheme from
the finite element displacements field $\boldsymbol{u}(\boldsymbol{x})$ at the integration point
of a finite element is (of course an iteration loop is needed for implicit
schemes)
\begin{equation}
\,_{\hspace{3.2ex}0}^{t+\Delta t}\boldsymbol{u}\rightarrow\,_{\hspace{3.2ex}0}^{t+\Delta t}%
\boldsymbol{X}\rightarrow \,^{tr}\boldsymbol{X}_e\rightarrow\,^{tr}%
\boldsymbol{S}^{|e}\rightarrow \,^{tr}\bar
\kappa_g\rightarrow\Delta\gamma^p_g\rightarrow \Delta t\boldsymbol{l}%
_p\rightarrow \,_{\hspace{3.2ex}0}^{t+\Delta t}\boldsymbol{X}%
_p^{-1}\rightarrow\{\,_{\hspace{3.2ex}0}^{t+\Delta t}\boldsymbol{X}%
_e,\,_{\hspace{3.2ex}0}^{t+\Delta t}\boldsymbol{A}%
_e,\,^{t+\Delta t}{\boldsymbol{S}^{|e}} \}
\end{equation}
with $\,_{\hspace{3.2ex}0}^{t+\Delta t}%
\boldsymbol{X}=d^{t+\Delta t}\boldsymbol{x}/d^0\boldsymbol{x}$ and $\,^{t+\Delta t}\boldsymbol{x}=\,^0\boldsymbol{x}+\,_{\hspace{3.2ex}0}^{t+\Delta t}\boldsymbol{u}$ are the updated coordinates. Note that this is an \textit{algorithmic} implementation of the material model motivated by the dissipation inequality with Mandel stress and plastic velocity gradient so
\begin{equation}
\mathcal{D}=\boldsymbol{\Xi}:\boldsymbol{l}_p\simeq\sum_{g=1}^{G}\boldsymbol{\Xi}:\boldsymbol{l}_{pg}=\sum_{g=1}^{G}\mathcal{D}%
_g\ge 0
\label{disipation1}
\end{equation}
The purpose of the present formulation is to depart from a more convenient
form of the dissipation equation, avoiding many of the algorithmic
approximations and arriving to an algorithmic implementation which is a
direct application of the plain Backward-Euler scheme to the continuum
theory. Furthermore, as in the continuum plasticity framework, the algorithm
will have the structure of a small strains algorithm, to which explicit
kinematic mappings are applied for the large strain case.

\section{Proposed framework: Continuum formulation}

\subsection{Kinematics of plastic deformation}

The stress power $\mathcal{P}$ may be split into a conservative part $\dot{%
\Psi}$ and a dissipative part $\mathcal{D}$ as
\begin{equation}
\mathcal{P}=\dot{\Psi}+\mathcal{D}
\label{eq11}\end{equation}%
In order to take natural advantage of this split, in our formulation we
consider the elastic gradient $\boldsymbol{X}_{e}$ as an internal \textit{variable of
state} which defines the intermediate stress-free local (incompatible)
configuration, meaning that only the current value is relevant to the state
of the solid. The variable of state $\boldsymbol{X}_{e}$,
considered as \textit{dependent}, may be written in terms of the \textit{%
independent} (driving) variables $\boldsymbol{X}$ and $\boldsymbol{X}_{p}$
as $\boldsymbol{X}_{e}\left( \boldsymbol{X},\boldsymbol{X}_{p}\right) $.
Note that by definition if $\boldsymbol{\dot X}_p=\boldsymbol{0}$, then $%
\boldsymbol{l}_p=\boldsymbol{0}$ and the dissipation $\mathcal{D}=0$
(regardless of the value of $\mathcal{P}=\dot{\Psi}$); and if $\boldsymbol{%
\dot X}=\boldsymbol{0}$, then the velocity gradient vanishes and the power $%
\mathcal{P}=0$ (regardless of the value of $\mathcal{D}=-\dot{\Psi}$), Then,
by immediate use of the chain rule, the rate is

\begin{equation}
\boldsymbol{\dot{X}}_{e}\left( \boldsymbol{X},\boldsymbol{X}_{p}\right)=%
\underbrace{\left. \frac{\partial\boldsymbol{X}_{e}}{\partial\boldsymbol{X}}%
\right|_{\boldsymbol{\dot X}_p=\boldsymbol{0}} :\boldsymbol{\dot{X}}}_{\text{%
conservative; external input}}+\underbrace{\left. \frac{\partial\boldsymbol{X%
}_{e}}{\partial \boldsymbol{X}_{p}}\right|_{\boldsymbol{\dot X}=\boldsymbol{0%
}} :\boldsymbol{\dot{X}}_{p}}_{\text{dissipative; internal evolution}%
}=:\,^{tr}\boldsymbol{\dot{X}}_{e}+\,^{ct}\boldsymbol{\dot{X}}_{e}
\end{equation}%
where the first addend $~^{tr}\boldsymbol{\dot{X}}_{e}$ represents the
partial \textit{continuum} rate of $\boldsymbol{X}_{e}$ when the plastic
flow (dissipation) is frozen (hence we name trial rate for similarity with
the algorithmic concept). The second addend $~^{ct}\boldsymbol{\dot{X}}_{e}$
represents the partial derivative of $\boldsymbol{X}_{e}$ when the external
deformation (i.e. $\boldsymbol{X}$) is frozen, so no external power input
takes place and there is just an internal evolution dissipating stored
energy. These concepts are typically employed in predictor-corrector
algorithms of computational plasticity, but we remark herein that we are
still dealing with the \textit{continuum} formulation; they are just the
partial derivatives when considering the explicit dependencies in $%
\boldsymbol{X}_{e}\left( \boldsymbol{X},\boldsymbol{X}_{p}\right) $. In
other words, they represent \textit{at every instant } the conservative and
dissipative fractions in $\boldsymbol{\dot{X}}_e$.

These concepts may be applied to all kinematic quantities. For example, the
spatial velocity gradient can be written as%
\begin{equation}
\boldsymbol{l} =\boldsymbol{\dot{X}X}^{-1}=\boldsymbol{\dot{X}}_{e}%
\boldsymbol{X}_{e}^{-1}+\boldsymbol{X}_{e}\left( \boldsymbol{\dot{X}}_{p}%
\boldsymbol{X}_{p}^{-1}\right) \boldsymbol{X}_{e}^{-1} =\boldsymbol{l}_{e}+%
\boldsymbol{X}_{e}\boldsymbol{l}_{p}\boldsymbol{X}_{e}^{-1}=\boldsymbol{l}%
_{e}+\left( \boldsymbol{X}_{e}\odot \boldsymbol{X}_{e}^{-T}\right) :%
\boldsymbol{l}_{p}  \label{l=le+lp}
\end{equation}%
so the two contributions to the elastic velocity gradient $\boldsymbol{l}%
_{e}\left( \boldsymbol{l},\boldsymbol{l}_{p}\right) $ are
\begin{equation}
\boldsymbol{l}_e=\boldsymbol{l}-\left( \boldsymbol{X}_{e}\odot \boldsymbol{X}%
_{e}^{-T}\right) :\boldsymbol{l}_{p}  \label{l=le-lp}
\end{equation}
with $\left[ \boldsymbol{Y}\odot \boldsymbol{Z}\right] _{ijkl}=Y_{ik}Z_{jl}$%
. Then, we can define a purely geometric mapping tensor  from
the elastic deformation gradient as $\left. \mathbb{M}_{l_{p}}^{l_e}\right|_{%
\boldsymbol{l}=\boldsymbol{0}}:=-\left( \boldsymbol{X}_{e}\odot \boldsymbol{X%
}_{e}^{-T}\right)$, so ---see details on this type of formalism in continuum
mechanics in \cite{latorre2016stress}
\begin{equation}
\boldsymbol{l}_{e}\left( \boldsymbol{l},\boldsymbol{l}_{p}\right) =%
\boldsymbol{l}+\left. \mathbb{M}_{l_{p}}^{l_{e}}\right|_{\boldsymbol{l}=%
\boldsymbol{0}}:\boldsymbol{l}_{p}=~\left. \boldsymbol{l}_{e}\right|_{%
\boldsymbol{l}_p=\boldsymbol{0}}+~\left. \boldsymbol{l}_{e}\right|_{%
\boldsymbol{l}=\boldsymbol{0}}=:\,^{tr}\boldsymbol{l}_{e}+\,^{ct}\boldsymbol{%
l}_{e}
\end{equation}%
where $^{tr}\left( \bullet \right) $ and $^{ct}\left( \bullet \right) $
refer again to trial (i.e. the rate with $\boldsymbol{\dot{X}}_{p}=%
\boldsymbol{l}_{p}=\boldsymbol{0}$) and corrector (i.e. the rate with $%
\boldsymbol{\dot{X}}=\boldsymbol{l}=\boldsymbol{0}$) contributions. The
tensor $\left. \mathbb{M}_{l_{p}}^{l_{e}}\right|_{\boldsymbol{l}=\boldsymbol{%
0}}$ performs the push-forward of the plastic velocity gradient $\boldsymbol{%
l}_{p}:=\mathbf{\dot{X}}_{p}\mathbf{X}_{p}^{-1}$, lying in the intermediate
configuration, to the spatial one where $\boldsymbol{l}_e$ and $\boldsymbol{l%
}$ live (so they can be added). The symmetric (strain rate) and
skew-symmetric (spin) parts are, respectively%
\begin{align}
\boldsymbol{d}_{e}\left( \boldsymbol{d},\boldsymbol{l}_{p}\right) &=%
\boldsymbol{d}-\mathsf{sym}\left( \boldsymbol{X}_{e}\boldsymbol{l}_{p}%
\boldsymbol{X}_{e}^{-1}\right) =:\,^{tr}\boldsymbol{d}_{e}+\,^{ct}%
\boldsymbol{d}_{e} \\
\boldsymbol{w}_{e}\left( \boldsymbol{w},\boldsymbol{l}_{p}\right) &=%
\boldsymbol{w}-\mathsf{skw}\left( \boldsymbol{X}_{e}\boldsymbol{l}_{p}%
\boldsymbol{X}_{e}^{-1}\right) =:\,^{tr}\boldsymbol{w}_{e}+\,^{ct}%
\boldsymbol{w}_{e}
\end{align}
with, for example
\begin{equation}
\,^{tr}\boldsymbol{d}_{e}=\left. \mathbb{M}_{d}^{d_{e}}\right|_{\boldsymbol{d%
}_p=\boldsymbol{0}}:\boldsymbol{d}=\mathbb{I}^{S}:\boldsymbol{d}=\boldsymbol{%
d}
\end{equation}
where $\mathbb{I}^{S}$ is the fully symmetric identity tensor ($\left.
\mathbb{M}_{d}^{d_{e}}\right|_{\boldsymbol{d}_p=\boldsymbol{0}}=\mathbb{I}%
^{s}$ because both $~^{tr}\boldsymbol{d}_{e}$ and $\boldsymbol{d}$ lie in
the same spatial configuration); and%
\begin{equation}
^{ct}\mathbf{d}_{e}=\left. \mathbb{M}_{l_{p}}^{d_{e}}\right|_{\boldsymbol{d}=%
\boldsymbol{0}}:\boldsymbol{l}_{p}=-\tfrac{1}{2}\left( \boldsymbol{X}%
_{e}\odot\boldsymbol{X}_{e}^{-T}+\boldsymbol{X}_{e}^{-T}\boxdot\boldsymbol{X}%
_{e}\right) :\boldsymbol{l}_{p}=-\mathsf{sym}\left( \boldsymbol{X}_{e}%
\boldsymbol{l}_{p}\boldsymbol{X}_{e}^{-1}\right)
\end{equation}
with $\left[ \boldsymbol{Y}\boxdot\boldsymbol{Z}\right] _{ijkl}=Y_{il}Z_{jk}$%
.

Quadratic (Green-Lagrange) strains are obtained directly from the
deformation gradients, and they are usual in finite element programs to
build the geometrical stiffness matrices, so they will be also important in
establishing links between our framework and the typical finite element
programs. Consider the Green-Lagrange strains obtained from the
corresponding deformation gradients as%
\begin{equation}
\boldsymbol{A}=\tfrac{1}{2}\left( \boldsymbol{X}^{T}\boldsymbol{X}-%
\boldsymbol{I}\right) \text{, \ \ }\boldsymbol{A}_{e}=\tfrac{1}{2}\left(
\boldsymbol{X}_{e}^{T}\boldsymbol{X}_{e}-\boldsymbol{I}\right) \text{, \ \ }%
\boldsymbol{A}_{p}=\tfrac{1}{2}\left( \boldsymbol{X}_{p}^{T}\boldsymbol{X}%
_{p}-\boldsymbol{I}\right)
\end{equation}%
Obviously $\boldsymbol{A}\neq \boldsymbol{A}_{e}+\boldsymbol{A}_{p}$ because
they lie in different configurations: $\boldsymbol{A}$ and $\boldsymbol{A}%
_{p}$ live in the reference, undeformed configuration, whereas $\boldsymbol{A%
}_{e}$ lives in the intermediate configuration. However, since $\boldsymbol{X%
}_{e}=\boldsymbol{XX}_{p}^{-1}$, it is straightforward to verify that

\begin{equation}
\boldsymbol{A}_{e}\left( \boldsymbol{A},\boldsymbol{X}_{p}\right) =\tfrac{1}{%
2}\left( \boldsymbol{X}_{e}^{T}\boldsymbol{X}_{e}-\boldsymbol{I}\right) =%
\boldsymbol{X}_{p}^{-T}\left( \boldsymbol{A}-\boldsymbol{A}_{p}\right)
\boldsymbol{X}_{p}^{-1}=\boldsymbol{X}_{p}^{-T}\odot \boldsymbol{X}%
_{p}^{-T}:\left( \boldsymbol{A}-\boldsymbol{A}_{p}\right)  \label{Ae(A,Xp)}
\end{equation}%
where the possibility of the operation $\left( \boldsymbol{A}-\boldsymbol{A}%
_{p}\right) $ results from the fact that they lie in the same, undeformed
configuration. However, $\boldsymbol{A}_{e}$ lies in the intermediate
configuration, so the mapping tensor $\boldsymbol{X}_{p}^{-T}\odot
\boldsymbol{X}_{p}^{-T}$ performs the proper push-forward operation. On the
other hand, from the dependencies $\boldsymbol{A}_{e}\left( \boldsymbol{A},%
\boldsymbol{X}_{p}\right) $ which have been explicitly obtained in Eq. %
\eqref{Ae(A,Xp)}, by use of the chain rule, the rate $\boldsymbol{\dot{A}}%
_{e}$ may be written as%
\begin{align}
\boldsymbol{\dot{A}}_{e}& =\left. \frac{\partial \boldsymbol{A}_{e}}{%
\partial \boldsymbol{A}}\right|_{\boldsymbol{\dot X}_p=\boldsymbol{0}}:%
\boldsymbol{\dot{A}}+\left. \frac{\partial \boldsymbol{A}_{e}}{\partial
\boldsymbol{X}_{p}}\right|_{\boldsymbol{\dot{A}}=\boldsymbol{\dot X}=%
\boldsymbol{0}}:\boldsymbol{\dot{X}}_{p}  \label{Aedot} \\
& =\left. \boldsymbol{\dot{A}}_{e}\right|_{\boldsymbol{\dot X}_p=\boldsymbol{%
0}}+\left. \boldsymbol{\dot{A}}_{e}\right|_{\boldsymbol{\dot{A}}=\boldsymbol{%
\dot X}=\boldsymbol{0}}=\,^{tr}\boldsymbol{\dot{A}}_{e}+\,^{ct}\boldsymbol{%
\dot{A}}_{e}  \label{AedotSplit}
\end{align}%
so identifying terms we find an interesting meaning for the previous mapping
tensor (known at any given instant from the multiplicative decomposition)
which will be used below%
\begin{equation}
\frac{\partial \boldsymbol{A}_{e}\left( \boldsymbol{A},\boldsymbol{X}%
_{p}\right) }{\partial \boldsymbol{A}}\equiv \left. \frac{\partial
\boldsymbol{A}_{e}}{\partial \boldsymbol{A}}\right|_{\boldsymbol{\dot X}_p=%
\boldsymbol{0}}=\boldsymbol{X}_{p}^{-T}\odot \boldsymbol{X}_{p}^{-T}\equiv
\left. \mathbb{M}_{\dot{A}}^{\dot{A}_{e}}\right|_{\boldsymbol{\dot X}_p=%
\boldsymbol{0}}  \label{dAe/dA _Xpdot=0}
\end{equation}%
Note that this tensor maps the rate $\boldsymbol{\dot{A}}$ in the material
configuration to the trial elastic one in the intermediate configuration $%
\,^{tr}\boldsymbol{\dot{A}}_{e}$ if we use $\,_0^t\boldsymbol{X}_p$.

\subsection{Dissipation equation and work-conjugate stresses¡}

The stress-strain work-conjugate measures for the formulation may be chosen
arbitrarily, because there is a one-to-one relation between them that makes
the stress power invariant to such choice (refer to \cite{latorre2016stress}
for relations and equivalences between measures, and to \cite{latorreAMP}
for full physical equivalence of the elastoplastic formulations if properly
transformed). For instance, the stress power per reference volume $\mathcal{P%
}$ may be written in any of the following equivalent forms, among others

\begin{equation}
\mathcal{P}=J\boldsymbol{\sigma}:\boldsymbol{d}=\boldsymbol{\tau }:%
\boldsymbol{d}=\boldsymbol{S}:\boldsymbol{\dot{A}}=\boldsymbol{T}:%
\boldsymbol{\dot{E}}  \label{EqPower}
\end{equation}%
where $\boldsymbol{\tau }$ is the spatial Kirchhoff stress tensor and $%
J=\det(\boldsymbol{X})$. The tensor $\boldsymbol{E}=\ln\boldsymbol{U}$ is
the logarithmic strain tensor in the reference configuration, $\boldsymbol{%
\dot{E}}$ is its rate, $\boldsymbol{U}$ is the right stretch tensor and $%
\boldsymbol{T}$ is the generalized Kirchhoff stress tensor, which is the
work conjugate of $\boldsymbol{E}$ in the most general anisotropic case (see
details and proof in \cite{latorreAMP}). To grasp a physical meaning for
this tensor we note that in the case of elastic isotropy (regardless of
strains being large) $\boldsymbol{T}=\boldsymbol{\tau}_R$ (rotated Kirchhoff
stresses), an equality which holds in any case for diagonal terms. In
anisotropy, or moderately large strains $\boldsymbol{T}\simeq \mathsf{sym}(%
\boldsymbol{\Xi})$. The relation between $\boldsymbol{A}$ and $\boldsymbol{E}
$ is simple when using the spectral decomposition

\begin{equation}
\boldsymbol{A}=\sum_{i=1}^{3}\tfrac{1}{2}\left( \lambda _{i}^{2}-1\right)
\boldsymbol{n}_{i}\otimes \boldsymbol{n}_{i}\text{ \ \ and }\boldsymbol{E}%
=\sum_{i=1}^{3}\ln \lambda _{i}\boldsymbol{n}_{i}\otimes \boldsymbol{n}_{i}
\end{equation}%
Although of little practical use, we can relate them conceptually through the fourth order mapping tensor $\mathbb{M}_{A}^{E}$ (note that this relation is nonlinear because the mapping depends on the stretches $\lambda_i$)

\begin{equation}
\boldsymbol{E}=\mathbb{M}_{A}^{E}:\boldsymbol{A}\text{ \ with }\mathbb{M}%
_{A}^{E}=\sum_{i=1}^{3}\frac{2\ln \lambda _{i}}{\lambda _{i}^{2}-1}%
\boldsymbol{M}_{i}\otimes \boldsymbol{M}_{i}
\end{equation}%
where $\boldsymbol{M}_{i}:=%
\boldsymbol{n}_{i}\otimes \boldsymbol{n}_{i}$ and $\mathbb{M}_{A}^{E}$ is
the corresponding mapping tensor. Then, in a similar way, the rates are also related through mapping tensors

\begin{equation}
\boldsymbol{\dot{E}}=\mathbb{M}_{\dot{A}}^{\dot{E}}:\boldsymbol{\dot{A}}%
\text{ \ or \ }\boldsymbol{\dot{A}}=\mathbb{M}_{\dot{E}}^{\dot{A}}:%
\boldsymbol{\dot{E}}
\end{equation}%
with the invertible mapping tensor, defined from the current deformation
state \cite{latorre2016stress}

\begin{equation}
\mathbb{M}_{\dot{A}}^{\dot{E}}\equiv \frac{d\boldsymbol{E}}{d\boldsymbol{A}}%
=\left( \mathbb{M}_{\dot{E}}^{\dot{A}}\right) ^{-1}=\sum_{i=1}^{3}\frac{1}{%
\lambda _{i}^{2}}\boldsymbol{M}_{i}\otimes \boldsymbol{M}_{i}+\sum_{i=1}^{3}%
\sum_{j\neq i}\frac{2\ln \left( \lambda _{j}/\lambda _{i}\right) }{\lambda
_{j}^{2}-\lambda _{i}^{2}}\boldsymbol{M}_{ij}\otimes \boldsymbol{M}_{ij}
\label{mapping}
\end{equation}%
where $\boldsymbol{M}_{ij}=\tfrac{1}{2}\left( \boldsymbol{n}_{i}\otimes
\boldsymbol{n}_{j}+\boldsymbol{n}_{j}\otimes \boldsymbol{n}_{i}\right)$.
Equivalent one-to-one relations apply between $\boldsymbol{\dot{E}}_{e}$ and
$\boldsymbol{\dot{A}}_{e}$ by simply changing the total principal stretches
and their directions by the principal elastic ones. Then, it is apparent
that similar relations to those given by Eqs. \eqref{Aedot} and %
\eqref{AedotSplit} apply also to logarithmic strains with the dependencies $%
\boldsymbol{E}_{e}\left( \boldsymbol{E},\boldsymbol{X}_{p}\right) $.
Applying the mappings and comparing with the rate relations from the chain
rule
\begin{align}
\boldsymbol{\dot{E}}_{e}& =\left[ \mathbb{M}_{\dot{A}_{e}}^{\dot{E}%
_{e}}:\left. \frac{\partial \boldsymbol{A}_{e}}{\partial \boldsymbol{A}}%
\right|_{\boldsymbol{\dot{X}}_p=\boldsymbol{0}}:\mathbb{M}_{\dot{E}}^{\dot{A}%
}\right] :\boldsymbol{\dot{E}}+\left[ \mathbb{M}_{\dot{A}_{e}}^{\dot{E}%
_{e}}:\left. \frac{\partial \boldsymbol{A}_{e}}{\partial \boldsymbol{X}_{p}}%
\right|_{\boldsymbol{\dot{X}}=\boldsymbol{0}}\right] :\boldsymbol{\dot{X}}%
_{p} \\
& =\left. \frac{\partial \boldsymbol{E}_{e}}{\partial \boldsymbol{E}}%
\right|_{\boldsymbol{\dot{X}}_p=\boldsymbol{0}}:\boldsymbol{\dot{E}}+\left.
\frac{\partial \boldsymbol{E}_{e}}{\partial \boldsymbol{X}_{p}}\right|_{%
\boldsymbol{\dot{X}}=\boldsymbol{0}}:\boldsymbol{\dot{X}}_{p} \\
& =\left. \boldsymbol{\dot{E}}_{e}\right|_{\boldsymbol{\dot{X}}_p=%
\boldsymbol{0}}+\left. \boldsymbol{\dot{E}}_{e}\right|_{\boldsymbol{\dot{X}}=%
\boldsymbol{0}}=\,^{tr}\boldsymbol{\dot{E}}_{e}+\,^{ct}\boldsymbol{\dot{E}}%
_{e}
\end{align}%
so we have the same split of the rate of the elastic logarithmic rate tensor
in a conservative (``trial'') part and a dissipative (``corrector'') one.
The specific properties of logarithmic strains will be exploited below.

If there is a one-to-one mapping between elastic strain measures, we can
define the stored energy in terms of any strain measure, for example in
terms of logarithmic strains. A classical approach in metal plasticity uses
quadratic energies in terms of logarithmic strains; see e.g. \cite%
{Anand79,Anand86,xiao1997hypo}, \cite{XiaoChen} and therein references.
Stored energies in terms of logarithmic strains are not only used in metal
plasticity, but also in soft materials (see e.g. \cite%
{CrespoLatMon,CrespoAuxetic,CrespoIJES,RomeroLatorreMontans}). Then, we
write the energy as $\Psi(\boldsymbol{E}_e,...)$, where we employ the same
symbol as before for the function $\Psi$ to simplify notation (if needed, we
will write dependencies explicitly to avoid confusion). The ellipsis denote
again the set of constant structural parameters for the symmetry group. Note
that the strain energy may be written equivalently as function of any
desired strain measure, because a one-to-one tensor mapping conversion
exists \cite{latorre2016stress}. The power balance, Eqs. \eqref{EqPower} and \eqref{eq11} may
be written as
\begin{equation}
\boldsymbol{T}:\boldsymbol{\dot{E}}=\mathcal{P}={\dot{\Psi}}+\mathcal{D}=%
\frac{d\Psi}{d\boldsymbol{E}_e}:\boldsymbol{\dot{E}}_e+\mathcal{D}=\frac{%
d\Psi}{d\boldsymbol{E}_e}:(\,^{tr}\boldsymbol{\dot{E}}_e+\,^{ct}\boldsymbol{%
\dot{E}}_e)+\mathcal{D}  \label{eqpower}
\end{equation}
Then, following standard Colemann arguments \cite{Truesdell-Nopll}, we
consider two cases. In the first one the internal evolution is frozen (i.e.
conservative with $\boldsymbol{\dot{X}}_p=\,^{ct}\boldsymbol{\dot{E}_e}=%
\boldsymbol{0}$ and $\mathcal{D}=0$), which gives
\begin{equation}
\boldsymbol{T}:\boldsymbol{\dot{E}}=\left.\frac{\partial\Psi(\boldsymbol{E}%
_e(\boldsymbol{E},\boldsymbol{X}_p))}{\partial\boldsymbol{E}}\right|_{%
\boldsymbol{\dot{X}}_p=\boldsymbol{0}}:\boldsymbol{\dot{E}}=\frac{d\Psi}{d%
\boldsymbol{E}_e}:\left.\frac{\partial\boldsymbol{E}_e(\boldsymbol{E},%
\boldsymbol{X}_p)}{\partial\boldsymbol{E}}\right|_{\boldsymbol{\dot{X}}_p=%
\boldsymbol{0}}:\boldsymbol{\dot{E}}
\end{equation}
Since this equation must hold for all symmetric $\boldsymbol{\dot{E}}$, we have
\begin{equation}
\boldsymbol{T}=\left.\frac{\partial\Psi(\boldsymbol{E}%
_e(\boldsymbol{E},\boldsymbol{X}_p))}{\partial\boldsymbol{E}}\right|_{%
\boldsymbol{\dot{X}}_p=\boldsymbol{0}}=\boldsymbol{T}^{|e}:\left.\frac{\partial\boldsymbol{E}_e}{%
\partial\boldsymbol{E}}\right|_{\boldsymbol{\dot{X}}_p=\boldsymbol{0}}%
\hspace{3ex}=:\boldsymbol{T}^{|e}:\mathbb{M}_{\hspace{1ex}E}^{^{tr}E_e}
\label{T=Te}
\end{equation}
where $\boldsymbol{T}^{|e}:=d\Psi/d\boldsymbol{E}_e$ is the generalized
Kirchhoff stress in the intermediate configuration (derived from the stored
energy), $\boldsymbol{T}$ is the generalized Kirchhoff stress tensor in the
material configuration and $\mathbb{M}_{\hspace{1ex}E}^{^{tr}E_e}:=\partial%
\boldsymbol{E}_e/\partial\boldsymbol{E}|_{{\boldsymbol{\dot{X}}_p=%
\boldsymbol{0}}}$ is a geometric mapping tensor converting the
energy-derived stress $\boldsymbol{T}^{|e}$ to the one obtained from equilibrium $\boldsymbol{T}$. In the second
case, the external power is frozen ($\boldsymbol{\dot{X}}=\boldsymbol{\dot{E}%
}=\boldsymbol{0})$, so Eq. \eqref{eqpower} results in---cf. Eq. %
\eqref{disipation1}
\begin{equation}
\mathcal{D}=-\boldsymbol{T}^{|e}:\,^{ct}\boldsymbol{\dot{E}}_e\ge0
\label{disipation2}
\end{equation}
This expression is identical to that obtained for anisotropic continuum
elastoplasticity using elastic corrector rates, see \cite%
{latorreAMP,Sanz2017}, and also identical to that obtained for anisotropic
finite nonlinear (non-equilibrium) viscoelasticity based on the Sidoroff
multiplicative decomposition \cite{latorre2015anisotropic}.
We remark that there is no approximation in Eq. \eqref{disipation2} and that both $\boldsymbol{T}^{|e}$ and $^{ct}\boldsymbol{\dot{E}}_e$ are symmetric tensors---cf. Eq. \eqref{disipation1}
\subsection{Plastic flow in terms of corrector rates\label%
{SECkinematics}}

Consider several consecutive steps (one may think of each step involving
just one sliding mechanism)%
\begin{equation}
\,_{\hspace{3.2ex}0}^{t+\Delta t}\boldsymbol{X}_{p}=\,_{\hspace{3.2ex}%
t}^{t+\Delta t}\boldsymbol{X}_{p}~_{t-\Delta t}^{\hspace{3.2ex}t}\boldsymbol{%
X}_{p}..._{\;0}^{\Delta t}\boldsymbol{X}_{p}
\end{equation}%
and the polar decompositions%
\begin{equation}
\,_{\hspace{3.2ex}0}^{t+\Delta t}\boldsymbol{R}_{p}\,_{\hspace{3.2ex}%
0}^{t+\Delta t}\boldsymbol{U}_{p}=\,_{\hspace{3.2ex}t}^{t+\Delta t}%
\boldsymbol{R}_{p}\,_{\hspace{3.2ex}t}^{t+\Delta t}\boldsymbol{U}%
_{p}~_{t-\Delta t}^{\hspace{3.2ex}t}\boldsymbol{R}_{p}~_{t-\Delta t}^{%
\hspace{3.2ex}t}\boldsymbol{U}_{p}~..._{\;0}^{\Delta t}\boldsymbol{R}%
_{p}~_{\;0}^{\Delta t}\boldsymbol{U}_{p}
\end{equation}%
Obviously
\begin{equation}
\,_{\hspace{3.2ex}0}^{t+\Delta t}\boldsymbol{R}_{p}\neq \,_{\hspace{3.2ex}%
t}^{t+\Delta t}\boldsymbol{R}_{p}~_{t-\Delta t}^{\hspace{3.2ex}t}\boldsymbol{%
R}_{p}..._{\;0}^{\Delta t}\boldsymbol{R}_{p}\text{ and }\,_{\hspace{3.2ex}%
0}^{t+\Delta t}\boldsymbol{U}_{p}\neq \,_{\hspace{3.2ex}t}^{t+\Delta t}%
\boldsymbol{U}_{p}~_{t-\Delta t}^{\hspace{3.2ex}t}\boldsymbol{U}%
_{p}...~_{\;0}^{\Delta t}\boldsymbol{U}_{p}
\end{equation}%
so from a constitutive standpoint $\boldsymbol{U}_p$ and $\boldsymbol{R}_p$
are meaningless \cite{MontansMRC} because a rotation would be changed in
character to stretch (and vice-versa) in subsequent steps, changing the nature of past deformations/dissipation. Hence, the polar decomposition for $\boldsymbol{X}_p$ is meaningful
only for incremental infinitesimal steps (or equivalently in rate form as $\boldsymbol{l}%
_p=\boldsymbol{d}_p+\boldsymbol{w}_p$), and the order in which the incremental $\boldsymbol{X}_p$ takes place is important in plastic gradients. Successive corrections to the elastic deformation gradient gives\begin{equation}
\,_{\hspace{3.2ex}0}^{t+\Delta t}\boldsymbol{X}_e=\,^{tr}\boldsymbol{X}_e\,^{ct}\boldsymbol{X}_e^{(1)}\,^{ct}\boldsymbol{X}_e^{(2)}...\,^{ct}\boldsymbol{X}_e^{(n)}
\label{Eq40}\end{equation}
In this case, the polar decomposition $\,_{\hspace{3.2ex}0}^{t+\Delta t}\boldsymbol{X}_e=\,_{\hspace{3.2ex}0}^{t+\Delta t}\boldsymbol{R}_e\,_{\hspace{3.2ex}0}^{t+\Delta t}\boldsymbol{U}_e$ is meaningful, because elastic deformations are path independent (function of state).
This observation, which is also   connected to the
property of weak-invariance \cite{Shutov2012,shutov2014analysis} because the plastic reference is not present,  is relevant when considering several slip systems \cite{RotersReview}, because if elastic deformations are employed, the order is not important. Hence, we can determine the final stretch $\boldsymbol{U}_e$ (which is all needed to compute the stresses) and then $\boldsymbol{R}_e$, from which $\boldsymbol{X}_e$ follows.

For simplicity consider for now a unique slip system, which plane is $%
\boldsymbol{m}$ and which direction is $\boldsymbol{s}$, both in the
reference and intermediate (isoclinic) configurations. We assume that the
plastic deformation gradient does not change the crystal directions. Since $%
\boldsymbol{s}\perp\boldsymbol{m}$, consider also the cartesian system of
representation defined by the slip direction and slip plane $\#\equiv\{%
\boldsymbol{s},\boldsymbol{m},\boldsymbol{s}\times\boldsymbol{m} \}$. If a
slip $\gamma(t)$ takes place in this slip mechanism, from a reference
configuration $\tau$, the plastic deformation gradient is
\begin{equation}
\,_{\tau}^{t}\boldsymbol{X}_{p}=\boldsymbol{I}+\gamma(t) \boldsymbol{s}%
\otimes \boldsymbol{m}=\left[
\begin{array}{ccc}
1 & \gamma \left( t\right) &  \\
& 1 &  \\
&  & 1%
\end{array}%
\right] _{\#}   \text{ and } \,_{\tau}^{t}\boldsymbol{X}_p^{-1}=\boldsymbol{I}-\gamma(t)\boldsymbol{s}\otimes\boldsymbol{m}\label{Xp}
\end{equation}%
where by $\left[ \bullet\right]_{\#}$ we imply matrix representation in the
system $\#$. Since in the intermediate configuration $\boldsymbol{\dot{s}}=%
\boldsymbol{\dot{m}}=\boldsymbol{0}$, its rate is%
\begin{equation}
^{t}\boldsymbol{\dot{X}}_{p}=\dot{\gamma}\boldsymbol{s}\otimes \boldsymbol{m}=\left[
\begin{array}{ccc}
0 & \dot{\gamma}\left( t\right) &  \\
& 0 &  \\
&  & 0%
\end{array}%
\right] _{\#}  \label{Xdotp}
\end{equation}%
Then, using Eqs. \eqref{Xp} and \eqref{Xdotp}, it is immediate to check that

\begin{equation}
^{t}\boldsymbol{l}_{p}=\,^{t}\boldsymbol{\dot{X}}_{p}\,_{\tau}^{t}%
\boldsymbol{X}_{p}^{-1}=[\dot{\gamma}\boldsymbol{s}\otimes \boldsymbol{m}]%
[\boldsymbol{I}-\gamma\boldsymbol{s}\otimes\boldsymbol{m}]=\left[
\begin{array}{ccc}
0 & \dot{\gamma}\left( t\right) &  \\
& 0 &  \\
&  & 0%
\end{array}%
\right] _{\#}\equiv \,^{t}\boldsymbol{\dot{X}}_{p}  \label{WI}
\end{equation}%
so $^{t}\boldsymbol{l}_{p}$ is independent of the referential state at $\tau$%
, i.e. of $\boldsymbol{X}_{p}$, a property which is useful for construction of w-invatiant models \cite%
{Shutov2012,shutov2014analysis}. We seek an equivalent velocity rate but based on the elastic gradient. To motivate it, now consider the internal \emph{corrector} phase during which the external
flow is frozen, i.e. $\boldsymbol{\dot{X}=0}$, and $\left. \boldsymbol{\dot{X%
}}_{e}\right|_{\boldsymbol{\dot{X}}=\boldsymbol{0}} \equiv \,^{ct}%
\boldsymbol{\dot{X}}_{e}$. In the incremental step we have
simultaneously $\,_{\hspace{3.2ex}0}^{t+\Delta t}\boldsymbol{X=}\,^{tr}%
\boldsymbol{X}_{e}\,_{0}^{t}\boldsymbol{X}_{p}$ and $\,_{\hspace{3.2ex}%
0}^{t+\Delta t}\boldsymbol{X=}\,_{\hspace{3.2ex}0}^{t+\Delta t}\boldsymbol{X}%
_{e}\,_{\hspace{3.2ex}0}^{t+\Delta t}\boldsymbol{X}_{p}$%
\begin{equation}
\,_{\hspace{3.2ex}0}^{t+\Delta t}\boldsymbol{X}=\,_{\hspace{3.2ex}%
0}^{t+\Delta t}\boldsymbol{X}_{e}\,_{\hspace{3.2ex}0}^{t+\Delta t}%
\boldsymbol{X}_{p}=\,^{tr}\boldsymbol{X}_{e}\underset{%
\begin{array}{c}
\boldsymbol{I}%
\end{array}%
}{\underbrace{\,^{ct}\boldsymbol{X}_{e}\,\,_{\hspace{3.2ex}t}^{t+\Delta t}%
\boldsymbol{X}_{p}}}\,_{0}^{t}\boldsymbol{X}_{p}
\end{equation}%
so $~\,^{ct}\boldsymbol{X}_{e}~\,_{\hspace{3.2ex}t}^{t+\Delta t}\boldsymbol{X%
}_{p}=\boldsymbol{I}$ results in the following expression, always valid%
\begin{equation}
\,^{ct}\boldsymbol{\dot{X}}_{e}\,\,_{\hspace{3.2ex}t}^{t+\Delta t}%
\boldsymbol{X}_{p}+\,^{ct}\boldsymbol{X}_{e}\,\,_{\hspace{3.2ex}t}^{t+\Delta t}%
\boldsymbol{\dot{X}}_{p}=\boldsymbol{0\;\Longrightarrow \;}^{t}\boldsymbol{l}_{p}\equiv\,_{\hspace{%
3.2ex}t}^{t+\Delta t}\boldsymbol{\dot{X}}_{p}\,\,_{\hspace{3.2ex}t}^{t+\Delta
t}\boldsymbol{X}_{p}^{-1}=-\,^{ct}\boldsymbol{X}_{e}^{-1}\,^{ct}\boldsymbol{%
\dot{X}}_{e} =:-\,^{ct}\boldsymbol{\bar{l}}_{e} \label{-le=lp}
\end{equation}%
which means that $^{t}\boldsymbol{l}_{p}=-\,^{ct}\boldsymbol{\bar{l}}_{e}$, but whereas one is defined in terms of plastic the gradient rate, the other one is defined in terms of the elastic gradient corrector. Note that in motivating  $\,^{ct}\boldsymbol{\bar{l}}_{e}$ we have omitted the reference configuration in the definition of $\,^{ct}\boldsymbol{%
\dot{X}}_{e} $, and obviously $\,^{ct}\boldsymbol{%
\dot{X}}_{e} $ is different for different reference configurations, the same way as $\,_{\hspace{3.2ex}t}^{t+\Delta t}%
\boldsymbol{\dot{X}}_{p}\ne\,_{\hspace{3.2ex}0}^{t+\Delta t}%
\boldsymbol{\dot{X}}_{p}$. However, this is irrelevant to the obtained result because both $^{t}\boldsymbol{l}_{p}$ and $\,^{ct}\boldsymbol{\bar{l}}_{e}$ are independent of such reference.

The use of $^{t}\boldsymbol{l}_{p}=-\,^{ct}\boldsymbol{\bar{l}}_{e}$ is convenient in the flow rule because it will result in a correction of elastic strains which avoids the direct integration of the plastic velocity gradient to yield an update gradient of the type $\,_{\hspace{%
3.2ex}t}^{t+\Delta t}\boldsymbol{{X}}_p$. Instead, we will obtain directly the elastic strains and, hence, the stretch tensor. To see the advantage, consider now two glide mechanisms occurring simultaneously. For simplicity
we consider the typical 2D think model in which $\boldsymbol{s}_1\otimes%
\boldsymbol{m}_1=\boldsymbol{s}\otimes\boldsymbol{m}$ and $\boldsymbol{s}%
_2\otimes\boldsymbol{m}_2=\boldsymbol{m}\otimes\boldsymbol{s}$.
Consider the exponential mapping
 so we can write the incremental gradient from $t$ to $t+\Delta t$, depending on the order, as different possibilities, for example
\begin{align}\exp{(\Delta t\dot\gamma_1^p\boldsymbol{s}\otimes%
\boldsymbol{m})}
\exp{(\Delta t\dot\gamma_2^p\boldsymbol{m}\otimes\boldsymbol{s})}&\ne\exp{(\Delta t\dot\gamma_2^p\boldsymbol{m}\otimes\boldsymbol{s})}\exp{(\Delta t\dot\gamma_1^p\boldsymbol{s}\otimes%
\boldsymbol{m})} \\ & \ne \exp{(\Delta t\dot\gamma_2^p\boldsymbol{m}\otimes\boldsymbol{s}+}{\Delta t\dot\gamma_1^p\boldsymbol{s}\otimes%
\boldsymbol{m})}\overset{?}{=}\,_{\hspace{%
3.2ex}t}^{t+\Delta t}\boldsymbol{{X}}_{p}\end{align}
Indeed, note that in the first two cases, the reference configuration for the following substep has been changed. Moreover, if we use the typical approximation \begin{equation} \,_{\hspace{%
3.2ex}t}^{t+\Delta t}\boldsymbol{{X}}_p=\exp{(\Delta t\,\boldsymbol{l}_p)}\simeq\boldsymbol{I}+\Delta t\,\boldsymbol{l}_p=\boldsymbol{I}+\Delta t\dot\gamma_2^p\boldsymbol{m}\otimes\boldsymbol{s}+\Delta t\dot\gamma_1^p\boldsymbol{s}\otimes%
\boldsymbol{m} \end{equation}
then, in general, for this approximation $\det({\,_{\hspace{%
3.2ex}t}^{t+\Delta t}\boldsymbol{{X}}_p})\ne 1$, so $\,_{\hspace{%
3.2ex}t}^{t+\Delta t}\boldsymbol{{X}}_p=\exp({\Delta t \boldsymbol{l}_p})\simeq\boldsymbol{I}+\Delta t \,\boldsymbol{l}_p$ results in non-isochoric flow when multiple slip mechanisms are involved.
Consider
 $\boldsymbol{X}_e=\,^{tr}%
\boldsymbol{X}_e\,^{ct}\boldsymbol{X}_e$ and $\boldsymbol{\dot{X}}_e=\,^{tr}%
\boldsymbol{\dot{X}}_e\,^{ct}\boldsymbol{X}_e+\,^{tr}\boldsymbol{X}_e\,^{ct}%
\boldsymbol{\dot{X}}_e$ as---see also Eq. \eqref{l=le+lp} and note that $\boldsymbol{l}=\,^{tr}\boldsymbol{l}_e$ \begin{align}
\boldsymbol{l}_e=\boldsymbol{\dot{X}}_e\boldsymbol{X}_e^{-1}=\,^{tr}%
\boldsymbol{l}_e+\,^{ct}\boldsymbol{l}_e=\,^{tr}\boldsymbol{l}_e\,+\,%
\boldsymbol{X}_e\,^{ct}\boldsymbol{\bar{l}}_e\boldsymbol{X}_e^{-1} &=\,^{tr}%
\boldsymbol{\dot{X}}_e\,^{tr}\boldsymbol{X}_e^{-1} +\,^{tr}\boldsymbol{X}%
_e\,^{ct}\boldsymbol{\dot{X}}_e\,^{ct}\boldsymbol{X}_e^{-1}\,^{tr}%
\boldsymbol{X}_e^{-1} \\
&=\,^{tr}\boldsymbol{l}_e\,\,\, \, +\,\,\, \,^{tr}\boldsymbol{X}_e\,^{ct}\boldsymbol{\tilde{l}}%
_e\,^{tr}\boldsymbol{X}_e^{-1}
\end{align}

\begin{figure}[h]
\center
\includegraphics[width=.6\textwidth]{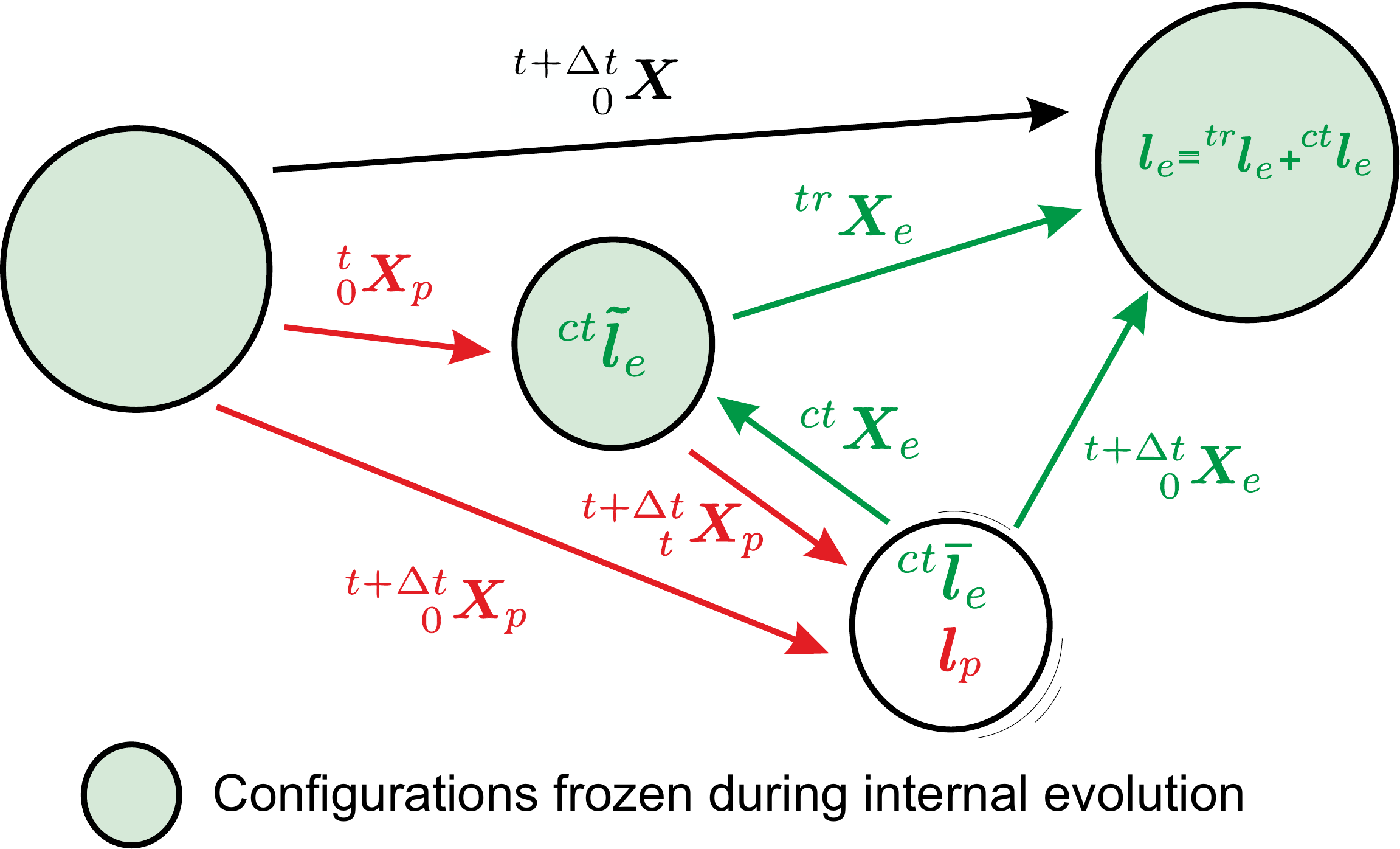}
\caption{Configurations during stress integration. Note that $\,^{ct}\boldsymbol{\tilde{l}}_e$ is computed in a configuration that is frozen if only internal evolution takes place (i.e. not considering external power input)}
\label{configurations.eps}
\end{figure}

\begin{figure}[h]
\center
\includegraphics[width=.75\textwidth]{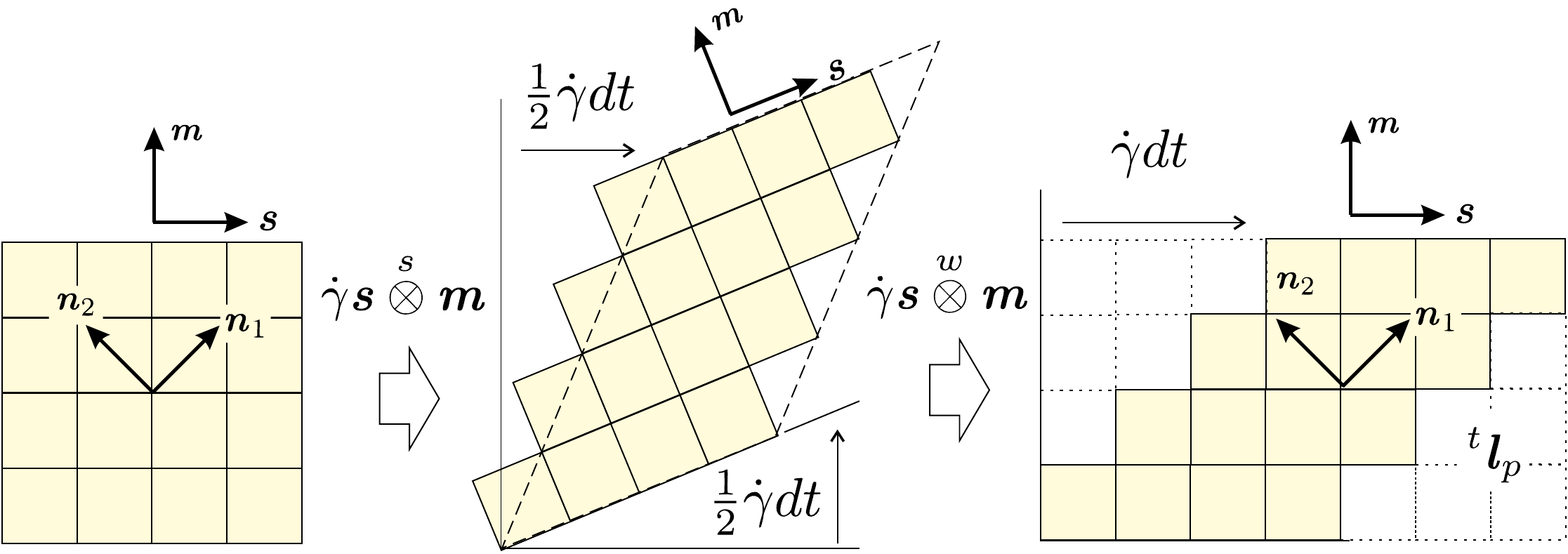}
\caption{Decomposition of the plastic rate in deformation rate and rotation
rate. The final intermediate configuration is isoclinic.}
\label{plastic_slip.eps}
\end{figure}
For a single mechanism we have $\,^{ct}\boldsymbol{\dot{X}}_e=\,^{ct}\boldsymbol{\bar{l}}_e=\,^{ct}\boldsymbol{\tilde{l}}_e $, but for multiple mechanisms and large steps this is only an approximation, i.e. if $\,^{ct}\boldsymbol{{X}}_e\simeq\boldsymbol{I}$ we can write $\,^{ct}\boldsymbol{\dot{X}}_e\simeq\,^{ct}\boldsymbol{\bar{l}}_e\simeq\,^{ct}\boldsymbol{\tilde{l}}_e $. The configurations are shown in Fig. \ref{configurations.eps}.

If we consider a single mechanism, we can take the \emph{current
intermediate configuration} as reference configuration and the following
infinitesimal strain and rotation rates over that configuration

\begin{equation}
^{ct}\boldsymbol{\dot{\varepsilon}}_{e}=-\dot{\gamma}\mathsf{sym}\left(
\boldsymbol{s}\otimes \boldsymbol{m}\right) =:-\dot{\gamma}\boldsymbol{s}%
\overset{s}{\otimes }\boldsymbol{m}=\left[
\begin{array}{ccc}
0 & -\frac{1}{2}\dot{\gamma} & 0 \\
-\frac{1}{2}\dot{\gamma} & 0 & 0 \\
0 & 0 & 0%
\end{array}%
\right] _{\#}
\end{equation}%
\begin{equation}
^{ct}\boldsymbol{\omega }_{e}\equiv ~^{ct}\boldsymbol{W}_{e}=-\dot{\gamma}%
\mathsf{skw}\left( \boldsymbol{s}\otimes \boldsymbol{m}\right) =:-\dot{\gamma%
}\boldsymbol{s}\overset{w}{\otimes }\boldsymbol{m}=\left[
\begin{array}{ccc}
0 & -\frac{1}{2}\dot{\gamma} & 0 \\
\frac{1}{2}\dot{\gamma} & 0 & 0 \\
0 & 0 & 0%
\end{array}%
\right] _{\#}
\end{equation}%
Remarkably, because $\boldsymbol{s}\otimes \boldsymbol{m}$ is constant
(recall that the intermediate configuration remains always isoclininc and $%
\boldsymbol{s}\perp\boldsymbol{m}$), these deformation rates take always
place in the same principal directions, at $45%
%TCIMACRO{\U{ba}}%
%BeginExpansion
{{}^o}%
%EndExpansion
$; i.e.

\begin{equation}
^{ct}\boldsymbol{\dot{\varepsilon}}_{e}=-\dot{\gamma}\boldsymbol{s}\overset{s%
}{\otimes }\boldsymbol{m}=-\tfrac{1}{2}\dot{\gamma}\boldsymbol{n}_{1}\otimes
\boldsymbol{n}_{1}+\tfrac{1}{2}\dot{\gamma}\boldsymbol{n}_{2}\otimes
\boldsymbol{n}_{2}
\end{equation}%
where $\boldsymbol{n}_{1}=1/\sqrt{2}\left( \boldsymbol{s}+\boldsymbol{m}%
\right) $ and $\boldsymbol{n}_{2}=1/\sqrt{2}\left( -\boldsymbol{s}+%
\boldsymbol{m}\right) $ are the principal directions, see Figure \ref%
{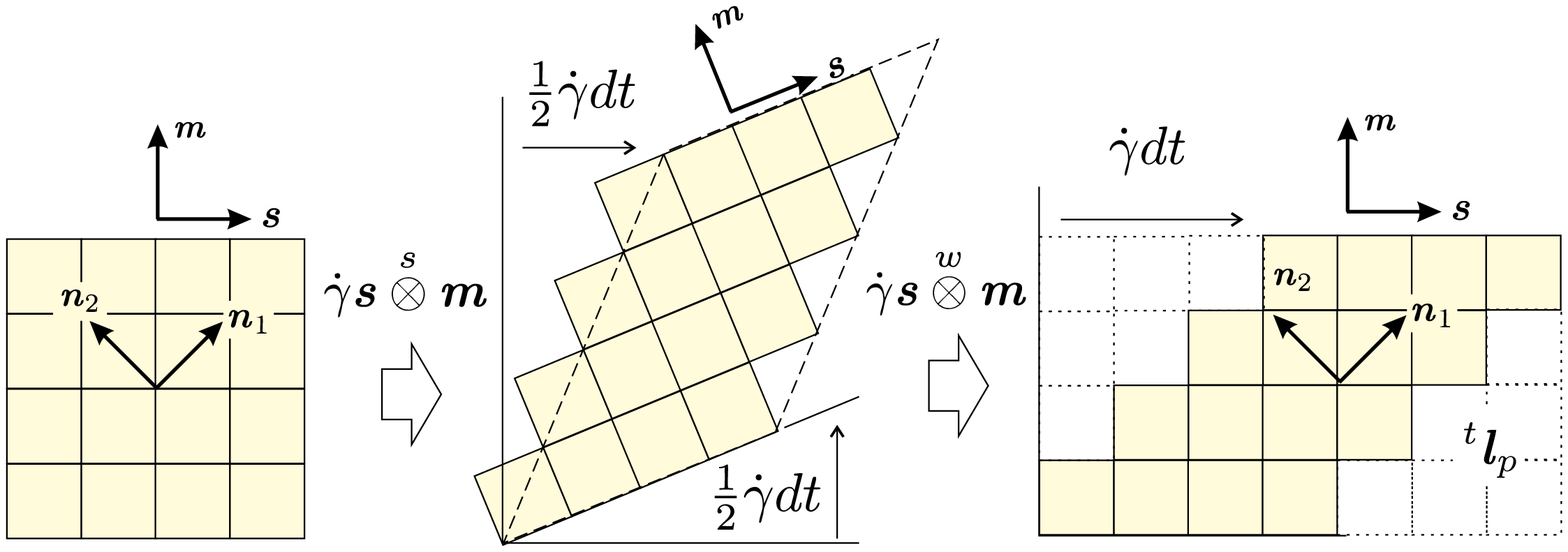}. In this case it can be shown that logarithmic strains
are the integral of infinitesimal strains; i.e. the rate of logarithmic
strains, the rate of the infinitesimal strains, and the deformation rate in
the intermediate configuration, are the same (see details and interpretation
in \cite{Latorre_logstrains})

\begin{equation}
\mathsf{sym}\left( ^{ct}\boldsymbol{\bar{l}}%
_{e}\right) \equiv ~^{ct}\boldsymbol{\dot{E}}_{e}\equiv ~^{ct}\boldsymbol{%
\dot{\varepsilon}}_{e}=-\dot{\gamma}\boldsymbol{s}\overset{s}{\otimes }%
\boldsymbol{m}
\end{equation}%
so we arrive at the following corrector identities, holding in crystal
plasticity \emph{without approximation}

\begin{equation}
^{ct}\boldsymbol{\bar{l}}_{e}= ~^{ct}\boldsymbol{\dot{X}}_{e}=%
\boldsymbol{~}^{ct}\boldsymbol{\dot{E}}_{e}+~^{ct}\boldsymbol{W}_{e}=-\dot{%
\gamma}\boldsymbol{s}\overset{s}{\otimes }\boldsymbol{m}-\dot{\gamma}%
\boldsymbol{s}\overset{w}{\otimes }\boldsymbol{m}
\end{equation}%
Then, the rate of the elastic deformation gradient is, without approximation

\begin{align}
\boldsymbol{\dot{X}}_{e}& =~^{tr}\boldsymbol{\dot{X}}_{e}+~^{ct}\boldsymbol{%
\dot{X}}_{e}  \notag \\
& =~^{tr}\boldsymbol{\dot{X}}_{e}+\boldsymbol{~}^{ct}\boldsymbol{\dot{E}}%
_{e}+~^{ct}\boldsymbol{W}_{e}  \notag \\
& =~^{tr}\boldsymbol{\dot{X}}_{e}-\dot{\gamma}\boldsymbol{s}\overset{s}{%
\otimes }\boldsymbol{m}-\dot{\gamma}\boldsymbol{s}\overset{w}{\otimes }%
\boldsymbol{m}
\end{align}%
Note that once $\boldsymbol{~}^{ct}\boldsymbol{\dot{E}}_{e}=-\dot{\gamma}%
\boldsymbol{s}\overset{s}{\otimes }\boldsymbol{m}$ is given (symmetric), we also have
immediately $~^{ct}\boldsymbol{W}_{e}=-\dot{\gamma}\boldsymbol{s}\overset{w}{%
\otimes }\boldsymbol{m}$ (skew-symmetric counterpart); it is the requirement to enforce that the
intermediate configuration is isoclinic. In a crystal there are several $%
g=1,...,G$ glide mechanisms, and we consider
\begin{equation}
\,^{ct}\boldsymbol{\tilde{l}}%
_{e}\simeq\,^{ct}\boldsymbol{\bar{l}}%
_{e}=-\sum_{g=1}^{G}\dot{\gamma}_{g}%
\boldsymbol{s}_{g}\otimes \boldsymbol{m}_g =\sum_{g=1}^{G}\left( ^{ct}\boldsymbol{\dot{E}}_{eg}+~^{ct}\boldsymbol{W}%
_{eg}\right) =\boldsymbol{~}^{ct}\boldsymbol{\dot{E}}_{e}+~^{ct}\boldsymbol{W%
}_{e}
\end{equation}
The resolved shear stress in the intermediate configuration is defined by
the dissipation power in Eq. \eqref{disipation2} and the condition of being
work-conjugate of $\dot{\gamma}_g$ so $\mathcal{D}_g=\tau|^e_g \dot{\gamma}%
_g $ with

\begin{equation}
\tau _{g}^{|e}:=\boldsymbol{s}_{g}\cdot \boldsymbol{T}^{|e}\boldsymbol{m}%
_{g}=\left( \boldsymbol{s}_{g}\otimes \boldsymbol{m}_{g}\right) :\boldsymbol{%
T}^{|e}\equiv \boldsymbol{s}_{g}\overset{s}{\otimes }\boldsymbol{m}_{g}:%
\boldsymbol{T}^{|e}\equiv \boldsymbol{T}^{|e}:\boldsymbol{s}_{g}\overset{s}{%
\otimes }\boldsymbol{m}_{g}
\end{equation}%
where the last identities hold because of the symmetry of $\boldsymbol{T}%
^{|e}$. If the glide strain rate is $\dot{\gamma}_{g}$, obviously when the
mechanism is not activated, we have $\dot{\gamma}_{g}=0$, and when the
mechanism is activated, we have $\dot{\gamma}_{g}\neq 0$ ($\dot{\gamma}%
_{g}>0 $ if the forward and backward slip directions are considered as two
mechanisms). The elastic corrector rates for each mechanism is written as:
\begin{align}
& ~^{ct}\boldsymbol{\dot{E}}_{eg}=-\dot{\gamma}_{g}\boldsymbol{s}_{g}\overset%
{s}{\otimes }\boldsymbol{m}_{g}=-\dot{\gamma}_{g}\boldsymbol{N}_{g}^{s}
\label{flowg} \\
& ~^{ct}\boldsymbol{W}_{eg}=-\dot{\gamma}_{g}\boldsymbol{s}_{g}\overset{w}{%
\otimes }\boldsymbol{m}_{g}=-\dot{\gamma}_{g}\boldsymbol{N}_{g}^{w}
\end{align}
with the definitions for convenience of notation: $\boldsymbol{N}_{g}^{s}:=%
\boldsymbol{s}_{g}\overset{s}{\otimes }\boldsymbol{m}_{g}$ and $\boldsymbol{N%
}_{g}^{w}:=\boldsymbol{s}_{g}\overset{w}{\otimes }\boldsymbol{m}_{g}$. The rate of the elastic strains is
\begin{equation}
\boldsymbol{\dot{E}}_e=\,^{tr}\boldsymbol{\dot{E}}_e+\,^{ct}\boldsymbol{\dot{E}}_e=\,^{tr}\boldsymbol{\dot{E}}_e-\sum_{g=1}^{G}\dot{\gamma}_{g}%
\boldsymbol{N}_{g}^{s}
\end{equation}

\begin{table}[tbp]
\caption{Continuum formulation and main assumptions}
\label{Table_form}\centering
\begin{tabular}{|p{3.5cm}|p{6cm}|p{6cm}|}
\hline
&  &  \\
\textbf{Variable/assumption} & \textbf{Classical framework} & \textbf{Proposed
formulation} \\[1ex] \hline\hline
&  &  \\
Multiplicative decomposition assumption& $\boldsymbol{X}=\boldsymbol{X}_e\boldsymbol{X}%
_p$ & $\boldsymbol{X}=\boldsymbol{X}_e\boldsymbol{X}_p $ \\[1ex]
Plastic dissipation \newline
$\dot\gamma^p_g\simeq\dot\gamma_g$ \& $\bar\kappa_g\simeq\tau_g^{|e}$ & $%
\mathcal{D}=\boldsymbol{\Xi}:\boldsymbol{l}_p=\sum\limits_{g=1}^{G}\bar%
\kappa_g\dot\gamma_g^p$ & $\mathcal{D}=-\boldsymbol{T}^{|e}:\,^{ct}%
\boldsymbol{\dot{E}}_e=\sum\limits_{g=1}^{G}\tau_g^{|e}\dot\gamma_g$ \\[1ex]Stress in dissipation & 2nd PK for energy and unsymmetric Mandel for dissipation:\newline
$\boldsymbol{\Xi}=\boldsymbol{C}_e\boldsymbol{S}^{|e}$  & Generalized Kirchhoff (symmetric) for
both energy and dissipation \newline
$\boldsymbol{T}^{|e}$ \\[1ex]Assumed stored energy &  $\boldsymbol{S}%
^{|e}={d\Psi}/{d\boldsymbol{A}_e}=\mathbb{C}^{|e}:\boldsymbol{A}_e$\newline (linear in Green-Lagrange strains)& Any hyperelastic relation. \newline For simplicity can be in terms of logarithmic strains:
$\boldsymbol{T}^{|e}={d\Psi}/{d\boldsymbol{E}_e}$\newline (which we assume linear in  examples) \\[1ex]
&  &  \\

Basic strain measure & Green-Lagrange (int. configuration): \newline
$\boldsymbol{A}_e=\tfrac{1}{2}(\boldsymbol{X}_e^T\boldsymbol{X}_e-%
\boldsymbol{I})$ & Logarithmic strains (int. config.): \newline
$\boldsymbol{E}_{e}=\tfrac{1}{2}\ln(\boldsymbol{X}_e^T\boldsymbol{X}_e)$ \\%
[1ex]
&  &  \\
Flow rule assumption (additive)\newline
 & Plastic velocity gradient\newline
$\boldsymbol{l}_p=\sum\limits_{g=1}^{G}\dot{\gamma}_g^p \boldsymbol{s}%
_g\otimes\boldsymbol{m}_g=\boldsymbol{\dot X}_p\boldsymbol{X}_p^{-1}$ & Elastic velocity corrector gradient\newline
$\,^{ct}\boldsymbol{\bar{l}}_e=-\sum\limits_{g=1}^{G}\dot{\gamma}_g
\boldsymbol{s}_g\otimes\boldsymbol{m}_g=\,^{ct}\boldsymbol{\dot{E}}_e+\,^{ct}%
\boldsymbol{W}_e$ \\[2ex]
Resolved stress & $\bar\kappa_g=\boldsymbol{s}_g\cdot\boldsymbol{C}_e%
\boldsymbol{S}^{|e}\cdot\boldsymbol{m}_g$ (conceptual, but not in practice; see next) & $\tau_{g}^{|e}=\boldsymbol{s}%
_g\cdot\boldsymbol{T}^{|e}\cdot\boldsymbol{m}_g$ \\[2ex]
Infinitesimal elasticity approximation & $\bar\kappa_g=\boldsymbol{s}_g\cdot\boldsymbol{C}_e\boldsymbol{S}^{|e}\cdot\boldsymbol{m}%
_g\simeq\boldsymbol{s}_g\cdot\boldsymbol{S}^{|e}\cdot\boldsymbol{m}_g\simeq
\boldsymbol{s}_g\cdot\boldsymbol{\tau}_R\cdot\boldsymbol{m}_g\simeq
\boldsymbol{s}_g\cdot\boldsymbol{\sigma}_R\cdot\boldsymbol{m}_g$ & Not needed; finite elastic strains considered without restriction\\[2ex] Hardening and rate sensitivity & Any relation, e.g. $\dot\gamma_g^p(...),\,\dot{\bar\kappa}_{cg}(...)$ &Any relation, e.g. $\dot\gamma_g(...),\,\dot{\tau}^{|e}_{cg}(...)$\\[2ex] & &
\\
 \hline
\end{tabular}%
\end{table}
\subsection{Summary of main assumptions and equations}
In Table \ref{Table_form} we summarize and compare the main assumptions and governning equations for both frameworks.
\section{Comparison with infinitesimal plasticity}

The algorithm is written in a more efficient and clean way when using matrix
notation by stacking all the mechanisms in a vector, and the continuum
elastoplastic matrix is better appreciated in this format. Let $\mathsf{N}%
_{a}^{s}$ be the Mandel pseudovector notation of the symmetric Schmid tensor
$\mathbf{N}_{a}^{s}$, and let $\mathsf{N}=\left[\mathsf{N}_{1}^{s},\mathsf{N}%
_{2}^{s},...\mathsf{N}_{A}^{s}\right]$ (dimension $6\times A$) be the
collection of the active ones, $a=1,...,A\in\mathcal{A}$, where $\mathcal{A}$
is the active set. The active slip systems are those such that the Schmid
stress $\tau_g^{|e}$ is greater than the resolved critical stress (CRSS) $%
\tau_{cg}^{|e}$ for the given slip system, i.e.
\begin{equation}
f_g=|\tau_g^{|e}|-\tau^{| e}_{cg} \ge0\,\,\,\text{ with }\,\tau_g^{|e}:=%
\boldsymbol{N}_{g}^{s}:\boldsymbol{T}^{|e}
\end{equation}
In rate-dependent plasticity, the inequality is chosen, and the absolute
value may be omitted if we considered signed directions (24 mechanisms in
FCC). Using Mandel notation also for the elastic strain rate $\mathsf{\dot{E}%
}_e$ (dimension $6\times1$), we have
\begin{equation}
\mathsf{\dot{E}}_{e}=\,^{tr}\mathsf{\dot{E}}_{e}+\,^{ct}\mathsf{\dot{E}}%
_{e}=\,^{tr}\mathsf{\dot{E}}_{e}-\mathsf{N\dot{\Gamma}}
\end{equation}%
where $\mathsf{\dot{\Gamma}=[\dot{\gamma}}_{1},...,\mathsf{\dot{\gamma}}%
_{A}]^{T}$ (dimension $A\times 1$). The rate of the stress tensor is
\begin{equation}
\boldsymbol{\dot{T}}^{|e} =\frac{d^2\Psi(\boldsymbol{E}_e,...)}{d\boldsymbol{%
E}_e\otimes d\boldsymbol{E}_e}:\boldsymbol{\dot{E}}_e=:\mathbb{A}^{|e}:%
\boldsymbol{\dot{E}}_e
\end{equation}
where $\mathbb{A}^{|e}$ is the elastic tangent (not necessarily constant,
although typically assumed constant in metal plasticity \cite%
{Anand79,Anand86}). This equation may be written again in Mandel matrix
notation as $\mathsf{\dot T}^{|e}=\mathsf{A}^{|e}:\mathsf{\dot{E}}_e$ so
\begin{eqnarray}
\mathsf{\dot{T}}^{|e} &=&\mathsf{A}^{|e}(^{tr}\mathsf{\dot{E}}_{e}-\mathsf{N%
\dot{\Gamma}})  \label{eqTdot}
\end{eqnarray}%
To perform simple intuitive comparisons with the classical infinitesimal
theory of plasticity, consider rate independent plasticity (we do not
discuss about the uniqueness of the solution; see e.g. \cite%
{KalidindiKotari,Miehe2004,BorjaBook}). In this case, the rates are given by
the condition that $f_g=0$. Considering the vector of active conditions as $%
\mathsf{F}=[f_1,...,f_A]^T$, we require $\mathsf{\dot{F}}=0$. Since $\mathsf{%
N}$ remains constant, this implies
\begin{equation}
\mathsf{\dot{F}} =\mathsf{N}^T\mathsf{A}^{|e}\,^{tr}\mathsf{\dot{E}}_{e}-%
\mathsf{N}^T\mathsf{A}^{|e}\mathsf{N\dot{\Gamma}}=0  \label{eqgamma}
\end{equation}
so

\begin{equation}
\underbrace{\mathsf{\dot{\Gamma}}}_{A\times 1}=\underbrace{[ \mathsf{N}^T\,\,%
\mathsf{A}^{|e}\,\,\mathsf{N}]^{-1}}_{(A\times 6) \times( 6 \times 6)
\times( 6 \times A) }\underbrace{\mathsf{N}^T\,\,\mathsf{A}^{|e}\,\,\,^{tr}%
\mathsf{\dot{E}}_{e}}_{(A\times 6)\times(6\times 6)\times(6\times1)}
\label{Eqgamma}
\end{equation}
This Equation may be compared to that of classical (unhardened)
infinitesimal continuum plasticity, e.g. Eq. (2.2.18) of \cite%
{simo2006computational}, Eq. (6.64) of \cite{SouzaNetoPericBook} and Eq.
(3.97) of \cite{BorjaBook}, among others, showing identical form despite
being a finite strain crystal plasticity model, unrestricted in the
magnitude or anisotropy of elastic and plastic responses. Of course, inside
this framework hardening may be equally considered. As well-known, the
solution of the system of equations may not be unique in the inviscid case,
and several solutions have been proposed as Moore-Penrose solutions, work
minimization \cite{Petryk} or the \textit{ultimate algorithm} \cite%
{KalidindiKotari,SchroderMiehe,Borja1993,BorjaBook}; however the examples
below will be performed with rate-dependent plasticity, as usual in
materials science, so we do not elaborate further. The tangent may be
obtained also as in small strains continuum plasticity, namely substituting
Eq. \eqref{eqgamma} into Eq. \eqref{eqTdot}
\begin{equation}
\mathsf{\dot{T}}^{|e}=[\mathsf{A}^{|e}-\mathsf{A}^{|e}\mathsf{N}\text{ } (
\mathsf{N}^T\,\,\mathsf{A}^{|e}\,\,\mathsf{N})^{-1}\mathsf{N}^T\,\,\mathsf{A}%
^{|e}]\,^{tr}\mathsf{\dot{E}}_{e} =:\mathsf{A}^{|e}_{ep}:\,^{tr}\mathsf{\dot{%
E}}_e
\end{equation}
Except for the obvious matrix form (because it includes all active systems
simultaneously), note that the expression for $\mathsf{A}^{|e}_{ep}$ is the
same as that of the continuum elastoplastic tangent for infinitesimal
continuum elastoplasticity, cf. Eq. (2.2.22) of \cite{simo2006computational}%
, Eq. (6.67) of \cite{SouzaNetoPericBook} or Eq. (3.98) of \cite{BorjaBook}.
Indeed, it is also the same layout as that obtained for continuum finite
strain elastoplasticity using this same framework; see Eq. (59) in \cite%
{Sanz2017}. The tensors $\boldsymbol{\dot{T}}^{|e}$ and $\,^{tr}\boldsymbol{%
\dot{E}}_e$ live in the isoclinic configuration for logarithmic strains. The
tensor $\mathbb{A}_{ep}^{|e}$ relating both rates is the main block of the
elastoplastic formulation because any other constitutive tensor between any
other stress/strain couple in any configuration is obtained from this one
using explicit geometrical mappings. For example, a natural elastoplastic
tangent may be $\mathbb{A}_{ep}$ such that $\boldsymbol{\dot{T}}=\mathbb{A}%
_{ep}:\boldsymbol{\dot{E}}$, relating the generalized Kirchhoff stress
tensor in the reference configuration with the logarithmic strain rate in
that configuration. In this case, the rate of Eq. \eqref{T=Te} may be used
\begin{equation}
\boldsymbol{\dot{T}}=\boldsymbol{\dot{T}}^{|e}:\mathbb{M}_{\hspace{1ex}%
E}^{^{tr}E_e}+\boldsymbol{T}^{|e}:\mathbb{\dot{M}}_{\hspace{1ex}E}^{^{tr}E_e}
\label{Tdot terms}
\end{equation}%
where

\begin{equation}
\mathbb{\dot{M}}_{\hspace{1ex}E}^{^{tr}E_e}=\frac{d}{dt}\left( \left. \frac{%
\partial \boldsymbol{E}_{e}}{\partial \boldsymbol{E}}\right|_{\boldsymbol{\dot X}_p=0}\right)
\end{equation}%
is the derivative of the mapping between both configurations in the
logarithmic space; it is elaborate but the term can be neglected (see
details in \cite{latorreAMP,Sanz2017}). Equation \eqref{Tdot terms} contains
two terms.\ The first term relates the rates $\boldsymbol{\dot{T}}^{|e}$ and
$\boldsymbol{\dot{T}}$ in the intermediate and the reference configuration
when these configurations are frozen. The second term in Eq.
\eqref{Tdot
terms} contains the influence of the continuously evolving configurations in
the mapping tensor when the plastic deformation gradient is fixed. However,
from a practical side, in numerical implementations it is customary to
develop the algorithm in a frozen intermediate configuration; a setting that
is similar in nature to updated Lagrangean finite element formulations.
Hence, we take%
\begin{align}
\boldsymbol{\dot{T}}& =\boldsymbol{\dot{T}}^{|e}:\mathbb{M}_{\hspace{1ex}%
E}^{^{tr}E_e}=\,^{tr}\boldsymbol{\dot{E}}_{e}:\mathbb{A}_{ep}^{|e}:\mathbb{M}%
_{\hspace{1ex}E}^{^{tr}E_e} \\
& ={\mathbb{M}_{\hspace{1ex}E}^{^{tr}E_e}}^{T}:\mathbb{A}_{ep}^{|e}:\,^{tr}%
\boldsymbol{\dot{E}}_{e}=[{\mathbb{M}_{\hspace{1ex}E}^{^{tr}E_e}}^{T}:%
\mathbb{A}_{ep}^{|e}:{\mathbb{M}_{\hspace{1ex}E}^{^{tr}E_e}}]:\boldsymbol{%
\dot{E}} =:\mathbb{A}_{ep}:\boldsymbol{\dot{E}}
\end{align}%
If, for example, we wish the tangent such that $\boldsymbol{\dot{S}}=\mathbb{%
C}_{ep}:\boldsymbol{\dot{A}}$, we just have to employ the proper mapping
tensors \cite{latorre2016stress}. We deal more in detail below for the
actual elastoplastic tangent for the specific Total Lagrangean finite
element formulation.

\section{Incremental computational formulation for rate-dependent p%
lasticity}

Since most algorithms and simulations of crystal plasticity in materials
science employ rate-dependent formulations, we develop here the specific
algorithm for this case and compare results in Section \ref{SEC examples}
with published results employing an equivalent classical crystal plasticity
formulation.

The computational algorithm integrates in two phases the trial rate and the
corrector rate. The first phase ``integrates'' the trial rate, i.e. the
partial derivatives for $\boldsymbol{\dot{X}}_p=\boldsymbol{0}$. The second
phase integrates the internal evolution by the corrector phase, i.e. the
partial derivatives with $\boldsymbol{\dot{X}}=\boldsymbol{0}$.

\subsection{Trial elastic state}

Since the first part of the integration process ``integrates'' the partial
derivatives with $\boldsymbol{\dot{X}}_p=\boldsymbol{0}$, this is a purely
conservative (hyperelastic) phase. Therefore, since the result is
path-independent we do not need to perform the integration, we just compute
the final state at the end of this substep, which is characterized by the
elastic gradient%
\begin{equation}
\,^{tr}\boldsymbol{X}_{e}=\,_{\hspace{3.2ex}t}^{t+\Delta t}\boldsymbol{X}%
\,_{0}^{t}\boldsymbol{X}_{e}
\end{equation}%
from which the trial elastic logarithmic strain and the trial stress are
immediately obtained%
\begin{equation}
^{tr}\boldsymbol{E}_{e}=\tfrac{1}{2}\ln \left( ^{tr}\boldsymbol{X}%
_{e}^{T}\,^{tr}\boldsymbol{X}_{e}\right) \text{ \ \ and }\,^{tr}\boldsymbol{T%
}^{|e}=\frac{d\Psi \left( ^{tr}\boldsymbol{E}_{e}\right) }{d^{tr}\boldsymbol{%
E}_{e}}  \label{Etr}
\end{equation}%
where $^{tr}\Psi:=\,\Psi \left( ^{tr}\boldsymbol{E}_{e}\right) $ is the
trial elastic strain energy. Note that $^{tr}\boldsymbol{E}_{e}$ is not the
total integral of $^{tr}\boldsymbol{\dot{E}}_{e}$ from $t=0$ (which we do not need to
perform). Instead, taking reference at time $t$, conceptually it would be%
\begin{equation}
^{tr}\boldsymbol{E}_{e}=\,_{0}^{t}\boldsymbol{E}_{e}+\int_{t}^{t+\Delta
t}\,^{tr}\boldsymbol{\dot{E}}_{e}d\tau
\end{equation}%
However, as mentioned,  Eq. \eqref{Etr} gives $^{tr}\boldsymbol{E}_{e}$ without any
approximation. Both $^{tr}\boldsymbol{E}_{e}$ and \ $^{tr}\boldsymbol{T}^{|e}$
tensors live in the trial elastic configuration defined by $\,^{tr}%
\boldsymbol{X}_{e}^{-1}$ from the spatial one at time $t$ or by $_{0}^{t}\boldsymbol{X}%
_{p}$ from the material one. The objective of the second phase of the stress
integration algorithm below is to obtain $\,_{\hspace{3.2ex}0}^{t+\Delta t}%
\boldsymbol{E}_{e}$ and the related stress $\,^{t+\Delta t}\boldsymbol{T}%
^{|e}$ upon knowledge of $^{tr}\boldsymbol{E}_{e}$, which is kept fixed
during the second substep.

\subsection{Example of rate-sensitive (viscous-type) constitutive equations}

The presented framework may be used both in rate-independent and
rate-dependent plasticity. To demonstrate an application we develop an
integration algorithm comparable to the typical rate-dependent plasticity, e.g. Ch. 6 in \cite{Rotersbook}.

For the shear rate function of each glide mechanism we choose the well-known
phenomenological power law proposed by Hutchinson (Eq. (2.4) in \cite%
{Hutchinson76}, see also Eq. (2.17) in \cite{Pierce83} and the similar Eq.
(23) in \cite{PanRice83}) motivated in the continuum mechanics power law for
creep

\begin{equation}
\dot{\gamma}_{g}(\tau_g^{|e}(\boldsymbol{T}^{|e}),\tau_{cg}^{|e})=\dot{\gamma}%
_{0}\left(\frac{|\tau_g^{|e}| }{\tau_{cg}^{|e}}\right)^{1/m}\mathsf{sgn}%
(\tau_g^{|e})  \label{power_law}
\end{equation}
where $\dot{\gamma}_{0}$ is a material parameter, a reference shear rate
that is related to the loading velocity, $\tau^{|e}_g$ is the resolved shear stress,  $\tau_{cg}^{|e}$ is the \textit{%
hardened} critical resolved shear stress of slip system $g$, and $m$ is the
rate sensitivity exponent (often $N:=1/m$ is used in the literature). We
also assume that the critical resolved shear stress hardens due to
dislocation pile-up and other related effects, following the
phenomenological relation ---e.g. Eq. (18) in \cite{RotersReview}
\begin{equation}
{\dot\tau_{cg}^{|e}}(\dot{\gamma}_j,\tau_{cj}^{|e},j=1,...,G)=\sum
\limits_{j=1}^{G} h_{gj} H_{j}|\dot{\gamma}_{j}| \left(1 - \dfrac{%
\tau_{cj}^{|e}}{\tau_{sj}^{|e}} \right)^{\alpha}  \label{latenth}
\end{equation}

The hardening rate of the critical resolved shear stress of each mechanism
accounts for the contribution of all the mechanisms; it has an isotropic
hardening form. $\tau_{cj}^{|e}$ is the current critical resolved shear
stress of slip system $j$, $h_{gj}$ is the lateral, latent hardening
parameter/ratio (typically $h_{gj}\in[1,1.4]$, with $1$ for coplanar and $%
1.4 $ otherwise; and $h_{gg}=1 $, see e.g. discussion in Sec. 3.2 of \cite%
{Pierce82}), and $H_{j}$ is the hardening modulus. The parameter ${%
\tau_{sj}^{|e}}$ is the saturated critical resolved shear stress for slip
system $j$ and $\alpha$ is the hardening exponent.

\subsection{Plastic correction. Example of implementation}

A backward Euler implementation of the flow equation is

\begin{equation}
\,_{\hspace{%
3.2ex}0}^{t+\Delta t}\boldsymbol{{E}}_e=\,^{tr}\boldsymbol{{E}}_e+\Delta\,^{ct}\boldsymbol{{E}}_e=\,^{tr}\boldsymbol{{E}}_e-\sum_{g=1}^{G}{\Delta\gamma}_{g}%
\boldsymbol{N}^s_{g}
\end{equation}
where the order in the addition is irrelevant,  no exponential mapping is employed, and the flow is fully isochoric because $\mathsf{trace}(\Delta ^{ct}\boldsymbol{E}_e)=0$. Of course, once $\,_{\hspace{%
3.2ex}0}^{t+\Delta t}\boldsymbol{{E}}_e$ is known, the elastic stretch tensor is immediately computed as indicated in Eq. \eqref{Ue} and the rotation as indicated in Eq. \eqref{Re} below.

There are several ways to build an iterative algorithm to solve the previous
equations in an implicit way. The scheme is the same as in infinitesimal
continuum plasticity with the obvious exception of the flow direction given
by the crystal slip mechanisms and the specific rate-dependent formulae
employed. We choose $\boldsymbol{E}_{e}$ and $\tau^{|e}_{cg}$ ($g=1,...,G$)
as the variables and the following $6+G$ residues as nonlinear equations to
solve. Classical plain Newton-Raphson iterations are used to solve the
system of equations. Using again Serif fonts to denote the Mandel vector/matrix
notation, the residuals enforcing a \textit{standard }backward-Euler
integration scheme are (the first set is a rate-dependent counterpart of the
typical yield conditions)
\begin{align}
^{t+\Delta t}R_{g}^{} &:=\Delta\tau^{|e}_{cg}-\Delta t\,^{t+\Delta t}\dot{%
\tau}^{|e}_{cg}\rightarrow 0,\hspace{5ex} i=1,...,G\,\text{
}  \label{EqR1} \\
\,^{t+\Delta t}\mathsf{R}_{E} &:=\,_{\hspace{%
3.2ex}0}^{t+\Delta t}\mathsf{E}_{e}-\,^{tr}%
\mathsf{E}_{e}+\underset{g=1}{\overset{G}{\sum }}\Delta \gamma
_{g}\mathsf{N}_{g}^{s}\rightarrow0  \label{EqR2}
\end{align}
The residual vector is $\mathsf{R}=[R_1,..,R_G,\mathsf{R}_E^T]^T$ with
dimension $G+6$. The tangent with dimension $(G+6)\times(G+6)$ for iteration
$[k]$ of step $t+\Delta t$ is:

\begin{equation}
\left[\dfrac{d\,^{t+\Delta t}\mathsf{R}}{d\,^{t+\Delta t}\mathsf{V}}\right]%
^{[k]}=%
\begin{bmatrix}
\dfrac{d^{t+\Delta t}R_{1}^{[k]}}{d^{t+\Delta t}\tau^{|e}_{c1}} & ... &
\dfrac{d^{t+\Delta t}R_{1}^{[k]}}{d^{t+\Delta t}\tau^{|e}_{cG}} & \dfrac{%
d^{t+\Delta t}R_{1}^{[k]}}{d^{t+\Delta t}\mathsf{E}_e} \\
\dfrac{}{} & ... & \dfrac{}{} & \dfrac{}{} \\
... & ... & ... & ... \\
\dfrac{d^{t+\Delta t}R_{G}^{[k]}}{d^{t+\Delta t}\tau^{|e}_{c1}} & ... &
\dfrac{d^{t+\Delta t}R_{G}^{[k]}}{d^{t+\Delta t}\tau^{|e}_{cG}} & \dfrac{%
d^{t+\Delta t}R_{G}^{[k]}}{d^{t+\Delta t}\mathsf{E}_e} \vspace{1ex} \\
\dfrac{d^{t+\Delta t}\mathsf{R}_{E}^{[k]}}{d^{t+\Delta t}\tau^{|e}_{c1}} &
... & \dfrac{d^{t+\Delta t}\mathsf{R}_{E}^{[k]}}{d^{t+\Delta t}\tau^{|e}_{cG}%
} & \dfrac{d^{t+\Delta t}\mathsf{R}_{E}^{[k]}}{d^{t+\Delta t}\mathsf{E}_e}%
\end{bmatrix}%
\end{equation}
with $\mathsf{V}=[\tau^{|e}_{c1},...,\tau^{|e}_{cG},\mathsf{E}_e^T]^T$ being
the iterative variables. Taking for simplicity in the exposition a constant
elastic behavior $d\boldsymbol{T}^{|e}/d\boldsymbol{E}_e=\mathbb{A}%
^{|e}\rightarrow\mathsf{A}^{|e}$, the terms of the tangent, in tensor
notation, are obtained from Eqs. \eqref{EqR1}, \eqref{EqR2} as ---we omit
the time and iteration indices for brevity
\begin{align}
\frac{\partial \boldsymbol{R}_{E}}{\partial \boldsymbol{E}_{e}}&=\mathbb{I}%
+\Delta t\underset{g=1}{\overset{G}{\sum }}\boldsymbol{N}%
_{g}^{s}\otimes \frac{\partial \dot{\gamma}_{g}}{\partial \boldsymbol{T}^{|e}%
}:\mathbb{A}^{|e} \\
\frac{\partial R_{g}^{}}{\partial \boldsymbol{E}_{e}}&=-\Delta
t\frac{\partial \dot{\tau}_{cg}^{|e}}{\partial \dot{\gamma%
}_{j}}\frac{\partial \dot{\gamma}_{j}}{\partial \boldsymbol{T}^{|e}}:\mathbb{%
A}^{|e}
\end{align}
and
\begin{align}
&\frac{\partial \boldsymbol{R}_{E}}{\partial \tau^{|e}_{cg}}=\boldsymbol{N}%
_{g}^{s}\Delta t\frac{\partial \dot{\gamma}_{g}}{\partial \tau^{|e}_{cg}} \\
&\frac{\partial R_{g}}{\partial \tau^{|e}_{cj}}=\delta_{gj}-\Delta t \left[%
\frac{\partial \dot{\tau}^{|e}_{cg}}{\partial \tau^{|e}_{cj}}+\frac{\partial
\dot{\tau}^{|e}_{cg}}{\partial\dot{\gamma}_{j}}\frac{\partial \dot{\gamma}%
_{j}}{\partial \tau^{|e}_{cj}}\right]\text{ }
\end{align}
which terms follow immediately from Eqs. \eqref{power_law} and Eq. %
\eqref{latenth}, and $\delta_{gj}$ is the Kronecker delta. During the
iterations, once $_{\hspace{3.2ex}0}^{t+\Delta t}\boldsymbol{E}_{e}$ and $%
^{t+\Delta t}\tau^{|e}_{cg}\,\,(g=1...G)$ are known, the generalized
Kirchhoff stress $^{t+\Delta t}\boldsymbol{T}^{|e}$ is known and used from
the hyperelasticity expression

\begin{equation}
^{t+\Delta t}\boldsymbol{T}^{|e}=\frac{d\Psi \left( _{\hspace{3.2ex}%
0}^{t+\Delta t}\boldsymbol{E}_{e}\right) }{d_{\hspace{3.2ex}0}^{t+\Delta t}%
\boldsymbol{E}_{e}}
\end{equation}
and the shear rates from Eq. \eqref{power_law} in backward-Euler form:

\begin{equation}
^{t+\Delta t}\dot{\gamma}_{g}=\dot{\gamma}_{0}\left(\frac{|^{t+\Delta t}%
\boldsymbol{T}^{|e}:\boldsymbol{N}_{g}^{s}|}{^{t+\Delta t}\tau^{|e}_{cg}}%
\right)^{1/m}\mathsf{sgn}(\,^{t+\Delta t}\boldsymbol{T}^{|e}:\boldsymbol{N}%
_{g}^{s})
\end{equation}
so the solutions for all $\Delta \gamma _{g}=\Delta t\,^{t+\Delta t}\dot{%
\gamma}_{g}$ are also known.

If the finite element program is running in infinitesimal strains mode (i.e.
a small strains formulation is pursued), the rest of the operations are
omitted; we simply accept $\boldsymbol{T}^{|e}\leftrightarrow\boldsymbol{%
\sigma}$ and $\boldsymbol{E}_{e}\leftrightarrow\boldsymbol{\varepsilon}_e$ and $\Delta t\boldsymbol{W}_e\leftrightarrow\Delta\boldsymbol{\omega}_e$.
For a large strains formulation, for the next step it is necessary to update
the elastic deformation gradient, which may be updated using the polar
decomposition, because it is a state variable (path independent)
\begin{equation}
\,_{\hspace{3.2ex}0}^{t+\Delta t}\boldsymbol{X}_{e}=~^{tr}\boldsymbol{X}%
_{e}~^{ct}\boldsymbol{X}_{e}=\,_{\hspace{3.2ex}0}^{t+\Delta t}\boldsymbol{R}%
_{e}\,_{\hspace{3.2ex}0}^{t+\Delta t}\boldsymbol{U}_{e}
\end{equation}%
Since the update should be consistent with the \emph{final }computed value
of $\,_{\hspace{3.2ex}0}^{t+\Delta t}\boldsymbol{E}_{e}$, we have%
\begin{equation}
\,_{\hspace{3.2ex}0}^{t+\Delta t}\boldsymbol{U}_{e}=\exp \left( \,_{\hspace{%
3.2ex}0}^{t+\Delta t}\boldsymbol{E}_{e}\right)
\label{Ue}\end{equation}%
and we take %
\begin{equation}
\,_{\hspace{3.2ex}0}^{t+\Delta t}\boldsymbol{R}_{e}= ~^{tr}\boldsymbol{R%
}_{e}~\exp \left( \Delta t~^{ct}\boldsymbol{W}_{e}\right)
\label{Re}\end{equation}%
with%
\begin{equation}
\Delta t~^{ct}\boldsymbol{W}_{e}=-\sum_{g=1}^{G}\Delta \gamma _{g}%
\boldsymbol{N}_{g}^{w}
\end{equation}%
Once the iterative solution is obtained, a last step is to perform the
mappings to the stress-strain couple used by the finite element program. In
total Lagrangean formulations it is typical that this stress-strain couple
is made by the second Piola Kirchhoff stress tensor $\boldsymbol{S}$ and the
Green-Lagrange strains $\boldsymbol{A}$ in the reference (initial,
undeformed) configuration \cite{Bathebook}. This operation is non-iterative and employs just
purely geometrical mappings \cite{latorre2016stress}.

\subsection{Geometric postprocessor}

With the final strains $\,_{\hspace{3.2ex}0}^{t+\Delta t}\boldsymbol{E}_{e}$%
, the generalized Kirchhoff stresses in the converged configuration are%
\begin{equation}
\,^{t+\Delta t}\boldsymbol{T}^{|e}=\frac{d\Psi }{d\,_{\hspace{3.2ex}%
0}^{t+\Delta t}\boldsymbol{E}_{e}}
\end{equation}%
By work conjugacy, the stresses in the trial configuration (with $%
\boldsymbol{\dot{X}}_{p}=\boldsymbol{0}$) are obtained from $\boldsymbol{T}%
^{|e}:\boldsymbol{\dot{E}}_{e}=~\boldsymbol{T}^{|tr}:~^{tr}\boldsymbol{\dot{E%
}}_{e}$, so%
\begin{equation}
\,^{t+\Delta t}\boldsymbol{T}^{|tr}=\left. \frac{\partial \Psi }{\partial
^{tr}\boldsymbol{E}_{e}}\right|_{\boldsymbol{\ \dot{X}}_p=\boldsymbol{0}}=%
\frac{d\Psi }{d\,_{\hspace{3.2ex}0}^{t+\Delta t}\boldsymbol{E}_{e}}:\left.
\frac{\partial \,_{\hspace{3.2ex}0}^{t+\Delta t}\boldsymbol{E}_{e}}{\partial
^{tr}\boldsymbol{E}_{e}}\right|_{\boldsymbol{\ \dot{X}}_p=\boldsymbol{0}%
}=\,^{t+\Delta t}\boldsymbol{T}^{|e}:\left. \mathbb{M}_{^{tr}E_{e}}^{E_{e}}%
\right|_{\boldsymbol{\ \dot{X}}_p=\boldsymbol{0}}\simeq\,^{t+\Delta t}\boldsymbol{T}^{|e}  \label{Ttr to Te}
\end{equation}%
where, if the steps are small or loading proportional for large steps, we
can consider $\mathbb{M}_{^{tr}E_{e}}^{E_{e}}|_{\boldsymbol{\ \dot{X}}_p=%
\boldsymbol{0}} \simeq \mathbb{I}^{s}$. Indeed, no distinction has been performed above between both configurations in the case of logarithmic strains and related stresses. Note that the stresses $^{t+\Delta t}%
\boldsymbol{T}^{|tr}\neq \,^{tr}\boldsymbol{T}^{|e}$. The latter $\,^{tr}%
\boldsymbol{T}^{|e}=d\,^{tr}\Psi /d^{tr}\boldsymbol{E}_{e}$ are the trial
stresses (independent of any plastic flow, corresponding and work-conjugate
to the trial strains $^{tr}\boldsymbol{E}_{e}$) whereas the former $%
\,^{t+\Delta t}\boldsymbol{T}^{|tr}=\left. \partial \,^{t+\Delta t}\Psi
/\partial ^{tr}\boldsymbol{E}_{e}\right|_{\boldsymbol{\ \dot{X}}_p=%
\boldsymbol{0}}$ are the final elastic stresses $^{t+\Delta t}\boldsymbol{T}%
^{|e}$ (corresponding to $\,_{\hspace{3.2ex}0}^{t+\Delta t}\boldsymbol{E}%
_{e} $, which depends on the plastic flow) mapped geometrically to the trial
configuration. In general $\,^{tr}\boldsymbol{T}^{|e}$ will be very
different from $^{t+\Delta t}\boldsymbol{T}^{|e}$ (they correspond to
different strains) but $^{t+\Delta t}\boldsymbol{T}^{|tr}$ will be similar
or equal to $\,^{t+\Delta t}\boldsymbol{T}^{|e}$ (they correspond to the
same strains, but lie in slightly different configurations); they are
coincident in the case of proportional loading or with infinitesimal steps,
because in such cases $\partial_0^t\boldsymbol{E}_e/\partial\,^{tr}%
\boldsymbol{E}_e|_{\boldsymbol{\dot{X}}_p=\boldsymbol{0}}\simeq\mathbb{I}^s$.

The purpose of employing the stresses $^{t+\Delta t}\boldsymbol{T}^{|tr}$
is to use the trial configuration and simpler mapping tensors to arrive at
the second Piola-Kirchhoff stress tensor in the reference configuration.
Consider the following expression immediate from work-conjugacy and the
chain rule
\begin{equation}
\,^{t+\Delta t}\boldsymbol{S}^{|tr}=\,^{t+\Delta t}\boldsymbol{T}^{|tr}:%
\frac{d^{tr}\boldsymbol{E}_{e}}{d^{tr}\boldsymbol{A}_{e}}
\end{equation}%
where $d\,^{tr}\boldsymbol{E}_{e}/d\,^{tr}\boldsymbol{A}_{e}$ is a mapping
tensor obtained immediately from the stretches and principal directions of
the trial state, see \cite{latorre2016stress} and Eq. \eqref{mapping}. Then,
the reference second Piola-Kirchhoff stress used in total Lagrangean
formulations is ---see Eq. \eqref{Ae(A,Xp)}

\begin{equation}
^{t+\Delta t}\boldsymbol{S=}\,^{t+\Delta t}\boldsymbol{S}^{|tr}:\frac{d^{tr}%
\boldsymbol{A}_{e}}{d\,_{\hspace{3.2ex}0}^{t+\Delta t}\boldsymbol{A}}%
=\,^{t+\Delta t}\boldsymbol{S}^{|tr}:\left( _{0}^{t}\boldsymbol{X}%
_{p}^{-T}\odot \,_{0}^{t}\boldsymbol{X}_{p}^{-T}\right)
\end{equation}

The derivation of the consistent tangent moduli are presented in the Appendix.
\subsection{Comparison of frameworks}

In Table \ref{Table_algos} we compare typical algorithmic settings

\begin{table}[tbp]
\caption{Scheme of typical computational steps using Newton algorithms (different implementations are possible)}
\label{Table_algos}\centering
\begin{tabular}{|p{4.5cm}|p{6cm}|p{6cm}|}
\hline
&  &  \\
\textbf{Step} & \textbf{Classical framework} & \textbf{Proposed
formulation} \\[1ex] \hline\hline
&  &  \\
(1) Given $\,^{t+\Delta t}_{\hspace{3ex}t}\boldsymbol{X}$ and $\,^{t}_{0}\boldsymbol{X}_e$, compute the trial elastic gradient   & $\,^{tr}\boldsymbol{X}_e=\,^{t+\Delta t}_{\hspace{3ex}t}\boldsymbol{X} \,_0^t%
\boldsymbol{X}_e$&$\,^{tr}\boldsymbol{X}_e=\,^{t+\Delta t}_{\hspace{3ex}t}\boldsymbol{X} \,_0^t%
\boldsymbol{X}_e$ \\ & &  \\
(2) Trial elastic strain & $\,^{tr}\boldsymbol{A}_e=\tfrac{1}{2}(\,^{tr}\boldsymbol{X}_e^T\,^{tr}\boldsymbol{X}_e-
\boldsymbol{I})$ \newline First iteration:\newline $\,^{t+\Delta t}_{\hspace{3ex}0}\boldsymbol{A}_e^{[0]}=\,^{tr}\boldsymbol{A}_e$  & $\,^{tr}\boldsymbol{E}_{e}=\tfrac{1}{2}\ln(\,^{tr}\boldsymbol{X}_e^T\,^{tr}\boldsymbol{X}_e)$ \newline First iteration:\newline $\,^{t+\Delta t}_{\hspace{3ex}0}\boldsymbol{E}_e^{[0]}=\,^{tr}\boldsymbol{E}_e$\\%
[1ex]
&  &  \\
(3) Stress & $\,^{t+\Delta t}\boldsymbol{S}^{|e}=\mathbb{C}^{|e}:\,^{t+\Delta t}_{%
\hspace{3ex}0}\boldsymbol{A}_e$ & $\,^{t+\Delta t}\boldsymbol{T}%
^{|e}=d\Psi/d\,^{t+\Delta t}_{\hspace{3ex}0}\boldsymbol{E}_e$ \\[1ex]
&  &  \\
(4) Resolved stresses  & $\,^{t+\Delta t}\bar\kappa_g\simeq\boldsymbol{s}_g\cdot%
\,^{t+\Delta t}\boldsymbol{S}^{|e}\cdot\boldsymbol{m}_g$ & $\,^{t+\Delta t}\tau_{g}^{|e}=\boldsymbol{s}%
_g\cdot\,^{t+\Delta t}\boldsymbol{T}^{|e}\cdot\boldsymbol{m}_g$ \\[2ex]
(5) Shear slips and critical resolved shear stresses\newline (solve nonlinear system of equations containing the shear slips and iterative tensorial kinematic variable) & Coupled iterations (symmetric + skew-symmetric contributions)\newline\vspace{0ex} Using (2) to (4), solve residuals by e.g. Newton-Raphson algoritm \newline $\Delta\gamma^p_g-\Delta t \dot\,^{t+\Delta t}\gamma_g^p(...)\rightarrow 0$ \vspace{1ex}\newline $\,^{t+\Delta t}_{\hspace{3ex}0}\boldsymbol{X}_e-\,^{tr}\boldsymbol{X}_e\exp(-\sum\Delta\gamma_g^p\boldsymbol{N}_g^s)\rightarrow 0$\vspace{1ex}\newline (This includes exponential mapping at each iteration, or linear approximation)\vspace{4ex} &   Uncoupled iterations (only symmetric contributions)\newline\vspace{0ex} Using (2) to (4), solve residuals by e.g. Newton-Raphson algoritm \newline $\Delta\tau_{cg}^{|e}-\Delta t\,^{t+\Delta t} \dot\tau_{cg}^{|e}(...)\rightarrow 0$\vspace{1ex}\newline  $\,^{t+\Delta t}_{\hspace{3ex}0}\boldsymbol{E}_e-\,^{tr}\boldsymbol{E}_e+\sum\Delta\gamma_g\boldsymbol{N}_g^s\rightarrow 0$ \vspace{1ex}\newline (Naturally additive, no need for approximation or exponential mapping) \\
(6) Update  & $\,^{t+\Delta t}_{\hspace{3ex}0}\boldsymbol{A}_e=\tfrac{1}{2}(\,^{t+\Delta
t}_{\hspace{3ex}0}\boldsymbol{X}_e^T \,^{t+\Delta t}_{\hspace{3ex}0}%
\boldsymbol{X}_e-\boldsymbol{I})$ \vspace{2ex}\newline  & $%
\,^{t+\Delta t}_{\hspace{3ex}0}\boldsymbol{U}_e=\exp(\,^{t+\Delta t}_{%
\hspace{3ex}0}\boldsymbol{E}_e)$\vspace{2ex}\newline $\Delta t\,^{ct}\boldsymbol{W}%
_e=\sum\Delta\gamma_g\boldsymbol{N}_g^w$ \vspace{2ex}\newline  $\,^{t+\Delta t}_{\hspace{3ex}0}%
\boldsymbol{X}_e=\,^{tr}\boldsymbol{R}_e\exp(\Delta t\,^{ct}\boldsymbol{W}%
_e)\,^{t+\Delta t}_{\hspace{3ex}0}\boldsymbol{U}_e$
 \\[2ex]
(7) Basic stress & $\,^{t+\Delta t}\boldsymbol{S}^{|e}=d\Psi/d\,^{t+\Delta t}_{%
\hspace{3ex}0}\boldsymbol{A}_e$ & $\,^{t+\Delta t}\boldsymbol{T}%
^{|e}=d\Psi/d\,^{t+\Delta t}_{\hspace{3ex}0}\boldsymbol{E}_e$ \\[1ex]
(8) Basic tangent & $\,^{t+\Delta t}\mathbb{C}_{ep}^{|e}= d\,^{t+\Delta t}%
\boldsymbol{S}^{|e}/d\,^{tr}\boldsymbol{A}_e$ & $\,^{t+\Delta t}\mathbb{A}%
_{ep}^{|e}= d\,^{t+\Delta t}\boldsymbol{T}^{|e}/d\,^{tr}\boldsymbol{E}_e$ \\%
[2ex]
(9) Final Mappings (e.g. TL formulation)& $\,^{t+\Delta t}\boldsymbol{S}%
^{|e}\rightarrow \,^{t+\Delta t}\boldsymbol{S} \newline
\,^{t+\Delta t}\mathbb{C}_{ep}^{|e}\rightarrow \,^{t+\Delta t}\mathbb{C}%
_{ep}$ & $\,^{t+\Delta t}\boldsymbol{T}^{|e}\rightarrow \,^{t+\Delta t}%
\boldsymbol{S} \newline
\,^{t+\Delta t}\mathbb{A}_{ep}^{|e}\rightarrow \,^{t+\Delta t}\mathbb{C}%
_{ep}$ \\[2ex] \hline
\end{tabular}%
\end{table}

\section{Numerical examples\label{SEC examples}}

The purpose of the examples in this section is (1) to show that the
formulation may be implemented successfully in an implicit finite element
code and (2) to compare results from our proposal with a classical
formulation. To this end, we demonstrate three finite element examples of
single crystals (so differences may not be attributed to polycrystal issues
or the specific RVE).

\subsection{Uniaxial compression of a cylinder made of FCC single crystal}

\begin{figure}[h!]
\centering
\includegraphics[width=\textwidth]{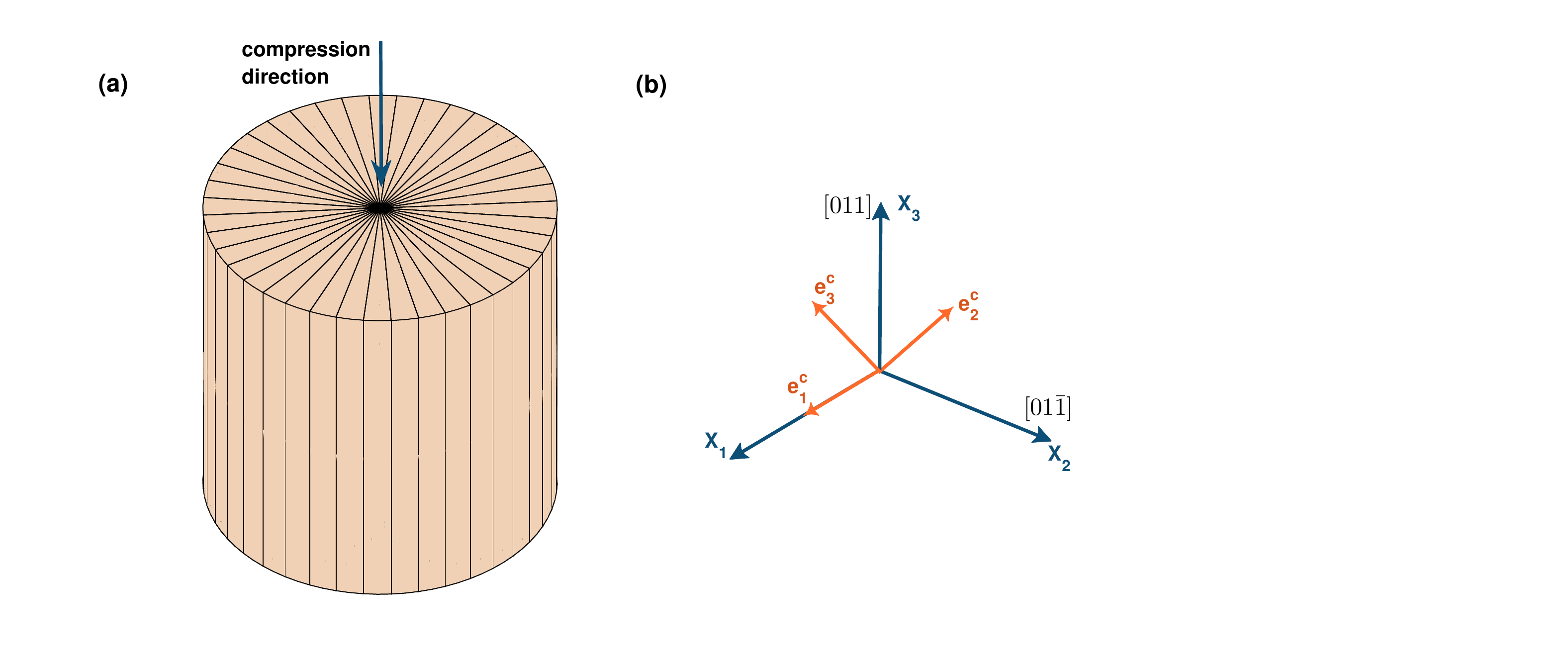}
\caption{(a) the initial finite element mesh and the loading direction; (b)
the crystal orientaion}
\label{cylinder_direction}
\end{figure}

\bigskip To compare the current framework with the classical formulation, we
selected an example from \cite{Kalidindi1993}. We performed the computations
with both frameworks and also compared our results to the results reported
therein. The material is the face-centered (FCC) single crystal copper. For
this type of crystal, there are $12$ slip systems. As in Ref. \cite%
{Kalidindi1993}, the loading direction is along the $\langle 011\rangle $
direction of the crystal and along the $X_{3}$ (axial) direction of the
specimen. The relation between the crystal orientation and the specimen
coordinate system is shown in Fig. \ref{cylinder_direction}. {
The parameters of the material are summarized in Table \ref{tab.parameters}}.
Since $\kappa _{g}\simeq \tau _{g}^{|e}$ and $\dot{\gamma}_{g}^{p}\simeq
\gamma _{g}$ and elastic strains are infinitesimal, we have used the same
parameters for both models.

\begin{table}[tbp]
\caption{Material paramters used for the copper single crystal}\centering%
\begin{tabular}{|cc|cc|}
\hline
Constant & value & Constant & Value \\ \hline
$C_{11}=A_{11}$ & $170$ GPa & $h_{gj}$ & $[1.4,1]$ \\
$C_{12}=A_{12}$ & $124$ GPa & $H_g$ & $250$ MPa \\
$C_{44}=A_{44}$ & $75$ GPa & $\alpha$ & $2.5$ \\
$\dot{\gamma}_{0}$ & $0.001$ s$^{-1}$ & $\tau _{cg}^{|e}$ & $16$ MPa \\
$m$ & $0.012$ & $\tau _{sg}^{|e}$ & $190$ MPa \\ \hline
\end{tabular}%
\label{tab.parameters}
\end{table}

For the finite element analysis, the exact specimen dimensions, crystal
orientation, loading direction as well as loading speed are prescribed. The
finite element model is meshed into $40$ elements, as in \cite{Kalidindi1993}%
. To run the example with the current framework, we used our in-house finite
element code Dulcinea (already employed in other works of the group, e.g.
\cite{Sanz2017,latorre2016fully,CrespoAuxetic,Minano}). For the purpose of
comparison, simulations with the conventional framework is performed with
the commercial FE code Abaqus, using user subroutines employed and validated
in other works, e.g. \cite{Ma2015,Ma2006,Ma2006a,Ma2004}.

\begin{figure}[h!]
\centering
\includegraphics[width=1.\textwidth]{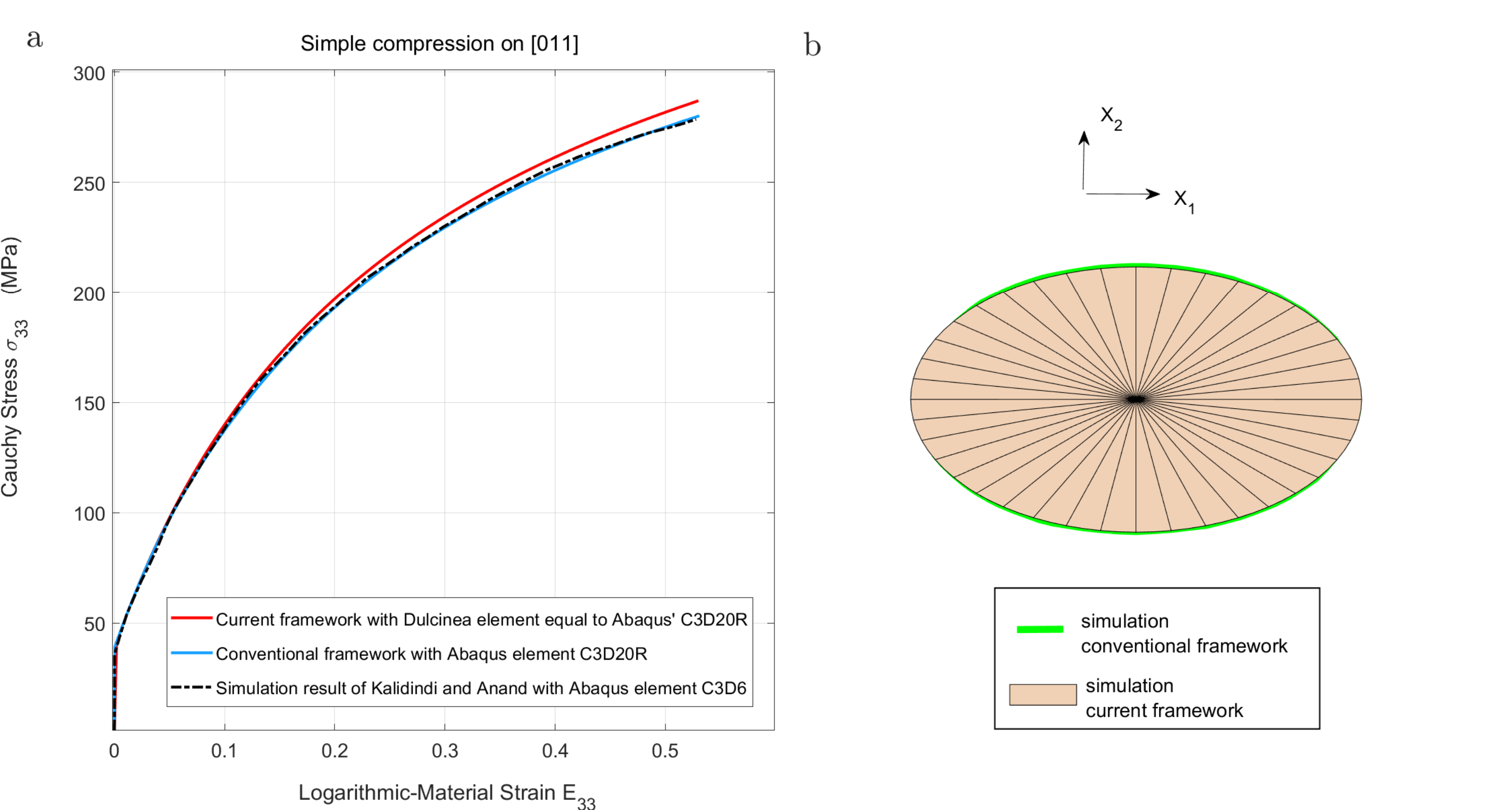}
\caption{a) Comparison of the stress-strain curve of the proposed and the
conventional frameworks, and with the
results in Fig. 2 of Ref. \protect\cite{Kalidindi1993}. b) Comparison of the deformed shapes between the classical and the present formulations.}
\label{cylinder_stress_strain}
\end{figure}

To facilitate comparisons eliminating possible differences due to the
specific element formulation, with the currrent framework in Dulcinea we
used an element equivalent to Abaqus' $C3D20R$: a quadratic brick element
with $20$ nodes and $8$ integration points which does not suffer volumetric
locking. Possible hourglass modes are not propagable. As shown in Fig. \ref%
{cylinder_direction}, the $20$ nodes element would degenerate\emph{\ }to the
prism geometry because of the nature of the mesh. With the conventional
framework in Abaqus, our simulations are run with the element type $C3D20R$
and in Ref. \cite{Kalidindi1993} they used element $C3D6$, to which results
from them we also compare in Fig.  \ref{cylinder_stress_strain}a.

If the material is isotropic, after the deformation, in both the simulated
result and the experimental result \cite{Kalidindi1993}, the dimension of
the cylinder in $x_{2}$-direction does not change, only the dimension in $%
x_{1}$-direction changes. A comparison of the simulated deformed shapes by
the current framework and the conventional framework in Kalidindi, Bronkhorst and Anand
is shown in Fig. \ref{cylinder_stress_strain}b. It is seen that differences are
small.
% \begin{figure}[htp!]
% \centering
% \includegraphics[width=0.7\textwidth]{deformed_cross}
% \caption{The deformed shape of the cross section of the cylinder by
% simulations of both frameworks}
% \label{deformed_cross}
% \end{figure}

During the loading process, there is no rotation of the crystals, and the
stress condition inside the cylinder is completely homogeneous. The
simulated stress-strain curve by the current framework and by the framework
of Kalidindi, Bronkhorst and Anand from Fig. 2 of Ref. \cite{Kalidindi1993} is given in
Fig. \ref{cylinder_stress_strain}a. As it can be seen, the curves
obtained with both frameworks are very similar, but differ slightly,
specially at large strains, when using the same element type. The reason
behind this difference at large strains may be due to several issues, as slightly different material parameters (for the classical framework they are interpreted in terms of Mandel stresses and for the present one are interpreted in terms of generalized Kirchhoff stresses),  to the approximation performed for the exponential mapping
in the multiplicative decomposition in the classical framework, or to the slightly different kinematics. In our
framework the exponential function for strains (symmetric tensors) are evaluated using the spectral
decomposition of the elastic deformations. For the rotations (skew-symmetric tensors), the Rodrigues formula is accurate and typical in computational mechanics, namely \cite{simo2006computational,Gallier2002}
\begin{equation}
\exp(\Delta t\boldsymbol{W})=\boldsymbol{I}+\frac{\sin\theta}{\theta}\Delta t\boldsymbol{W}+\frac{(1-\cos\theta)}{\theta^2}(\Delta t\boldsymbol{W})^2
\end{equation}
with $\theta=\Delta t|\boldsymbol{w}|$, being $\boldsymbol{w}$ the dual vector of the antisymmetric tensor $\boldsymbol{W}$, and $\exp(\Delta t\boldsymbol{W})\simeq\boldsymbol{I}+\Delta t\boldsymbol{W}$ for $\Delta t\rightarrow 0$.

\subsection{Simple shear test}

\begin{figure}[h!]
\centering
\includegraphics[width=1.0\textwidth]{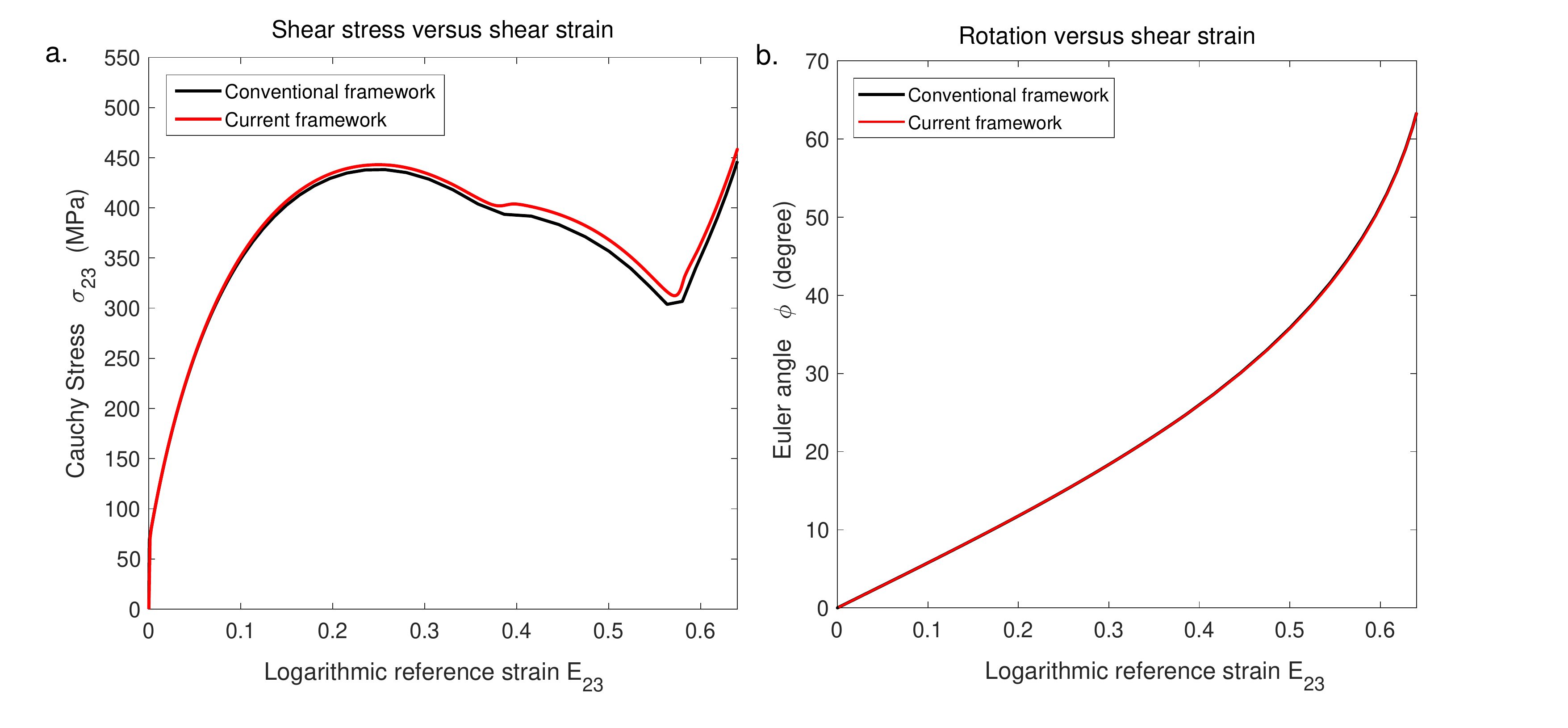}
\caption{a) Shear stress-strain curve of the current and the conventional
frameworks. b) Rotation (second Bunge Euler angle; the first and third ones remain constant) }
\label{simple_shear_test}
\end{figure}

The simple shear test has been frequently used to check the consistency of
large strain formulations and of the stress-integration algorithm (\cite%
{Dettmer2004,Brepols2014}). Therefore, we also selected this example to
validate our proposed framework. The material parameters of the single
copper crystal from Ref. \cite{Kalidindi1993} are used and summarized in Table %
\ref{tab.parameters}. For the example, since deformations are homogeneous,
any finite element may be employed (e.g. an 8-node brick). Displacements have been imposed via penalization. The shear
stress-strain curves obtained from both frameworks are represented in Fig. %
\ref{simple_shear_test}a. It can be seen that the results are similar for
both frameworks, although there is again a small difference when strains
become very large. As expected, fast global convergence is
 obtained (see Table \ref%
{tab.convergence_rate}), even
though the important presence of rotation in this test, see Figure %
\ref{simple_shear_test}b.

In this example, comparative execution times facilitate an approximation of the efficiency of the computational schemes, because even though both models are implemented in different codes,  most computational time of the example is mainly due to the execution of the material subroutine. To obtain an accurate result, $1000$ steps have been used, one element with $8$ nodes and $8$ integration points (although just one could have been used). The proposed model used $31$ seconds in Dulcinea, whereas the classical framework used  $95$ seconds in the same computer and conditions. In the case of large finite element problems, Dulcinea is significantly less efficient because the system of equations is stored and solved in skyline format, whereas Abaqus uses sparse systems (an issue not related to the material models).

\begin{table}[tbp]
\caption{Simple shear test: global equilibrium convergence values at steps
50, 200, 400, 600}
\label{tab.convergence_rate}\centering%
\begin{tabular}{lcccc}
\hline\hline
& \multicolumn{4}{c}{Residual norm} \\
Iteration & Step $50$ & Step $200$ & Step $400$ & Step $600$ \\ \hline
\multicolumn{1}{c}{$\left( 1\right) $} & \multicolumn{1}{l}{$1.982E+02$} &
\multicolumn{1}{l}{$2.193E+02$} & \multicolumn{1}{l}{$2.705E+02$} & $%
4.002E+02$ \\
\multicolumn{1}{c}{$\left( 2\right) $} & \multicolumn{1}{l}{$2.366E+01$} &
\multicolumn{1}{l}{$1.472E+01$} & \multicolumn{1}{l}{$1.051E+01$} & $%
4.377E+01$ \\
\multicolumn{1}{c}{$\left( 3\right) $} & \multicolumn{1}{l}{$1.422E-06$} &
\multicolumn{1}{l}{$1.355E-07$} & \multicolumn{1}{l}{$5.151E-07$} & $%
1.743E-06$ \\ \hline
& \multicolumn{4}{c}{Energy norm} \\
Iteration & Step $50$ & Step$150$ & Step $250$ & Step $350$ \\ \hline
\multicolumn{1}{c}{$\left( 1\right) $} & \multicolumn{1}{l}{$1.585E+00$} &
\multicolumn{1}{l}{$1.754E+00$} & \multicolumn{1}{l}{$2.154E+00$} & $%
3.125E+00$ \\
\multicolumn{1}{c}{$\left( 2\right) $} & \multicolumn{1}{l}{$2.236E-07$} &
\multicolumn{1}{l}{$7.738E-08$} & \multicolumn{1}{l}{$2.999E-08$} & $%
3.211E-07$ \\
\multicolumn{1}{c}{$\left( 3\right) $} & \multicolumn{1}{l}{$8.082E-22$} &
\multicolumn{1}{l}{$6.485E-24$} & \multicolumn{1}{l}{$7.201E-23$} & $%
5.089E-22$ \\ \hline\hline
\end{tabular}%
\end{table}

\subsection{Drawing of a thin circular flange}

This well-known example is a simplification for deep drawing of a cup \cite%
{Miehe2004,Caminero2011}. The purpose is to simulate the deformation of the
outer part of a circular sheet in the first phase of deep drawing process
without using the contact elements. This simulation can be used to study the
anisotropic elastoplasticity of the material, and the deformed shape
reflects the earring condition in deep drawing.

\begin{figure}[htbp!]
\centering
\includegraphics[width=\textwidth]{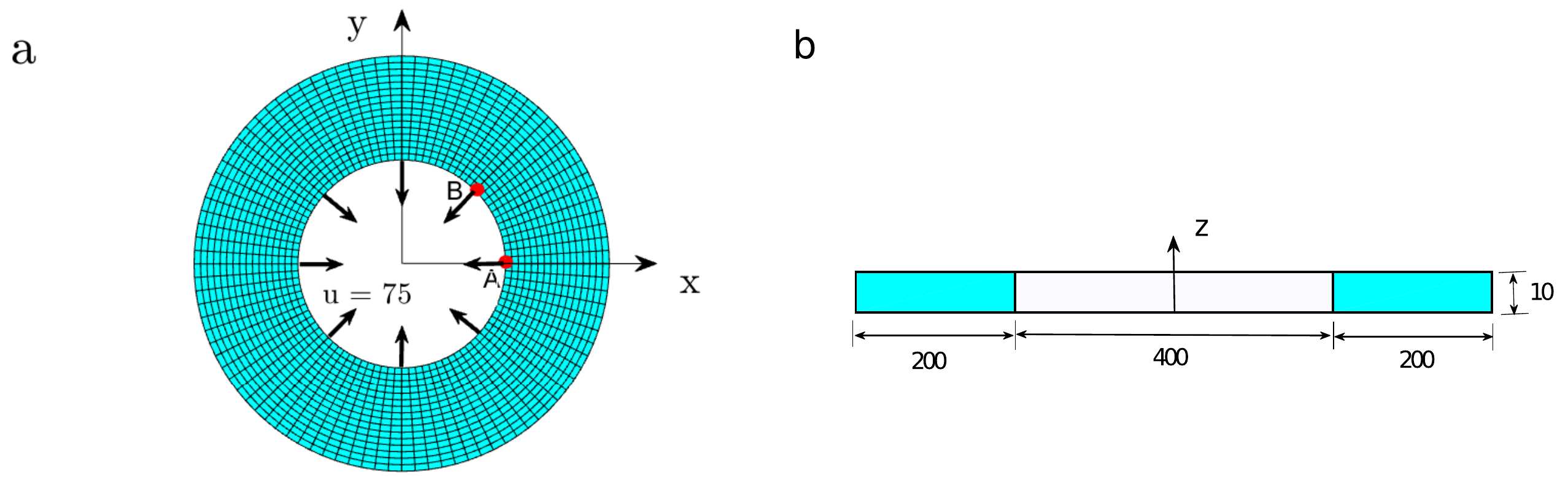}
\caption{Drawing of a thin circular flange : (a) finite element model; (b)
geometric dimensions }
\label{flange_graphic}
\end{figure}

A radial displacement up to $75mm$ is applied to the inner circle of the
flange. The nodal forces of two nodes are recorded during the loading. The
geometric shape, the loading condition\ and the locations of the two nodes
are shown in Fig. \ref{flange_graphic}. To avoid buckling, the vertical
displacement of the flange is fully supported in one side and a vanishing
rotation of the inner rim around the vertical axis is prescribed.

\begin{figure}[htbp]
\centering
\subfloat[crystal direction: $x\rightarrow\left<1,0,0\right>$;
$z\rightarrow\left<0,0,1\right>$]{\includegraphics[width=0.45\textwidth]{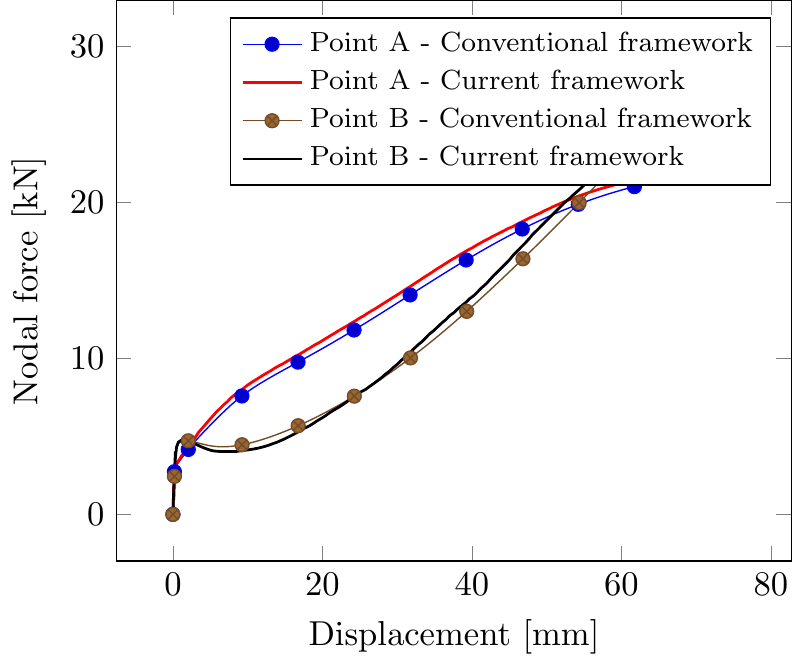}}
%-----
\subfloat[crystal direction: $x\rightarrow\left<1,\bar{1},0\right>$;
$z\rightarrow\left<0,0,1\right>$]{\includegraphics[width=0.45\textwidth]{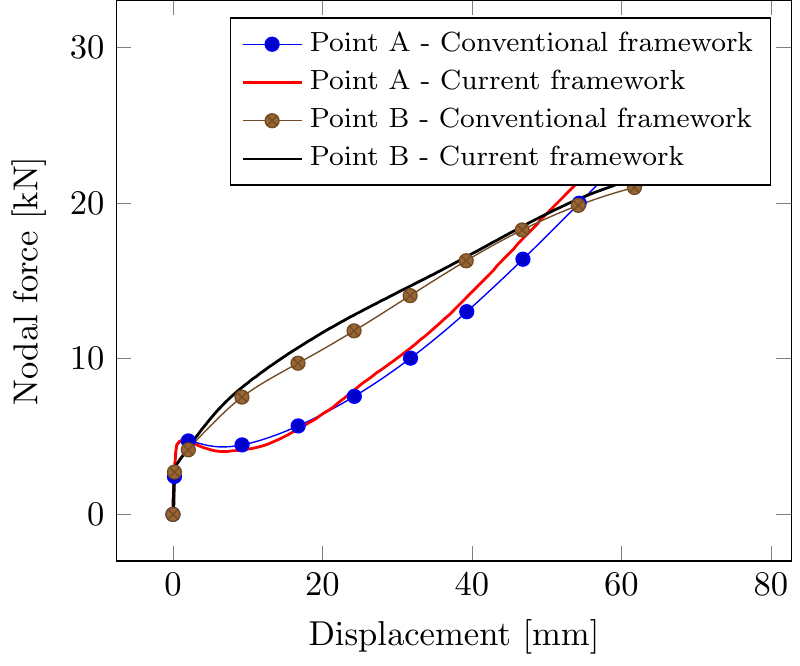}}
\newline
\subfloat[crystal direction: $x\rightarrow\left<0,\bar{1},1\right>$;
$z\rightarrow\left<1,0,0\right>$]{\includegraphics[width=0.45\textwidth]{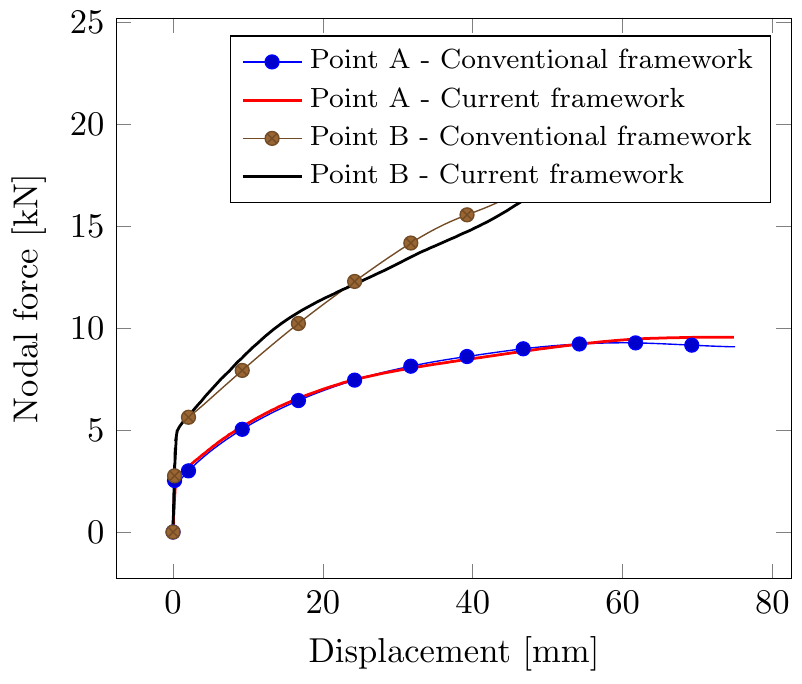}}
\caption{Comparison of nodal forces at point A and B}
\label{fig.comparison1}
\end{figure}

\begin{figure}[h!]
\centering
\includegraphics[width=1.0\textwidth]{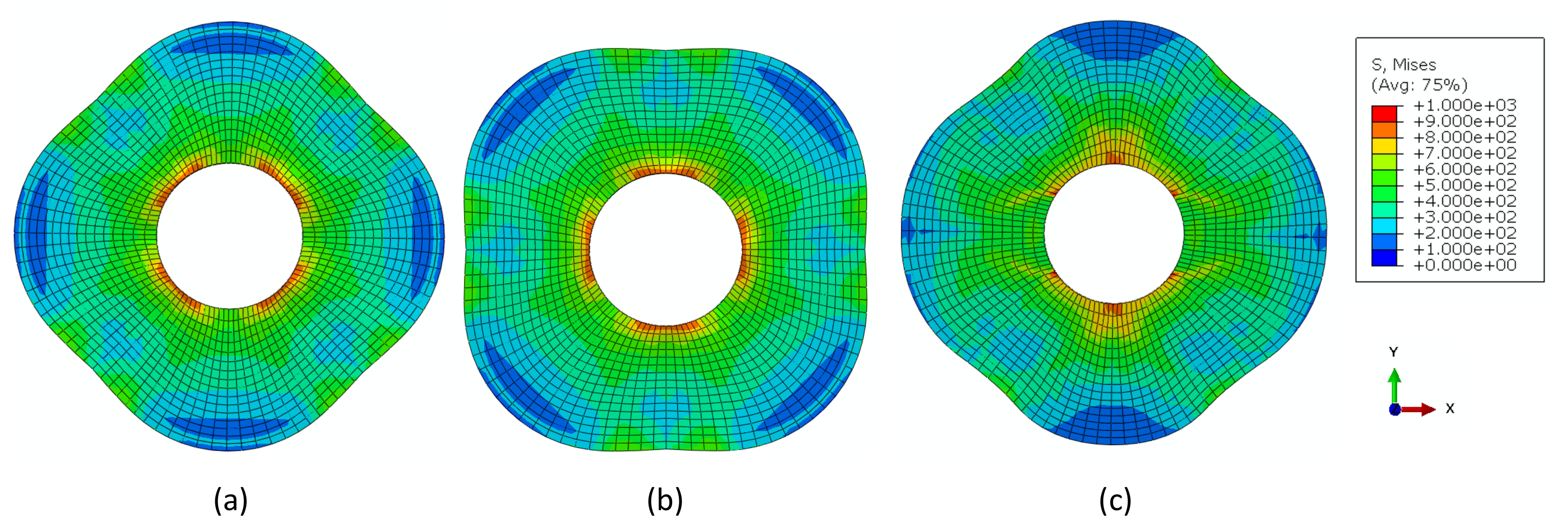}
\caption{Von Mises stress contour [MPa] for three crystal orientations with the
conventional framework performed in Abaqus}
\label{von_mises_contour_abaqus}
\end{figure}

\begin{figure}[h!]
\centering
\includegraphics[width=0.85\textwidth]{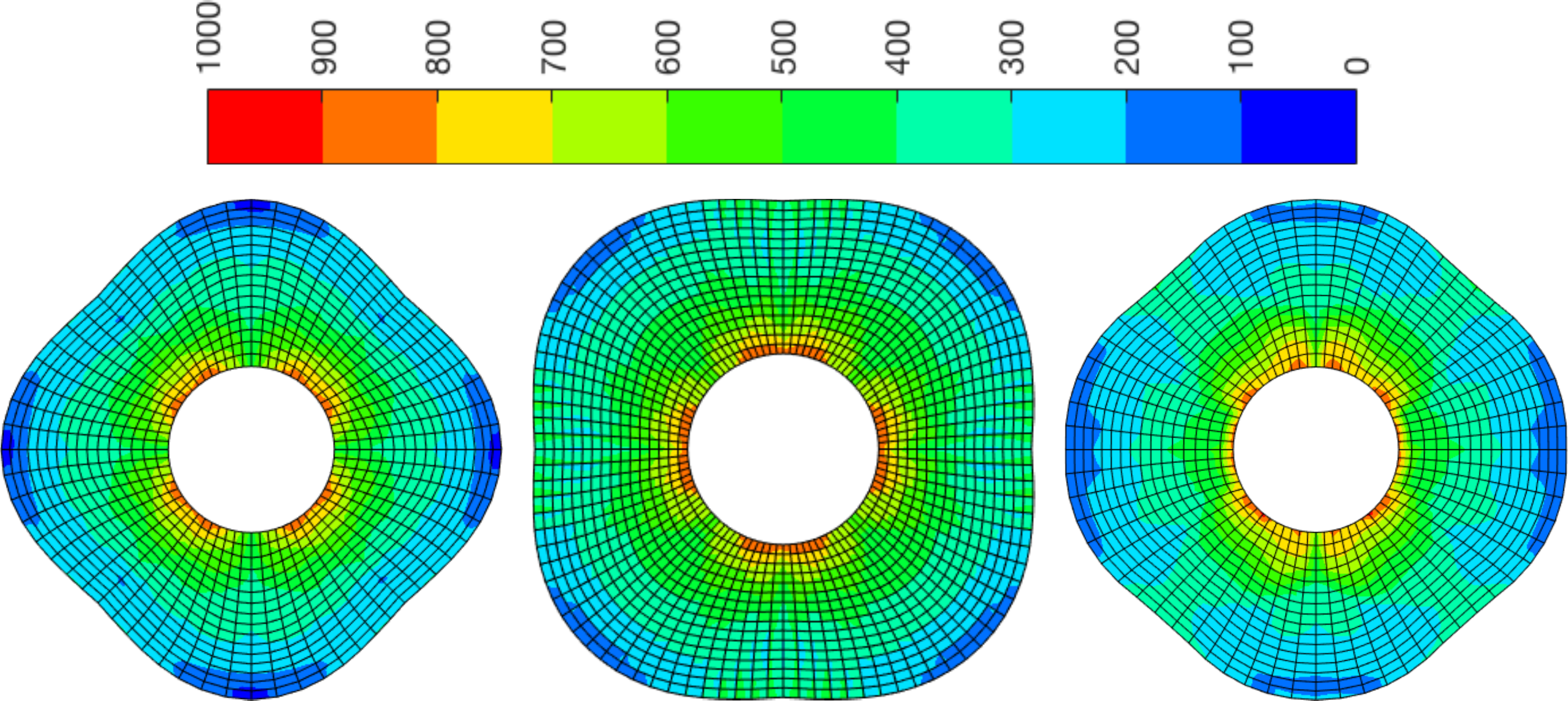}
\caption{Von Mises stress contour [MPa] for three crystal orientations with the
proposed model performed in Dulcinea}
\label{von_mises_contour}
\end{figure}

This example is used to compare the results obtained by the current framework
with those from the conventional framework. Since the problem is nearly under
plane stress conditions, supported in one side, the simulations have been
performed with Abaqus standard elements $C3D8$ (fully integrated 8-node
bricks) and Dulcinea's equivalent elements. Band-plot checks have been
performed to discard any relevant locking phenomena \cite{Bathebook}. Both models have $1536$
elements. In the direction of the thickness, one layer of elements is used.

The purpose of this example is to compare both frameworks, so we have
employed the material parameters shown in \ref{tab.parameters}. The material
parameters of the single crystal Al from \cite{Kalidindi1993} are used,
except for the sensitivity exponent $m$ which has been $m=0.1$ to reduce the
computational time of the examples.

To show the different deformed shapes due to different crystal orientations,
the example has been computed with three single crystal orientations:

\begin{enumerate}[label=(\alph*)]

\item Euler angles $(0,0,0) $, whose crystal direction $\langle 001\rangle $
aligns with the axial direction $z$ of the flange, and the crystal direction
$\langle 100\rangle $ aligns with the $x$ axial of the specimen coordinate
system.

\item Euler angles $(45^{\circ},0,0)$, whose crystal direction $\langle
001\rangle $ aligns with the axial direction of the flange, and crystal
direction $\langle 1\bar{1}0\rangle $ aligns with the $x$ axial of the
specimen coordinate system

\item Euler angles $(180^{\circ},45^{\circ},180^{\circ})$, whose crystal
direction $\langle 0\bar{1}1\rangle $ aligns with the axial direction of the
flange, and crystal direction $\langle 100\rangle $ aligns with the $x$
axial of the specimen coordinate system.
\end{enumerate}

If the material were isotropic, the outer rim of the flange would stay as a
regular circle during the loading, but because of the crystallography of the
FCC crystal, the outer rim shows different ``wave" forms depending on
different crystal orientations and loading directions, which is the same
phenomenon as earring in deep drawing of an anisotropic (rolled) sheet
metal. For all the three crystal orientations, the deformed shapes are shown
in Figs. \ref{von_mises_contour_abaqus} and \ref{von_mises_contour} along
von Mises stresses. It is seen again that only small differences are
observed; we note that stress smoothing employed in both programs is different. The nodal forces of two nodes in the top surface marked in Fig. %
\ref{flange_graphic} are obtained with both frameworks and plotted in Fig. %
\ref{fig.comparison1} for all three crystal orientations. It is seen once
more that, as expected and in line with the previous results, both
frameworks give very similar results, but differ slightly as plastic
deformations increase substantially.

All the simulations are done with total loading time set as $1$ $s$, and
with a fixed step size employing plain
Newton-Raphson iterations. In Table \ref{globalconv}, we list the global
convergence iteration values of several typical steps for the simulation
with initial crystal Euler angles $(0,0,0)$. It is shown that the
convergence is asymptotically quadratic as expected from Newton iterations,
and the desirable residue is typically reached in $4$ steps.

\begin{table}[tbp]
\centering%
\begin{tabular}{lcccc}
\hline\hline
& \multicolumn{4}{c}{Residual norm} \\
Iteration & Step $50$ & Step $150$ & Step $250$ & Step $350$ \\ \hline
\multicolumn{1}{c}{$\left( 1\right) $} & \multicolumn{1}{l}{$1.189E+06$} &
\multicolumn{1}{l}{$1.227E+06$} & \multicolumn{1}{l}{$1.241E+06$} & $%
1.257E+06$ \\
\multicolumn{1}{c}{$\left( 2\right) $} & \multicolumn{1}{l}{$6.785E+03$} &
\multicolumn{1}{l}{$5.185E+03$} & \multicolumn{1}{l}{$4.272E+03$} & $%
3.784E+03$ \\
\multicolumn{1}{c}{$\left( 3\right) $} & \multicolumn{1}{l}{$1.641E+01$} &
\multicolumn{1}{l}{$5.417E+00$} & \multicolumn{1}{l}{$2.335E+00$} & $%
1.238E+00$ \\
\multicolumn{1}{c}{$\left( 4\right) $} & \multicolumn{1}{l}{$2.040E-01$} &
\multicolumn{1}{l}{$3.079E-02$} & \multicolumn{1}{l}{$8.899E-03$} & $%
9.737E-03$ \\ \hline
& \multicolumn{4}{c}{Energy norm} \\
Iteration & Step $50$ & Step$150$ & Step $250$ & Step $350$ \\ \hline
\multicolumn{1}{c}{$\left( 1\right) $} & \multicolumn{1}{l}{$2.057E+06$} &
\multicolumn{1}{l}{$2.123E+06$} & \multicolumn{1}{l}{$2.145E+06$} & $%
2.168E+06$ \\
\multicolumn{1}{c}{$\left( 2\right) $} & \multicolumn{1}{l}{$4.741E+01$} &
\multicolumn{1}{l}{$3.162E+01$} & \multicolumn{1}{l}{$2.071E+01$} & $%
1.538E+01$ \\
\multicolumn{1}{c}{$\left( 3\right) $} & \multicolumn{1}{l}{$2.150E-03$} &
\multicolumn{1}{l}{$2.058E-04$} & \multicolumn{1}{l}{$3.504E-05$} & $%
8.734E-06$ \\
\multicolumn{1}{c}{$\left( 4\right) $} & \multicolumn{1}{l}{$3.357E-07$} &
\multicolumn{1}{l}{$5.729E-09$} & \multicolumn{1}{l}{$2.952E-10$} & $%
2.858E-11$ \\ \hline\hline
\end{tabular}%
\caption{Global equilibrium convergence values at steps 50, 150, 150, 350 of
the finite element simulation}
\label{globalconv}
\end{table}
%\afterpage{\clearpage}

\subsection{Texture evolution  example under axial load}
\begin{figure}[htbp!]
\centering
\includegraphics[width=1.0\textwidth]{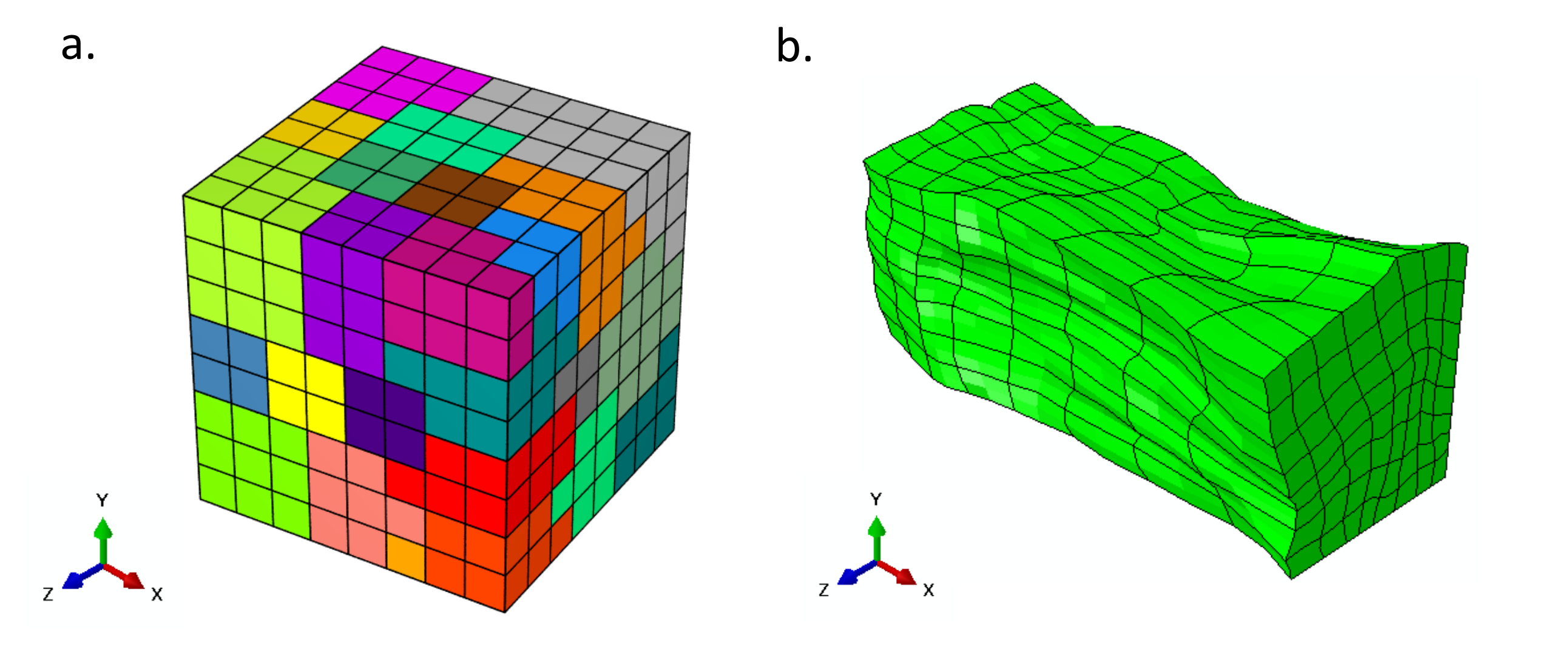}
\caption{a) Polycrystal cubic sample with random texture. b) Deformed specimen (visually indistinguishable using either formulation)}
\label{cubic_structure}
\end{figure}

\begin{figure}[htbp!]
\centering
\includegraphics[clip, trim=4cm 4cm 4cm 4cm, width=0.7\textwidth]{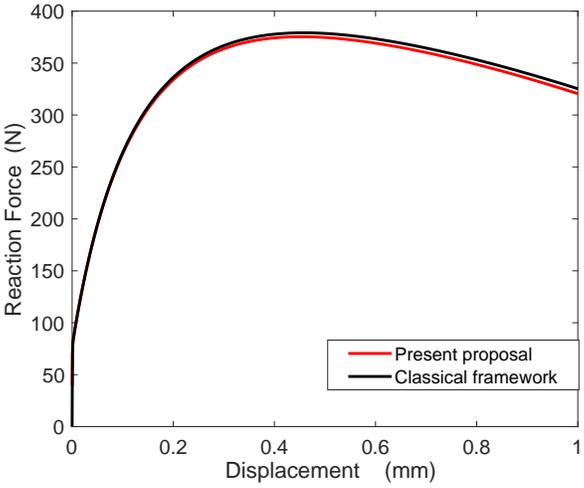}
\caption{Comparison of the reaction force versus displacement in the polycrystal cubic sample using both the classical and the proposed formulation}
\label{reactionforce_displacement}
\end{figure}

\begin{figure}[htbp!]
\centering
\includegraphics[width=0.85\textwidth]{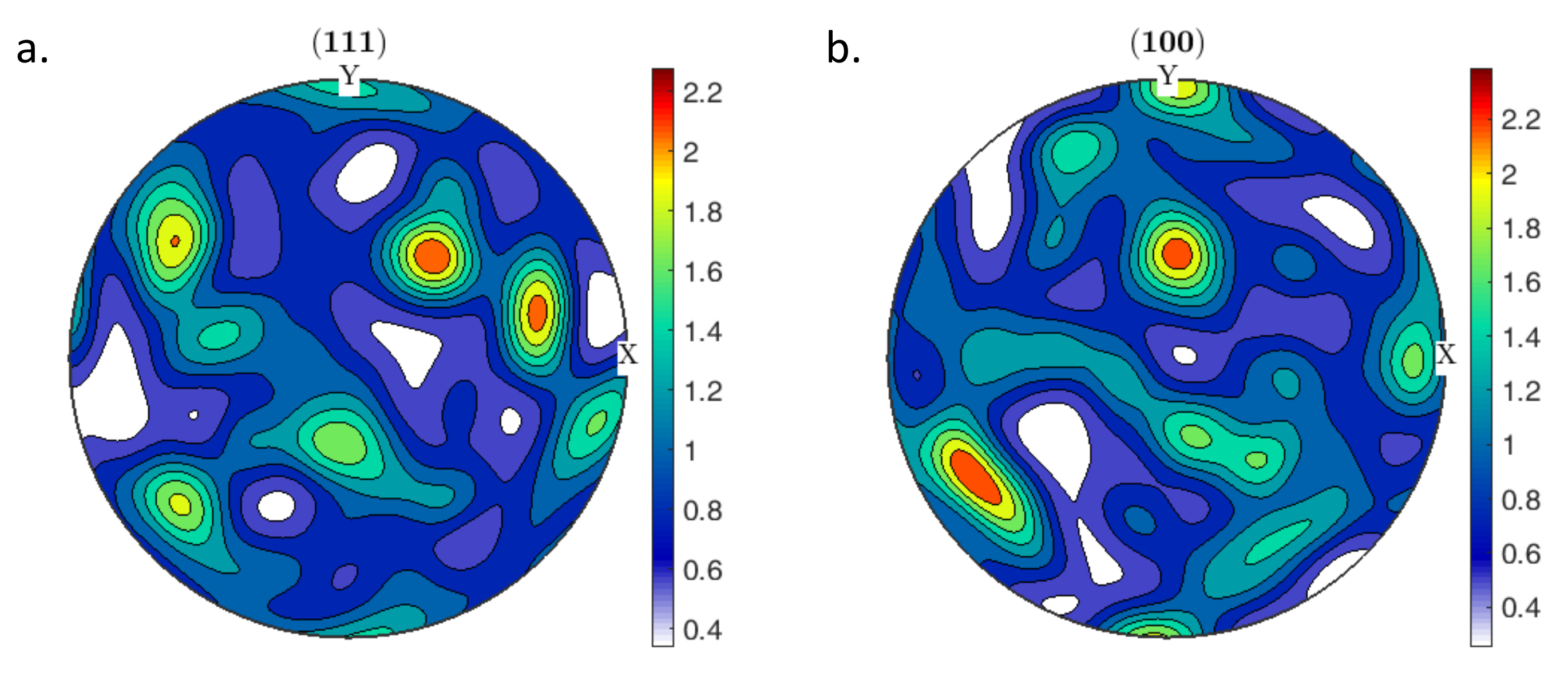}
\caption{Initial crystallographic texture pole plots for the cube example for directions (a) (111) and (b) (100). }
\label{pole_figure_before}
\end{figure}

\begin{figure}[htbp!]
\centering
\includegraphics[width=0.9\textwidth]{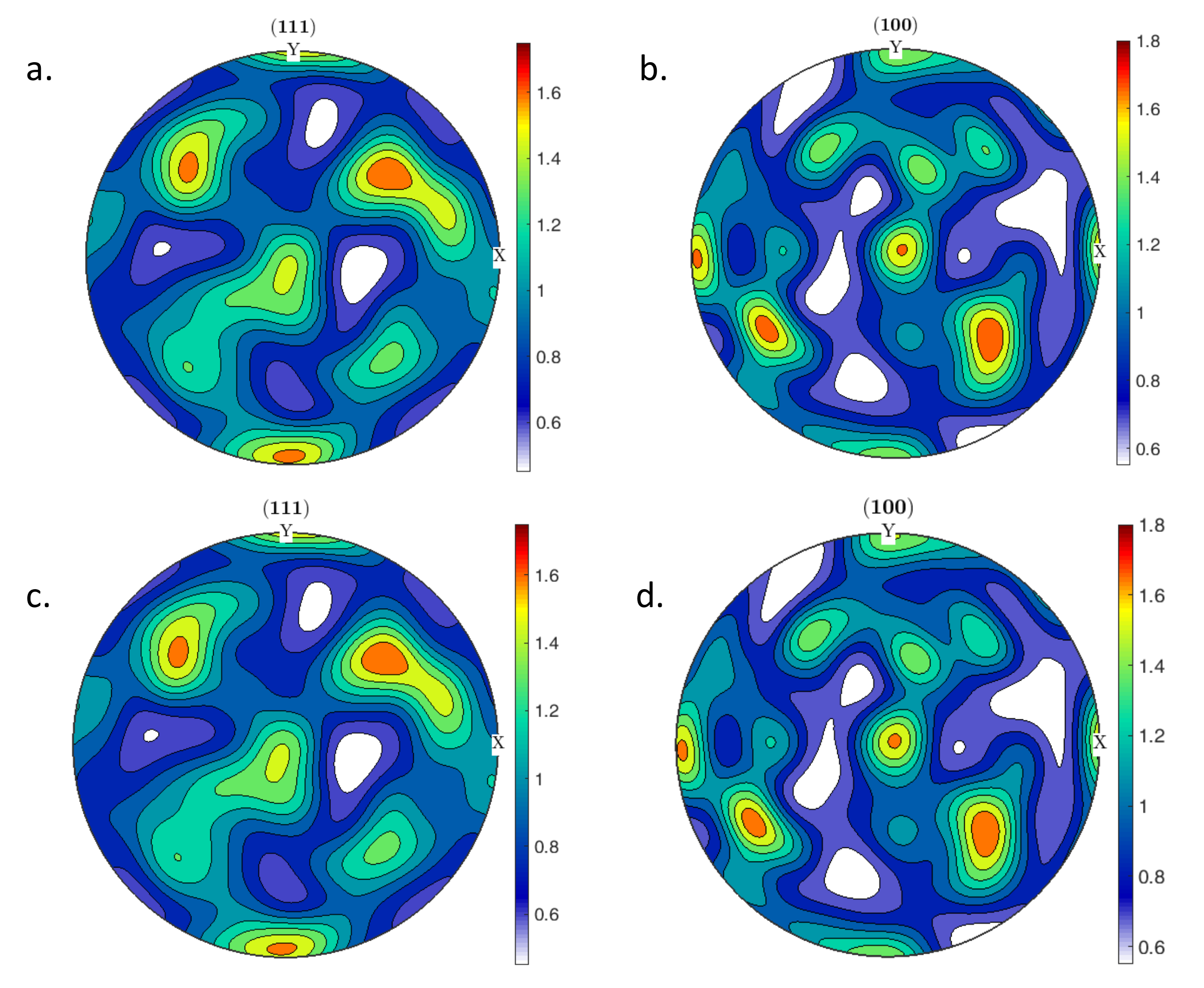}
\caption{Final crystallographic texture pole plots for the cube example for directions (111) and (100). Top row, (a) and (b), have been obtained using the  classical framework. Bottom row, (c) and (d) have been obtained using the proposed framework. }
\label{pole_figure_after}
\end{figure}

In this example we compare the texture evolution obtained from both approaches for a polycrystal under axial loading. The same copper parameters of the previous examples (Table \ref{tab.parameters}) are used. A random initial texture with $50$ orientations is used.  The process of axial tension is simulated with a cubic structure that contains $512$ elements, and the standard element that has $20$ nodes and $8$ integration points is used. The mesh is shown in Fig. \ref{cubic_structure}a. The applied tensile displacement is of the length of the dimension of the cube ($1 mm$), and the obtained deformed mesh is plotted in Fig. \ref{cubic_structure}b, which is visually indistinguishible using either formulation, so only the Abaqus result is shown. Resultant reaction force versus imposed displacement is compared in Fig. \ref{reactionforce_displacement}, where again it is observed that differences are minor. The original texture is shown in Fig. \ref{pole_figure_before}. The final textures using both formulations are shown in Fig. \ref{pole_figure_after}.  All the pole figures are calculated by assuming that every integration point has an equal volume. It is observed that the differences between both frameworks are very small.

\section{Conclusions}

In this paper, we present a new framework for crystal plasticity based on
elastic corrector rates, motivated in a similar successful formulation for
continuum anisotropic hyperelasto-plasticity at large strains. The
formulation uses the Kr\"oner-Lee multiplicative decomposition, but does not
need an explicit, algorithmically motivated, exponential mapping. Instead,
the use of logarithmic strains result in an additive structure parallel to
the classical continuum and algorithmic formulation of infinitesimal
plasticity. Noteworthy, the integration is fully performed in terms of elastic variables instead of plastic ones. The formulation is also fully comparable to our continuum framework, with the natural exception of using the specific flow conditions for crystal plasticity. As in the continuum formulation, the unsymmetric
Mandel stress plays no role in the formulation; any hyperelastic (e.g.
nonquadratic, anisotropic) stored energy may be employed; the flow rule is
conventional; a plain backward-Euler algorithm preserves volume during
plastic flow; both the continuum and algorithmic formulation are parallel to that of infinitesimal plasticity and only explicit purely kinematic tensors
transform the working stress-strain couple (logarithmic strains and
generalized Kirchhoff stresses) to any desired one (e.g. Green-Lagrange
strains and second Piola-Kirchhoff stresses). Skew-symmetric flow is uncoupled and the elastic rotations are computed explicitly from the solution of the symmetric part. Hence, now two fully equivalent sound and robust frameworks for continuum and crystal plasticity are available, and may facilitate comparisons between simulations using both approaches.

We have performed comparisons with the classical crystal plasticity approach
(Kalidindi-Bornkhorst-Anand algorithm). For the comparisons performed, as expected, we
have observed only some differences when the plastic deformations are
significatively large. We attribute those differences to the different
approximations employed in adding simultaneous slips and in the exponential
mapping.

We have also developed a fully implicit integration algorithm, consistently
linearized, based on a plain backward-Euler algorithm. It is observed that,
as expected, asymptotic quadratic convergence is obtained in both the local
and global (equilibrium) iterations.
%\end{linenumbers}
\section*{Acknowledgements}

{ Partial financial support for this work has been given by Agencia Estatal de
Investigaci\'{o}n of Spain under grant PGC2018-097257-B-C32.}
\appendix
\section{Consistent elastoplastic tangent moduli}

%\section{Consistent elastoplastic tangent moduli}

In order to preserve the asymptotic quadratic convergence of Newton schemes,
a consistent linearization of the stress integration algorithm is needed. As
in the continuum case, we develop the basic tensor using logarithmic stress
and strain measures in the trial configuration, which is the configuration
in which all the stress integration takes place. This is due to the fact
that the global integration algorithm just changes the displacements and,
hence, gives at each global iteration a modified trial elastic gradient $%
^{tr}\boldsymbol{X}_{e}$ for the element integration point. Then, for
performing the output of the material subroutine, the tensor may be mapped
to any other configuration or stress-strain couple, typically second
Piola-Kirchhoff stresses and Green-Lagrange strains in the material
configuration used in Total Lagrangean finite element formulations.

\subsection{Kinematics}

The basic tangent relating the stress change for that change is%
\begin{equation}
^{t+\Delta t}\mathbb{A}_{ep}^{|tr}:=\frac{d\,^{t+\Delta t}\boldsymbol{T}%
^{|tr}}{d\,^{tr}\boldsymbol{E}_{e}}
\end{equation}%
which, recalling the arguments in Eq. \eqref{Ttr to Te}, may be approximated
by%
\begin{equation}
^{t+\Delta t}\mathbb{A}_{ep}^{|tr}=\frac{d\,^{t+\Delta t}\boldsymbol{T}^{|tr}%
}{d\,^{tr}\boldsymbol{E}_{e}}\simeq \frac{d\,^{t+\Delta t}\boldsymbol{T}^{|e}%
}{d\,^{tr}\boldsymbol{E}_{e}}=\frac{d\,^{t+\Delta t}\boldsymbol{T}^{|e}}{%
d\,_{\hspace{3.2ex}0}^{t+\Delta t}\boldsymbol{E}_{e}}:\frac{d\,_{\hspace{%
3.2ex}0}^{t+\Delta t}\boldsymbol{E}_{e}}{d\,^{tr}\boldsymbol{E}_{e}}%
=\,^{t+\Delta t}\mathbb{A}^{|e}:\frac{d\,_{\hspace{3.2ex}0}^{t+\Delta t}%
\boldsymbol{E}_{e}}{d\,^{tr}\boldsymbol{E}_{e}}  \label{Atr}
\end{equation}%
where $\,^{t+\Delta t}\mathbb{A}^{|e}$ is the elasticity tensor (in terms of
$\boldsymbol{T}^{|e}$ and $\boldsymbol{E}_e$), evaluated at $t+\Delta t$, but
assumed constant in the examples. The tensor $d\,_{\hspace{3.2ex}%
0}^{t+\Delta t}\boldsymbol{E}_{e}/d\,^{tr}\boldsymbol{E}_{e}$ is the \textit{%
total} derivative, result of the integration algorithm, and relates the
trial elastic strain $^{tr}\boldsymbol{E}_{e}$ with the final one$\,_{%
\hspace{3.2ex}0}^{t+\Delta t}\boldsymbol{E}_{e}$. This relation is given by
the residue function Eq. \eqref{EqR2}, which in tensor format is%
\begin{equation}
^{t+\Delta t}\boldsymbol{R}_{E}:=\,_{\hspace{3.2ex}0}^{t+\Delta t}%
\boldsymbol{E}_{e}-~^{tr}\boldsymbol{E}_{e}+\sum_{g=1}^{G}%
\Delta \gamma _{g}\,\boldsymbol{N}_{g}^{s}=\boldsymbol{0}  \label{residueEe}
\end{equation}
so%
\begin{equation}
\frac{d\,_{\hspace{3.2ex}0}^{t+\Delta t}\boldsymbol{E}_{e}}{d^{tr}%
\boldsymbol{E}_{e}}=\mathbb{I}^s-\sum\limits_{g=1}^{G}%
\boldsymbol{N}_{g}^{s}\otimes \frac{d\Delta \gamma _{g}}{d^{tr}\boldsymbol{E}%
_{e}}  \label{dEdE}
\end{equation}
The total derivative ${d\Delta \gamma _{g}}/{d^{tr}\boldsymbol{E}_{e}}$
depends on the specific model and desired algorithm (semi-implicit or fully
implicit).

\subsection{Example: viscous-type hardening equations}

At the convergence step, for each glide mechanism $g$, the backward-Euler
approximation gives $\Delta \gamma_g = \Delta t\,^{t+\Delta t}\dot{\gamma}_g$%
, so
\begin{equation}
\frac{d\Delta \gamma _{g}}{d^{tr}\boldsymbol{E}_{e}}=\Delta t\frac{%
d^{t+\Delta t}\dot{\gamma}_{g}}{d^{tr}\boldsymbol{E}_{e}}
\end{equation}
To use Eqs. \eqref{power_law} and \eqref{latenth}, we perform a change of
variables to the one used therein by applying the chain rule---the sum is
due to latent hardening in Eq. \eqref{latenth}; an approximation may be
obtained accounting only for self-hardening terms

\begin{equation}
\frac{d\Delta \gamma _{g}}{d^{tr}\boldsymbol{E}_{e}}=\Delta t\underbrace{\left[{}\frac{%
\partial^{t+\Delta t}\dot{\gamma}_{g}}{\partial ^{t+\Delta t}\tau_g%
^{|e}}\frac{\partial ^{t+\Delta t}\tau_g%
^{|e}}{\partial \,_{\hspace{%
3.2ex}}^{t+\Delta t}\boldsymbol{T}_{e}}:\frac{d^{t+\Delta t} \boldsymbol{T}^{|e}}{d \,_{\hspace{%
3.2ex}0}^{t+\Delta t}\boldsymbol{E}_{e}}+\frac{\partial^{t+\Delta t}\dot{\gamma}_{g}}{%
\partial^{t+\Delta t} \tau_{cg}^{|e}}{}\Delta t\frac{\partial ^{t+\Delta
t} \dot\tau_{cg}^{|e}}{\partial \,_{\hspace{%
3.2ex}0}^{t+\Delta t}\boldsymbol{E}_{e}}\right]}_{d^{t+\Delta t}\dot{\gamma}_{g}/d \,_{\hspace{%
3.2ex}0}^{t+\Delta t}\boldsymbol{E}_{e}}:\frac{d \,_{\hspace{3.2ex}%
0}^{t+\Delta t}\boldsymbol{E}_{e}}{d ^{tr}\boldsymbol{E}_{e}}
\end{equation}%
where  ${\partial ^{t+\Delta t}\tau_g%
^{|e}}/{\partial \,_{\hspace{%
3.2ex}}^{t+\Delta t}\boldsymbol{T}_{e}}=\boldsymbol{N}^s_g$ and ${d^{t+\Delta t} \boldsymbol{T}^{|e}}/{d \,_{\hspace{%
3.2ex}0}^{t+\Delta t}\boldsymbol{E}_{e}}=\mathbb{A}^{|e} $ (both constants), and ${%
\partial^{t+\Delta t}\dot{\gamma}_{g}}/{\partial ^{t+\Delta t}\tau_g%
^{|e}} $ and ${%
\partial^{t+\Delta t}\dot{\gamma}_{g}}/{\partial ^{t+\Delta t}\tau_{cg}%
^{|e}} $ are scalar values obtained immediately from Eq. \eqref{power_law}. The remaining term still to compute may be neglected in the tangent because it results in terms of the order of $(\Delta t)^2$. If included, a system of equations must be solved because it includes effects of cross-hardening; other option is to include only the self-hardening term. The term is

\begin{equation}
\frac{\partial ^{t+\Delta
t} \dot\tau_{cg}^{|e}}{\partial \,_{\hspace{%
3.2ex}0}^{t+\Delta t}\boldsymbol{E}_{e}}=\sum_{j=1}^{G}\frac{%
\partial ^{t+\Delta
t}\dot\tau_{cg}^{|e}}{\partial^{t+\Delta t}\dot{\gamma}_{j}}\frac{d^{t+\Delta t}\dot{\gamma}_{j}}{d \,_{\hspace{%
3.2ex}0}^{t+\Delta t}\boldsymbol{E}_{e}}
+\Delta t\frac{%
\partial ^{t+\Delta
t}\dot\tau^{|e}_{cg}}{\partial ^{t+\Delta
t}\tau^{|e}_{cj}}\frac{\partial ^{t+\Delta
t} \dot\tau_{cj}^{|e}}{\partial \,_{\hspace{%
3.2ex}0}^{t+\Delta t}\boldsymbol{E}_{e}}
 \end{equation}
where again the last term is of higher order. This is a system of equations which may be solved in matrix format to obtain
all ${d\Delta \gamma _{g}}/{d^{tr}\boldsymbol{E}_{e}}$. These may be
substituted in Eq. \eqref{dEdE}, and thereafter the result in Eq. \eqref{Atr}
to obtain $\,^{t+\Delta t}\mathbb{A}_{ep}^{|tr}$

\subsection{Final mappings}Once the elastoplastic tangent in the trial intermediate configuration is
known, it can be mapped into the reference configuration to the desired
format, see, e.g. \cite{latorre2015anisotropic,latorre2016fully,Sanz2017}.
In particular, for total Lagrangean formulations we map the tangent to the
usual material second Piola-Kirchhoff stress and Green-Lagrange strains,
following the path%
\begin{equation}
^{t+\Delta t}\mathbb{A}_{ep}^{|tr}=\frac{d\,^{t+\Delta t}\boldsymbol{T}^{|tr}%
}{d^{tr}\boldsymbol{E}_{e}}\longrightarrow \,^{t+\Delta t}\mathbb{C}%
_{ep}^{|tr}=\frac{d\,^{t+\Delta t}\boldsymbol{S}^{|tr}}{d^{tr}\boldsymbol{A}%
_{e}}\longrightarrow \,^{t+\Delta t}\mathbb{C}_{ep}=\frac{d\,^{t+\Delta t}%
\boldsymbol{S}}{d\,_{\hspace{3.2ex}0}^{t+\Delta t}\boldsymbol{A}}
\end{equation}%
where the first mapping performs the conversion to quadratic measures and
the second one transforms the tensor from the intermediate to the reference
configuration.

\section*{References}

\bibliographystyle{elsarticle-num}
\bibliography{reference_MJ}

\begin{thebibliography}{100}
\expandafter\ifx\csname url\endcsname\relax
  \def\url#1{\texttt{#1}}\fi
\expandafter\ifx\csname urlprefix\endcsname\relax\def\urlprefix{URL }\fi
\expandafter\ifx\csname href\endcsname\relax
  \def\href#1#2{#2} \def\path#1{#1}\fi

\bibitem{KhanHuangbook}
A.~S. Khan, S.~Huang, Continuum Theory of Plasticity, John Wiley \& Sons, New
  York, 1995.

\bibitem{Kangbook}
G.~Kang, Q.~Kan, Cyclic Plasticity of Engineering Materials, John Wiley \&
  Sons, New Jersey, 2017.

\bibitem{Bathebook}
K.-J. Bathe, Finite Element Procedures, 2nd Ed., Klaus-J\"urgen Bathe,
  Watertown, 2014.

\bibitem{Kojic2005}
M.~Kojic, K.-J. Bathe, {Inelastic Analysis of Solids and Structures},
  Springer-Verlag Berlin Heidelberg, 2005.

\bibitem{Rotersbook}
F.~Roters, P.~Eisenlohr, T.~R. Bieler, D.~Raabe, Crystal Plasticity Finite
  Element Methods, Wiley-VCH Verlag, Weinheim, 2010.

\bibitem{simo2006computational}
J.~C. Simo, T.~J. Hughes, Computational inelasticity, Vol.~7, Springer Science
  \& Business Media, 2006.

\bibitem{hughes1980finite}
T.~J. Hughes, J.~Winget, Finite rotation effects in numerical integration of
  rate constitutive equations arising in large-deformation analysis,
  International journal for numerical methods in engineering 15~(12) (1980)
  1862--1867.

\bibitem{rolph1984large}
W.~Rolph~III, K.-J. Bathe, On a large strain finite element formulation for
  elasto-plastic analysis, Constitutive Equations: Macro and Computational
  Aspects (1984).

\bibitem{simo1985computational}
J.~Simo, On the computational significance of the intermediate configuration
  and hyperelastic stress relations in finite deformation elastoplasticity,
  Mechanics of Materials 4~(3-4) (1985) 439--451.

\bibitem{simo1985unified}
J.~Simo, M.~Ortiz, A unified approach to finite deformation elastoplastic
  analysis based on the use of hyperelastic constitutive equations, Computer
  Methods in Applied Mechanics and Engineering 49~(2) (1985) 221--245.

\bibitem{bernstein1960hypo}
B.~Bernstein, Hypo-elasticity and elasticity, Archive for Rational Mechanics
  and Analysis 6~(1) (1960) 89--104.

\bibitem{bernstein1960relations}
B.~Bernstein, Relations between hypo-elasticity and elasticity, Transactions of
  the Society of Rheology 4~(1) (1960) 23--28.

\bibitem{Kroner}
E.~Kro\"oner, Allgemeine kontinuumstheorie der versetzungen und
  eigenspannungen, Archive for Rational Mechanics and Analysis 4 (1959) 273.

\bibitem{Lee1967}
E.~H. Lee, D.~T. Liu, {Finite-strain elastic - Plastic theory with application
  to plane-wave analysis}, Journal of Applied Physics 38~(1) (1967) 19--27.
\newblock \href {https://doi.org/10.1063/1.1708953}
  {\path{doi:10.1063/1.1708953}}.

\bibitem{simo1988framework1}
J.~C. Simo, A framework for finite strain elastoplasticity based on maximum
  plastic dissipation and the multiplicative decomposition: Part i. continuum
  formulation, Computer Methods in Applied Mechanics and Engineering 66~(2)
  (1988) 199--219.

\bibitem{simo1988framework2}
J.~C. Simo, A framework for finite strain elastoplasticity based on maximum
  plastic dissipation and the multiplicative decomposition. part ii:
  computational aspects, Computer Methods in Applied Mechanics and Engineering
  68~(1) (1988) 1--31.

\bibitem{Bilby55}
B.~Bilby, R.~Bullough, E.~Smith, Continuous distributions of dislocations: a
  new application of the methods of non-riemannian geometry, Proceedings of the
  Royal Society of London A 231 (1955) 263--273.

\bibitem{Vladimirov2009}
I.~N. Vladimirov, M.~P. Pietryga, S.~Reese, {Prediction of springback in sheet
  forming by a new finite strain model with nonlinear kinematic and isotropic
  hardening}, Journal of Materials Processing Technology 209~(8) (2009)
  4062--4075.

\bibitem{Vladimirov2010}
I.~N. Vladimirov, M.~P. Pietryga, S.~Reese, {Anisotropic finite
  elastoplasticity with nonlinear kinematic and isotropic hardening and
  application to sheet metal forming}, International Journal of Plasticity
  26~(5) (2010) 659--687.

\bibitem{Caminero2011}
M.~{\'{A}}. Caminero, F.~J. Mont{\'{a}}ns, K.~J. Bathe, {Modeling large strain
  anisotropic elasto-plasticity with logarithmic strain and stress measures},
  Computers and Structures 89~(11-12) (2011) 826--843.

\bibitem{neff2016loss}
P.~Neff, I.-D. Ghiba, Loss of ellipticity for non-coaxial plastic deformations
  in additive logarithmic finite strain plasticity, International Journal of
  Non-Linear Mechanics 81 (2016) 122--128.

\bibitem{shutov2014analysis}
A.~Shutov, J.~Ihlemann, Analysis of some basic approaches to finite strain
  elasto-plasticity in view of reference change, International Journal of
  Plasticity 63 (2014) 183--197.

\bibitem{Shutov2012}
A.~Shutov, S.~Pfeiffer, J.~Ihlemann, On the simulation of multi-stage forming
  processes: invariance under change of the reference configuration,
  Materialwissenschaft und Werkstofftechnik 43 (2012) 617--625.

\bibitem{Shutov15}
A.~Shutov, On exploiting the weak invariance of multiplicative
  elasto-plasticity for efficient numerical integration, in: COMPLAS
  XIII--Proceedings of the XIII International Conference on Computational
  Plasticity: fundamentals and applications, CIMNE, 2015, pp. 272--283.

\bibitem{Brepols2014}
T.~Brepols, I.~N. Vladimirov, S.~Reese,
  \href{http://dx.doi.org/10.1016/j.ijplas.2014.06.003}{{Numerical comparison
  of isotropic hypo- and hyperelastic-based plasticity models with application
  to industrial forming processes Dedicated to Prof. Dr.-Ing. Otto Timme Bruhns
  on the occasion of his 70th birthday}}, International Journal of Plasticity
  63 (2014) 18--48.
\newblock \href {https://doi.org/10.1016/j.ijplas.2014.06.003}
  {\path{doi:10.1016/j.ijplas.2014.06.003}}.
\newline\urlprefix\url{http://dx.doi.org/10.1016/j.ijplas.2014.06.003}

\bibitem{loblein2003application}
J.~L{\"o}blein, J.~Schr{\"o}der, F.~Gruttmann, Application of generalized
  measures to an orthotropic finite elasto-plasticity model, Computational
  Materials Science 28~(3-4) (2003) 696--703.

\bibitem{miehe1998formulation}
C.~Miehe, A formulation of finite elastoplasticity based on dual co-and
  contra-variant eigenvector triads normalized with respect to a plastic
  metric, Computer Methods in Applied Mechanics and Engineering 159~(3-4)
  (1998) 223--260.

\bibitem{miehe2002anisotropic}
C.~Miehe, N.~Apel, M.~Lambrecht, Anisotropic additive plasticity in the
  logarithmic strain space: modular kinematic formulation and implementation
  based on incremental minimization principles for standard materials, Computer
  Methods in Applied Mechanics and Engineering 191~(47-48) (2002) 5383--5425.

\bibitem{papadopoulos1998general}
P.~Papadopoulos, J.~Lu, A general framework for the numerical solution of
  problems in finite elasto-plasticity, Computer Methods in Applied Mechanics
  and Engineering 159~(1-2) (1998) 1--18.

\bibitem{papadopoulos2001formulation}
P.~Papadopoulos, J.~Lu, On the formulation and numerical solution of problems
  in anisotropic finite plasticity, Computer Methods in Applied Mechanics and
  Engineering 190~(37-38) (2001) 4889--4910.

\bibitem{sansour2003viscoplasticity}
C.~Sansour, W.~Wagner, Viscoplasticity based on additive decomposition of
  logarithmic strain and unified constitutive equations: Theoretical and
  computational considerations with reference to shell applications, Computers
  \& Structures 81~(15) (2003) 1583--1594.

\bibitem{Wilkins}
M.~Wilkins, Calculation of elastic-plastic flow, Tech. rep., California
  University-Livermore Radiation Lab. (1963).

\bibitem{weber1990finite}
G.~Weber, L.~Anand, Finite deformation constitutive equations and a time
  integration procedure for isotropic, hyperelastic-viscoplastic solids,
  Computer Methods in Applied Mechanics and Engineering 79~(2) (1990) 173--202.

\bibitem{eterovic1990hyperelastic}
A.~L. Eterovic, K.-J. Bathe, A hyperelastic-based large strain elasto-plastic
  constitutive formulation with combined isotropic-kinematic hardening using
  the logarithmic stress and strain measures, International Journal for
  Numerical Methods in Engineering 30~(6) (1990) 1099--1114.

\bibitem{simo1992algorithms}
J.~C. Simo, Algorithms for static and dynamic multiplicative plasticity that
  preserve the classical return mapping schemes of the infinitesimal theory,
  Computer Methods in Applied Mechanics and Engineering 99~(1) (1992) 61--112.

\bibitem{cuitino1992material}
A.~Cuitino, M.~Ortiz, A material-independent method for extending stress update
  algorithms from small-strain plasticity to finite plasticity with
  multiplicative kinematics, Engineering Computations 9~(4) (1992) 437--451.

\bibitem{SouzaNetoPericBook}
E.~A. de~Souza~Neto, D.~Peric, D.~R.~J. Owen, Computational Methods for
  Plasticity: Theory and Applications, John Wiley \& Sons, Chichester, 2008.

\bibitem{simo1992associative}
J.~Simo, C.~Miehe, Associative coupled thermoplasticity at finite strains:
  Formulation, numerical analysis and implementation, Computer Methods in
  Applied Mechanics and Engineering 98~(1) (1992) 41--104.

\bibitem{SimoChapter}
J.~C. Simo, Handbook of Numerical Analysis, North-Holland, Elsevier Science
  B.V., 1998, Ch. Numerical Analysis and Simulation of Plasticity; Ciarlet,
  P.G. and Lions, J.L. (Eds), pp. 183--499.

\bibitem{Miehe1996a}
C.~Miehe, {Exponential map algorithm for stress updates in anisotropic
  multiplicative elastoplasticity for single crystals}, International Journal
  for Numerical Methods in Engineering 39 (1996) 3367--3390.

\bibitem{Badreddine2010}
H.~Badreddine, K.~Saanouni, A.~Dogui, {On non-associative anisotropic finite
  plasticity fully coupled with isotropic ductile damage for metal forming},
  International Journal of Plasticity 26~(11) (2010) 1541--1575.

\bibitem{Ewing1899}
J.~A. Ewing, W.~Rosenhain, {Experiments in micro-metallography: Effects of
  strain, preliminary notice}, Proceedings of the Royal Society of London 65
  (1899) 85--90.

\bibitem{Ewing1900}
J.~A. Ewing, W.~Rosenhain, {The crustalline structure of metals}, Philosophical
  Transactions of the Royal Society 193 (1900) 353--375.

\bibitem{Taylor1923}
G.~I. Taylor, C.~F. Elam, {The distorsion of an aluminium crystal during a
  tensile test}, Proceeding of the Royal Society of London A: Mathematical,
  Physical and Engineering Sciences 102~(719) (1923) 643--667.

\bibitem{Taylor1925}
G.~I. Taylor, C.~F. Elam, {The plastic extension and fracture of aluminium
  crystals}, Proceedings of the Royal Society of London. Series A, Containing
  Papers of a Mathematical and Physical Character 108~(745) (1925) 28--51.
\newblock \href {https://doi.org/10.1098/rspa.1925.0057}
  {\path{doi:10.1098/rspa.1925.0057}}.

\bibitem{Taylor1938}
G.~I. Taylor, {Plastic strain in metals}, Journal of the Institute of Metals 62
  (1938) 307--324.

\bibitem{Bishop1951}
J.~Bishop, R.~Hill, {XLVI. A theory of the plastic distortion of a
  polycrystalline aggregate under combined stresses.}, The London, Edinburgh,
  and Dublin Philosophical Magazine and Journal of Science 42~(327) (1951)
  414--427.
\newblock \href {https://doi.org/10.1080/14786445108561065}
  {\path{doi:10.1080/14786445108561065}}.

\bibitem{Bishop1951a}
J.~Bishop, R.~Hill, {CXXVIII. A theoretical derivation of the plastic
  properties of a polycrystalline face-centred metal}, The London, Edinburgh,
  and Dublin Philosophical Magazine and Journal of Science 42~(334) (1951)
  1298--1307.
\newblock \href {https://doi.org/10.1080/14786444108561385}
  {\path{doi:10.1080/14786444108561385}}.

\bibitem{Rice1971}
J.~Rice, {Inelastic constitutive relations for solids: an internal-variable
  theory and its application to metal plasticity}, Journal of Mechanics and
  Physics of Solids 19 (1971) 433--455.
\newblock \href {https://doi.org/10.1038/nbt0910-877}
  {\path{doi:10.1038/nbt0910-877}}.

\bibitem{Hill1972}
R.~Hill, J.~R. Rice, {Constitutive analysis of elastic-plastic crystals at
  arbitrary strain}, Journal of the Mechanics and Physics of Solids 20~(6)
  (1972) 401--413.
\newblock \href {https://doi.org/10.1016/0022-5096(72)90017-8}
  {\path{doi:10.1016/0022-5096(72)90017-8}}.

\bibitem{Peirce1982}
D.~Peirce, R.~J. Asaro, A.~Needleman, {An analysis of nonuniform and localized
  deformation in ductile single crystals}, Acta Metallurgica 30~(6) (1982)
  1087--1119.
\newblock \href {https://doi.org/10.1016/0001-6160(82)90005-0}
  {\path{doi:10.1016/0001-6160(82)90005-0}}.

\bibitem{Peirce1983}
D.~Peirce, R.~J. Asaro, A.~Needleman, {Material rate dependence and localized
  deformation in crystalline solids}, Acta Metallurgica 31~(12) (1983)
  1951--1976.
\newblock \href {https://doi.org/10.1016/0001-6160(83)90014-7}
  {\path{doi:10.1016/0001-6160(83)90014-7}}.

\bibitem{Asaro1985}
R.~J. Asaro, A.~Needleman, J.~Lemonds, D.~Peirce, {Texture development and
  strain hardening in rate dependent polycrystals}, Acta Metallurgica 33 (1985)
  923--953.

\bibitem{Needleman1985}
A.~Needleman, R.~J. Asaro, J.~Lemonds, D.~Peirce, {Finite element analysis of
  crystalline solids}, Computer Methods in Applied Mechanics and Engineering
  52~(1-3) (1985) 689--708.
\newblock \href {https://doi.org/10.1016/0045-7825(85)90014-3}
  {\path{doi:10.1016/0045-7825(85)90014-3}}.

\bibitem{Rashid1992}
M.~M. Rashid, S.~Nemat-Nasser, {A constitutive algorithm for rate-dependent
  crystal plasticity}, Computer Methods in Applied Mechanics and Engineering
  94~(2) (1992) 201--228.
\newblock \href {https://doi.org/10.1016/0045-7825(92)90147-C}
  {\path{doi:10.1016/0045-7825(92)90147-C}}.

\bibitem{Tome1984}
C.~Tome, G.~R. Canova, U.~F. Kocks, N.~Christodoulou, J.~J. Jonas, {The
  relation between macroscopic and microscopic strain hardening in F.C.C.
  polycrystals}, Acta Metallurgica 32~(10) (1984) 1637--1653.
\newblock \href {https://doi.org/10.1016/0001-6160(84)90222-0}
  {\path{doi:10.1016/0001-6160(84)90222-0}}.

\bibitem{Bassani1991}
J.~L. Bassani, T.-Y. Wu, {Latent hardening in single crystals. II. Analytical
  characterization and predictions}, Proceedings of the Royal Society of
  London. Series A: Mathematical and Physical Sciences 435 (1991) 21--41.
\newblock \href {https://doi.org/10.1098/rspa.1991.0128}
  {\path{doi:10.1098/rspa.1991.0128}}.

\bibitem{Meric1991}
L.~M{\'{e}}ric, P.~Poubanne, G.~Cailletaud, {Single crystal modeling for
  structural calculations: Part 1-model presentation}, Journal of Engineering
  Materials and Technology, Transactions of the ASME 113~(1) (1991) 162--170.
\newblock \href {https://doi.org/10.1115/1.2903374}
  {\path{doi:10.1115/1.2903374}}.

\bibitem{Cailletaud1992}
G.~Cailletaud, {A micromechanical approach to inelastic behaviour of metals},
  International Journal of Plasticity 8~(1) (1992) 55--73.
\newblock \href {https://doi.org/10.1016/0749-6419(92)90038-E}
  {\path{doi:10.1016/0749-6419(92)90038-E}}.

\bibitem{Hasija2003}
V.~Hasija, S.~Ghosh, M.~J. Mills, D.~S. Joseph, {Deformation and creep modeling
  in polycrystalline Ti-6Al alloys}, Acta Materialia 51~(15) (2003) 4533--4549.
\newblock \href {https://doi.org/10.1016/S1359-6454(03)00289-1}
  {\path{doi:10.1016/S1359-6454(03)00289-1}}.

\bibitem{Venkatramani2007}
G.~Venkatramani, S.~Ghosh, M.~Mills, {A size-dependent crystal plasticity
  finite-element model for creep and load shedding in polycrystalline titanium
  alloys}, Acta Materialia 55~(11) (2007) 3971--3986.
\newblock \href {https://doi.org/10.1016/j.actamat.2007.03.017}
  {\path{doi:10.1016/j.actamat.2007.03.017}}.

\bibitem{Cruzado2017}
A.~Cruzado, J.~LLorca, J.~Segurado, {Modeling cyclic deformation of inconel 718
  superalloy by means of crystal plasticity and computational homogenization},
  International Journal of Solids and Structures 122-123 (2017) 148--161.
\newblock \href {https://doi.org/10.1016/j.ijsolstr.2017.06.014}
  {\path{doi:10.1016/j.ijsolstr.2017.06.014}}.

\bibitem{Arsenlis2004}
A.~Arsenlis, D.~M. Parks, R.~Becker, V.~V. Bulatov, {On the evolution of
  crystallographic dislocation density in non-homogeneously deforming
  crystals}, Journal of the Mechanics and Physics of Solids 52~(6) (2004)
  1213--1246.
\newblock \href {https://doi.org/10.1016/j.jmps.2003.12.007}
  {\path{doi:10.1016/j.jmps.2003.12.007}}.

\bibitem{Cheong2004}
K.~S. Cheong, E.~P. Busso, {Discrete dislocation density modelling of single
  phase FCC polycrystal aggregates}, Acta Materialia 52~(19) (2004) 5665--5675.
\newblock \href {https://doi.org/10.1016/j.actamat.2004.08.044}
  {\path{doi:10.1016/j.actamat.2004.08.044}}.

\bibitem{Ma2004}
A.~Ma, F.~Roters, {A constitutive model for fcc single crystals based on
  dislocation densities and its application to uniaxial compression of
  aluminium single crystals}, Acta Materialia 52~(12) (2004) 3603--3612.
\newblock \href {https://doi.org/10.1016/j.actamat.2004.04.012}
  {\path{doi:10.1016/j.actamat.2004.04.012}}.

\bibitem{Ma2006}
A.~Ma, F.~Roters, D.~Raabe, {A dislocation density based constitutive model for
  crystal plasticity FEM including geometrically necessary dislocations}, Acta
  Materialia 54~(8) (2006) 2169--2179.
\newblock \href {https://doi.org/10.1016/j.actamat.2006.01.005}
  {\path{doi:10.1016/j.actamat.2006.01.005}}.

\bibitem{Ma2006a}
A.~Ma, F.~Roters, D.~Raabe, {On the consideration of interactions between
  dislocations and grain boundaries in crystal plasticity finite element
  modeling - Theory, experiments, and simulations}, Acta Materialia 54~(8)
  (2006) 2181--2194.
\newblock \href {https://doi.org/10.1016/j.actamat.2006.01.004}
  {\path{doi:10.1016/j.actamat.2006.01.004}}.

\bibitem{Dunne2007}
F.~P. Dunne, D.~Rugg, A.~Walker, {Lengthscale-dependent, elastically
  anisotropic, physically-based hcp crystal plasticity: Application to
  cold-dwell fatigue in Ti alloys}, International Journal of Plasticity 23~(6)
  (2007) 1061--1083.
\newblock \href {https://doi.org/10.1016/j.ijplas.2006.10.013}
  {\path{doi:10.1016/j.ijplas.2006.10.013}}.

\bibitem{Rodriguez-Galan2015}
D.~Rodr{\'{i}}guez-Gal{\'{a}}n, I.~Sabirov, J.~Segurado, {Temperature and stain
  rate effect on the deformation of nanostructured pure titanium},
  International Journal of Plasticity 70 (2015) 191--205.
\newblock \href {https://doi.org/10.1016/j.ijplas.2015.04.002}
  {\path{doi:10.1016/j.ijplas.2015.04.002}}.

\bibitem{Shahba2016}
A.~Shahba, S.~Ghosh,
  \href{http://dx.doi.org/10.1016/j.ijplas.2016.09.002}{{Crystal plasticity FE
  modeling of Ti alloys for a range of strain-rates. Part I: A unified
  constitutive model and flow rule}}, International Journal of Plasticity 87
  (2016) 48--68.
\newblock \href {https://doi.org/10.1016/j.ijplas.2016.09.002}
  {\path{doi:10.1016/j.ijplas.2016.09.002}}.
\newline\urlprefix\url{http://dx.doi.org/10.1016/j.ijplas.2016.09.002}

\bibitem{Berdichevsky}
V.~L. Berdichevsky, Continuum theory of dislocation revisited, Continuum
  Mechanics and Thermodynamics 18 (2006) 195--222.

\bibitem{Forhmeister}
V.~Fohrmeister, G.~D\'iaz, J.~Mosler, Classic crystal plasticity theory vs
  crystal plasticity based on strong discontinuities---theoretical and
  algorithmic aspects., International Journal for Numerical Methods in
  Engineering 117 (2019) 1283--1303.

\bibitem{Cuitino1993}
A.~M. Cuiti{\~{n}}o, M.~Ortiz, {Computational modelling of single crystals},
  Modelling and Simulation in Materials Science and Engineering 1~(3) (1993)
  225--263.
\newblock \href {https://doi.org/10.1088/0965-0393/1/3/001}
  {\path{doi:10.1088/0965-0393/1/3/001}}.

\bibitem{Borja1993}
R.~I. Borja, J.~R. Wren, {Discrete micromechanics of elastoplastic crystals},
  International Journal for Numerical Methods in Engineering 36~(22) (1993)
  3815--3840.
\newblock \href {https://doi.org/10.1002/nme.1620362205}
  {\path{doi:10.1002/nme.1620362205}}.

\bibitem{Kalidindi1992}
S.~Kalidindi, C.~Bronkhorst, L.~Anand, {Crystallographic texture evolution in
  bulk deformation processing of FCC metals}, Journal of Mechanics and Physics
  of Solids 40~(3) (1992) 537--569.
\newblock \href {https://doi.org/10.1007/BF01176030}
  {\path{doi:10.1007/BF01176030}}.

\bibitem{Miehe1996}
C.~Miehe, {Multisurface thermoplasticity for single crystals at large strains
  in terms of Eulerian vector updates}, International Journal of Solids and
  Structures 33~(20-22) (1996) 3103--3130.
\newblock \href {https://doi.org/10.1016/0020-7683(95)00274-X}
  {\path{doi:10.1016/0020-7683(95)00274-X}}.

\bibitem{Kuroda}
M.~Kuroda, On large-strain finite element solutions of higher order gradient
  crystal plasticity, International Journal of Solids and Structures 48 (2011)
  3382--3394.

\bibitem{Izadbakhsh}
A.~Izadbakhsh, K.~Inal, R.~K. Mishra, M.~Niewczas, New crystal plasticity
  constitutive model for large strain deformation in single crystals of
  magnesium, Computational Materials Science 20 (2011) 2185--2202.

\bibitem{Zhou19}
R.~Zhou, A.~Roy, V.~V. Silberschmidt, A crystal-plasticity model of extruded
  am30 magnesium alloy, Computational Materials science 170 (2019) 109140.

\bibitem{Sakaguchi}
M.~Sakaguchi, R.~Komamura, X.~Chen, M.~Higaki, H.~Inoue, Crystal plasticity
  assessment of crystallographic {S}tage {I} crack propagation in a {N}i-based
  single crystal superalloy, International Journal of Fatigue 123 (2019)
  10--21.

\bibitem{Zecevic}
M.~Zezevic, M.~Knezevic, An implicit formulation of the elasto-plastic
  self-consistent polycrystal plasticity model and its implementation in
  implicit finite elements, Mechanics of Materials 136 (2019) 103065.

\bibitem{KimKim}
J.~H. Kim, D.~Kim, F.~Barlat, M.-G. Lee, Crystal plasticity approach for
  predicting the bauschinger effect in dual-phase steels, Materials Science and
  Engineerinng A 539 (2012) 259--270.

\bibitem{Li19}
J.~Li, I.~Romero, J.~Segurado, Development of a thermo-mechanically coupled
  crystal plasticity modeling framework: Application to polycrystalline
  homogenization, International Jounal of Plasticity 119 (2019) 313--330.

\bibitem{Lu20}
X.~Lu, J.~Zhao, Z.~Wang, B.~Gan, J.~Zhao, G.~Kang, X.~Zhang, Crystal plasticity
  finite element analysis of gradient nanostructured twip steel, International
  Jounal of Plasticity 130 (2020) 102703.

\bibitem{Guo20}
H.-J. Guo, C.~Ling, E.~Busso, Z.~Zhong, D.-F. Li, Crystal plasticity based
  investigation of micro-void evolution under multi-axial loading conditions,
  International Journal of Plasticity 129 (2020) 102673.

\bibitem{KaiserMenzel}
T.~Kaiser, A.~Menzel, A dislocation density tensor-based crystal plasticity
  framework, Journal of the Mechanics and Physics of Solids 131 (2019)
  276--302.

\bibitem{Anand79}
L.~Anand, On {H.} hencky's approximate strain-energy function for moderate
  deformations, Journal of Applied Mechanics 46~(78-82) (1979).

\bibitem{Anand86}
L.~Anand, Moderate deformations in extension-torsion of incompressible
  isotropic elastic materials, Journal of the Mechanics and Physics of Solids
  34 (1986) 293--304.

\bibitem{Baudoin19}
P.~Baudoin, T.~Hama, H.~Takuda, Influence of critical resolved shear stress
  ratios on the response of a commercially pure titanium oligocrystal: crystal
  plasticity simulations and experiment, International Journal of Plasticity
  115 (2019) 111--131.

\bibitem{Farooq20}
H.~Farooq, G.~Cailletaud, S.~Forest, D.~Ryckelynck, Crystal plasticity modeling
  of the cyclic behavior of polycrystalline aggregates under non-symmetric
  uniaxial loading: Global and local analyses, International Journal of
  Plasticity 126 (2020) 102619.

\bibitem{latorreAMP}
M.~Latorre, F.~J. Mont\'ans, A new class of plastic flow evolution equations
  for anisotropic multiplicative elastoplasticity based on the notion of a
  corrector elastic strain rate, Applied Mathematical Modelling 55 (2018)
  716--740.

\bibitem{latorre2015anisotropic}
M.~Latorre, F.~J. Mont{\'a}ns, Anisotropic finite strain viscoelasticity based
  on the sidoroff multiplicative decomposition and logarithmic strains,
  Computational Mechanics 56~(3) (2015) 503--531.

\bibitem{latorre2016fully}
M.~Latorre, F.~J. Mont{\'a}ns, Fully anisotropic finite strain viscoelasticity
  based on a reverse multiplicative decomposition and logarithmic strains,
  Computers \& Structures 163 (2016) 56--70.

\bibitem{Sanz2017}
M.~Sanz, F.~J. Mont{\'{a}}ns, M.~Latorre, {Computational anisotropic hardening
  multiplicative elastoplasticity based on the corrector elastic logarithmic
  strain rate}, Computer Methods in Applied Mechanics and Engineering 320
  (2017) 82--121.

\bibitem{zhang2019simple}
M.~Zhang, F.~J. Mont{\'a}ns, A simple formulation for large-strain cyclic
  hyperelasto-plasticity using elastic correctors. theory and algorithmic
  implementation, International Journal of Plasticity 113 (2019) 185--217.

\bibitem{nguyen2019}
K.~Nguyen, M.~Sanz, F.~J. Mont\'ans, Plane-stress constrained multiplicative
  hyperelasto-plasticity with nonlinear kinematic hardening. consistent theory
  based on elastic corrector rates and algorithmic implementation,
  International Journal of Plasticity 128 (2020) 102592.
\newblock \href {https://doi.org/10.1016/j.ijplas.2019.08.017}
  {\path{doi:10.1016/j.ijplas.2019.08.017}}.

\bibitem{Kalidindi1993}
S.~Kalidindi, L.~Anand, {Large deformation simple compression of a copper
  single crystal}, Metallurgical Transactions A 24~(A) (1993) 989--992.

\bibitem{KalidindiKotari}
L.~Anand, M.~Kothari, A computational procedure for rate-independent crystal
  plasticity, Journal of the Mechanics and Physics of Solids 44 (1996)
  525--558.

\bibitem{DvorkinBook}
E.~N. Dvorkin, M.~B. Goldschmit, Nonlinear Continua, Springer-Verlag, Berlin
  Heidelberg, 2006.

\bibitem{latorre2016stress}
M.~Latorre, F.~J. Mont{\'a}ns, Stress and strain mapping tensors and general
  work-conjugacy in large strain continuum mechanics, Applied Mathematical
  Modelling 40~(5-6) (2016) 3938--3950.

\bibitem{xiao1997hypo}
H.~Xiao, O.~Bruhns, A.~Meyers, Hypo-elasticity model based upon the logarithmic
  stress rate, Journal of Elasticity 47~(1) (1997) 51--68.

\bibitem{XiaoChen}
H.~Xiao, L.~Chen, Hencky's elasticity model and linear stress-strain relations
  in isotropic finite hyperelasticity, Acta Mechanica 157 (2002) 51--60.

\bibitem{CrespoLatMon}
J.~Crespo, M.~Latorre, F.~Mont\'{a}ns, {WYPIWYG} hyperelasticity for isotropic,
  compressible materials, Computational Mechanics 59 (2017) 73--92.

\bibitem{CrespoAuxetic}
J.~Crespo, F.~Mont\'{a}ns, A continuum approach for the large strain finite
  element analysis of auxetic materials, International Journal of the
  Mechanical Sciences 135 (2018) 441--457.

\bibitem{CrespoIJES}
J.~Crespo, F.~Mont\'{a}ns, General solution procedures to compute the stored
  energy density of conservative solids directly from experimental data,
  International Journal of Engineering Science 141~(16-34) (2019).

\bibitem{RomeroLatorreMontans}
X.~Romero, M.~Latorre, F.~Mont\'{a}ns, Determination of the wypiwyg strain
  energy density of skin through finite element analysis of the experiments on
  circular specimens, Finite Elements in Analysis and Design 134~(1-15) (2017).

\bibitem{Truesdell-Nopll}
C.~Truesdell, W.~Noll, The Nonlinear Field Theories of Mechanics (3rd Edition),
  Springer-Verlag, Berlin, 2004.

\bibitem{MontansMRC}
F.~Mont\'{a}ns, J.~Ben\'{i}tez, M.~A. Caminero, A large strain anisotropic
  elastoplastic continuum theory for nonlinear kinematic hardening and texture
  evolution, Mechanics Research Communications 43 (2012) 50--56.

\bibitem{RotersReview}
F.~Roters, P.~Eisenlohr, L.~Hantcherli, D.~Tjahjanto, T.~Bieler, D.~Raabe,
  Overview of constitutive laws, kinematics, homogenization and multiscale
  methods in crystal plasticity finite-element modeling: Theory, experiments,
  applications, Acta Materialia 58 (2010) 1152--1211.

\bibitem{Latorre_logstrains}
M.~Latorre, F.~Mont\'{a}ns, On the interpretation of the logarithmic strain
  tensor in an arbitrary system of representation, International Journal of
  Solids and Structures 51 (2014) 1507--1515.

\bibitem{Miehe2004}
C.~Miehe, J.~Schotte, {Anisotropic finite elastoplastic analysis of shells:
  Simulation of earing in deep-drawing of single- and polycrystalline sheets by
  Taylor-type micro-to-macro transitions}, Computer Methods in Applied
  Mechanics and Engineering 193~(1-2) (2004) 25--57.
\newblock \href {https://doi.org/10.1016/j.cma.2003.07.012}
  {\path{doi:10.1016/j.cma.2003.07.012}}.

\bibitem{BorjaBook}
R.~I. Borja, Plasticity: Modeling \& Computation, Springer-Verlag, Berlin,
  2013.

\bibitem{Petryk}
H.~Petryk, M.~Kursa, Incremental work minimization algorithm for
  rate-independent plasticity of single crystals, International Journal for
  Numerical Methods in Engineering 104 (2015) 157--184.

\bibitem{SchroderMiehe}
J.~Schr\"oder, C.~Miehe, Aspects of computational rate-independent crystal
  plasticity, Computational Materials Science 9 (1997) 168--176.

\bibitem{Hutchinson76}
J.~W. Hutchinson, Bounds of self-consistent estimates for creep of
  polycrystalline materials, Proceedings of the Royal Society of London, A 348
  (1976) 101--127.

\bibitem{Pierce83}
D.~Pierce, R.~Asaro, A.~Needleman, Material rate dependence and localized
  deformation in crystalline solids, Acta Metallurgica 31~(12) (1983)
  1951--1976.

\bibitem{PanRice83}
J.~Pan, J.~Rice, Rate sensitivity of plastic flow and implications for
  yield-surface vertices, International Journal of Solids and Structures
  19~(11) (1983) 973--987.

\bibitem{Pierce82}
D.~Pierce, R.~Asaro, A.~Needleman, An analysis of nonuniform and localized
  deformation in ductile single crystals, Acta Metallurgica 30~(6) (1982)
  1087--1119.

\bibitem{Minano}
M.~Mi\~nano, F.~Mont\'ans, {WYPiWYG} damage mechanics for soft materials: A
  data-driven approach, Archives of Computational Methods in Engineering 25
  (2018) 165--193.

\bibitem{Ma2015}
S.~Gao, M.~Fivel, A.~Ma, A.~Hartmaier, {Influence of misfit stresses on
  dislocation glide in single crystal superalloys: A three-dimensional discrete
  dislocation dynamics study}, Journal of the Mechanics and Physics of Solids
  76 (2015) 276--290.
\newblock \href {https://doi.org/10.1016/j.jmps.2014.11.015}
  {\path{doi:10.1016/j.jmps.2014.11.015}}.

\bibitem{Gallier2002}
J.~Gallier, D.~Xu, Computing exponentials of skew-symmetric matrices and
  logarithms of orthogonal matrices, International Journal of Robotics and
  Automation 17~(4) (2002) 1--11.

\bibitem{Dettmer2004}
W.~Dettmer, S.~Reese, {On the theoretical and numerical modelling of
  Armstrong-Frederick kinematic hardening in the finite strain regime},
  Computer Methods in Applied Mechanics and Engineering 193~(1-2) (2004)
  87--116.
\newblock \href {https://doi.org/10.1016/j.cma.2003.09.005}
  {\path{doi:10.1016/j.cma.2003.09.005}}.

\end{thebibliography}
% name your BibTeX data base

\end{document}